\documentclass[%
 reprint,
superscriptaddress,
 amsmath,amssymb,
 aps,
]{revtex4-2}

\usepackage{graphicx}% Include figure files
\usepackage{dcolumn}% Align table columns on decimal point
\usepackage{bm}% bold math
\usepackage{hyperref}% add hypertext capabilities
\hypersetup{
    colorlinks=true,
    linkcolor= blue,
    filecolor= black, 
    citecolor = blue,
    urlcolor= blue,
    pdftitle={Overleaf Example},
    pdfpagemode=FullScreen,
    }
\usepackage{xcolor}
\usepackage{comment}
\usepackage{subfigure}
\usepackage[utf8]{inputenc}
\usepackage{upgreek}
\usepackage{physics}
\usepackage{multirow}
\UseRawInputEncoding

\begin{document}

\title{Quantifying electron-nuclear spin entanglement dynamics in central-spin systems using one-tangles}

\author{Isabela Gnasso}
\email{ignasso@vt.edu}
\affiliation{Department of Physics, Virginia Tech, Blacksburg, Virginia 24061, USA}
\affiliation{Virginia Tech Center for Quantum Information Science and Engineering,
Blacksburg, Virginia 24061, USA}
\author{Khadija Sarguroh}
\affiliation{Department of Engineering Science, University of Oxford, Parks Road, Oxford OX1 3PJ, United Kingdom}
\author{Dorian Gangloff}
\affiliation{Cavendish Laboratory, University of Cambridge, J.J. Thomson Avenue, Cambridge, CB3 0HE, UK}
\author{Sophia E. Economou}
\affiliation{Department of Physics, Virginia Tech, Blacksburg, Virginia 24061, USA}
\affiliation{Virginia Tech Center for Quantum Information Science and Engineering,
Blacksburg, Virginia 24061, USA}
\author{Edwin Barnes}
\email{efbarnes@vt.edu}
\affiliation{Department of Physics, Virginia Tech, Blacksburg, Virginia 24061, USA}
\affiliation{Virginia Tech Center for Quantum Information Science and Engineering,
Blacksburg, Virginia 24061, USA}

\date{\today}

\begin{abstract}
    Optically-active solid-state systems such as self-assembled quantum dots, rare-earth ions, and color centers in diamond and SiC are promising candidates for quantum network, computing, and sensing applications. Although the nuclei in these systems naturally lead to electron spin decoherence, they can be repurposed, if they are controllable, as long-lived quantum memories. Prior work showed that a metric known as the one-tangling power can be used to quantify the entanglement dynamics of sparse systems of spin-$1/2$ nuclei coupled to color centers in diamond and SiC. Here, we generalize these findings to a wide range of electron-nuclear central-spin systems, including those with spin $ >1/2$ nuclei, such as in III-V quantum dots (QDs), rare-earth ions, and some color centers. Focusing on the example of an (In)GaAs QD, we offer a procedure for pinpointing physically realistic parameter regimes that yield maximal entanglement between the central electron and surrounding nuclei. We further harness knowledge of naturally-occurring degeneracies and the tunability of the system to generate maximal entanglement between target subsets of spins when the QD electron is subject to dynamical decoupling. We also leverage the one-tangling power as an exact and immediate method for computing QD electron spin dephasing times with and without the application of spin echo sequences, and use our analysis to identify coherence-sustaining conditions within the system. 
\end{abstract}

\maketitle

\section{Introduction}

Spin-photon interfaces are a crucial component in many approaches to quantum computing, networks, and sensing. Solid-state systems such as color centers in diamond \cite{bernien2013heralded, pfaff2014unconditional, iwasaki2017tin, kalb2017entanglement, nguyen2019integrated, nguyen2019quantum, trusheim2020transform, pompili2021realization, rugar2021quantum} or SiC \cite{widmann2015coherent, nagy2019high, bourassa2020entanglement, babin2022fabrication, ecker2024quantum}, rare-earth ions \cite{wu2023near, ruskuc2025multiplexed} and electron spins in semiconductor quantum dots (QDs) \cite{stockill2017phase, chekhovich2020nuclear, bodey2019optical} are particularly promising for their photon emission capabilities \cite{ding2016demand, buckley2012engineered, gao2015coherent} and the relatively long coherence times of their proximal spins. Recent experimental techniques have achieved reliable emission of indistinguishable photons along with optical control of a central QD electron \cite{koong2025coherent}. Several works also demonstrated entanglement generation between spatially-separated nodes \cite{hermans2022qubit, stas2022robust, delteil2016generation, bhaskar2017quantum} and production of multi-photon entangled states from quantum emitters \cite{lago2023long, krutyanskiy2024multimode, hartung2024quantum, minnella2025single-gate}, with recent developments in fabrication \cite{cheng2024laser, wu2022hybrid, yu2023telecom, schimpf2025optical} and efficient heralding techniques \cite{hermans2023entangling, ruskuc2024scalable} that promote further scalability and robustness of network technologies. Optically active solid-state spin systems are also widely used for quantum sensing \cite{abobeih2019atomic, gross2017real, van2024mapping, zhou2020quantum, maze2008nanoscale, degen2017quantum, van2025improved}. All these applications either require \cite{bradley2022robust} or benefit from \cite{zaiser2016enhancing} quantum memories, which can be realized using proximal nuclear spins in the surrounding material. 

There is an extensive collection of works that study and model the interplay between nuclear spin dynamics and central spin dephasing due to hyperfine interactions \cite{schlosshauer2007quantum, stockill2016quantum, maziero2010system, kleinjohann2018magnetic, gammon1996fine, gammon2001electron, urbaszek2013nuclear, khaetskii2002electron, merkulov2002electron, khaetskii2003electron, grishin2005low, johnson2005triplet, witzel2006quantum, yao2006theory, lutchyn2008quantum, cywinski2009electron, cywinski2009pure, huang2010theoretical, chekhovich2013element, chekhovich2013nuclear, welander2014influence, bechtold2015three, wust2016role, krzywda2024decoherence, roszak2021qubit, witzel2012quantum, de2003theory, hanson2007spins, bortz2007exact, latta2011hyperfine, hogele2012dynamic, barnes2011master, barnes2012nonperturbative}. These approaches all agree that fluctuations in nuclear spin orientations effectively shift the electron spin precession frequency and rapidly lead to dephasing when left untreated. However, there is also well-supported evidence that this destructive effect can be mitigated, and the nuclei can potentially serve as quantum memories provided one has sufficient control over them \cite{taylor2003controlling, erlingsson2001nuc, shenvi2005universal, klauser2006nuclear, koppens2006driven, vink2009locking, ladd2010pulsed, chekhovich2010dynamics, chekhovich2010pumping, krebs2010anomalous, cywinski2010dephasing, kloeffel2011controlling, yang2012collective, munsch2014manipulation, chekhovich2015suppression, beugeling2016quantum, prechtel2016decoupling, abobeih2018one, sharma2019enhancement, khaetskii2001spin, reiner2024high, bartling2022entanglement, maurer2012room, press2010ultrafast, liu2010quantum, burkard2023semiconductor, hung2013hyperfine, cywinski2014dynamical, bragar2015dynamics}. Methods have been established for collectively polarizing nuclear spins via periodic trains of pulses \cite{greilich2006mode,greilich2007robust,petta2008dynamic, schering2021nuclear, economou2014theory} which result in coherent feedback on the central electron, as well as decoupling spins by applying spin echo sequences \cite{petta2005coherent, bluhm2011dephasing, malinowski2017notch, Taminiau, medford2012scaling, zaporski2023ideal} that protect against unwanted interactions. More recently, preparation and control of collective nuclear states has been demonstrated for a variety of platforms for the purpose of developing functional quantum memories \cite{senkalla2024germanium, ruskuc2022nuclear, bradley2022robust, cai2025formation}. Specifically for GaAs QDs, breakthroughs have been made with single-excitation states \cite{gangloff2019quantum, denning2019collective, gangloff2021witnessing, jackson2021quantum, jackson2022optimal, zaporski2023many} that have been implemented for an electron-mediated, write-store-retrieve-readout protocol characterizing a working nuclear memory \cite{shofer2024tuning, appel2025many}. Memory qubits have also been harnessed for reaching milestones in network generation \cite{knaut2024entanglement}. 

Understanding and controlling electron-nuclear spin interactions is challenging because we have relatively few tools to study many-body entanglement dynamics in such systems. Recently, it was shown that an entanglement metric called the one-tangling power provides a useful method to characterize electron-nuclear dynamics in color centers coupled to spin-$1/2$ nuclei \cite{linowski2020entangling, Eva2023, dakis2024high}. However, many important systems involve higher-spin nuclei, such as QDs in GaAs or InGaAs, a qubit in $^{171}$Yb$^{3+}$ coupled to $^{51}$V$^{5+}$ ions \cite{ruskuc2022nuclear}, and Ge defects in diamond \cite{senkalla2024germanium}, and so existing results for the one-tangling power do not apply to these cases. 

In this paper, we derive general formulas for the one-tangling power of central-spin systems in which the nuclei have arbitrary total spin. This formalism covers a large class of physical systems, including color centers coupled to nuclei via dipolar hyperfine interactions and electron spins in strained QDs with hyperfine and nuclear quadrupolar interactions, as well as those with electronic g-factor anisotropy. Focusing on the latter system, we use the one-tangling power to study the growth of electron-nuclear entanglement in QDs with and without dynamical decoupling, and we identify the conditions under which maximal or minimal entanglement is achieved. We further leverage this entanglement metric to study hyperfine-induced decoherence of the electron spin. Our results show that the one-tangling power allows us to access detailed information about many-body entanglement dynamics in central-spin systems that play an important role in quantum information technologies.

The paper is organized as follows. In Sec.\ \ref{sec:Quantifying entanglement} we introduce the existing formalism for quantifying entanglement dynamics in central-spin systems via the one-tangling power. We then present our derived expressions for the one-tangling power that quantify the amount of entanglement that can be generated by a given unitary acting on a system of arbitrary size and total spin in Sec.\ \ref{Sec:generalizedexprs}. We give the explicit form of these one-tangling power expressions when applied to the example of a singly-charged QD in Sec.\ \ref{Sec:QDexample}. We demonstrate the use of these expressions for studying the growth of entanglement in this QD when it is freely evolving in Sec.\ \ref{Sec:freeevolution}. This section includes analysis of the effects of inhomogeneities in the system on the growth of entanglement between spins, as well as an interpretation of these correlations in the context of electron spin decoherence. In Sec.\ \ref{Sec:ddsequences}, we study the one-tangling power when dynamical decoupling sequences are applied to the electronic spin and we show how to identify control parameters that yield desired entanglement between a single spin and the rest of the ensemble when the electron is subject to simple $\pi$ pulse sequences. This section ends with a study of the application of additional iterations of the Carr-Purcell-Meiboom-Gill sequence in order to grow entanglement between the electron and a target subset of nuclei. We conclude in Sec.\ \ref{Sec:conclusion}. 

\section{Quantifying entanglement between spins}
\label{sec:Quantifying entanglement}

In this section, we introduce the one-tangles formalism and present expressions for analyzing many-body entanglement dynamics via the one-tangling power. These expressions can immediately be applied as they are given here to a wide range of physical platforms. In subsequent sections, we demonstrate the use of these expressions via the example of a strained QD to show the applicability of the one-tangling power to systems subject to collinear and non-collinear hyperfine, as well as quadrupolar interactions \cite{gangloff2019quantum}.

As previously mentioned, we study entanglement using the metric known as the one-tangling power, which provides dynamic insight into the correlations that can arise across a quantum system  \cite{zanardi2000entangling, linowski2020entangling}. It involves the averaging of an entanglement measure, called the one-tangle, over all possible initial product states for a given bipartition of the whole system \cite{zanardi2000entangling}. The one-tangle is defined in this case by the linear entropy, or one minus the subsystem purity. For a pure state $|\psi\rangle$ and a bipartition between subsystems $p$ and $q$, this is given as $\tau_{p|q}(|\psi\rangle) = 1 - \text{Tr}[\rho_{q}^2]$, where $\rho_{q} = \text{Tr}_{p}[|\psi\rangle\langle \psi|]$ \cite{zanardi2000entangling}. Note that for our purposes, we choose to omit the overall factor of $2$ typically included in the linear entropy. We are interested in the general entangling capabilities of the evolution operator, so we can instead consider the one-tangle of the unitary of interest, $U$, as it acts on a generic initial state, $|\Psi\rangle =  \bigotimes_i |\psi_i\rangle = \bigotimes_iV_i|\psi_{0}\rangle$ \cite{linowski2020entangling}. This general product state $|\Psi\rangle$ is defined in terms of single-qubit unitaries acting on a fixed initial state. These single-qubit unitaries purely serve the purpose of generating all locally equivalent possibilities for our initial product state. This grants us generality in the use of the one-tangling power for quantifying the entanglement that can be generated by a given evolution, regardless of how the system is initialized. 

We obtain the one-tangling power by averaging the one-tangle of the time-evolved state over all possible initial product states: $\epsilon_{p|q}(U) = \langle \tau_{p|q}( U|\Psi\rangle) \rangle_{V_i} $ \cite{linowski2020entangling}. $\epsilon_{p|q}(U)$ quantifies the ability of $U$ to create entanglement between subsystems $p$ and $q$. It vanishes for unitaries that cannot create entanglement, while for maximally entangling unitaries, it attains a finite maximal value that depends on the sizes of $p$ and $q$ and on the way in which $U$ couples these subsystems. It can be expressed as \cite{linowski2020entangling}
\begin{equation}
\label{eq:epq}
\begin{aligned}
    \hspace{-5 pt} \epsilon_{p|q}(U) = 1 - \big(\prod_{i = 0}^n \frac{d_i}{d_i + 1}\big)\sum_{x'|y'}\text{Tr}\{(\text{Tr}_{px'}[|U\rangle \langle U|])^2\}, \\
\end{aligned}
\end{equation}
\noindent where $i = 0,...,n$ labels individual subsystems (e.g., single spins) each of dimension $d_i$, and $U$ has been vectorized through the Choi-Jamio\l kowski isomorphism \cite{choi1975completely, jamiolkowski1972linear}. In accordance with this basis-independent mapping, $|U\rangle$ exists in a composite Hilbert space made up of the space of the real spin system and its identical copy ($|U\rangle \in \mathcal{H}\otimes\mathcal{H'}$). We can utilize this extended Hilbert space to compute the one-tangling power by considering all possible divisions across $\mathcal{H}\otimes\mathcal{H'}$. The bipartition $p|q$ indicates which part of the real subsystem has been isolated from the rest \cite{linowski2020entangling, Eva2023}. Specifically, the possibilities for $p|q$ include the ordered divisions in which $q$ is a single spin separated from the remaining ones, denoted by $p$ (e.g., for a three-spin system, $p|q \in \{01|2, 12|0, 02|1\}$). Note that we label the central electronic spin as $0$ and the remaining nuclear spins by $1...n$. In order to properly account for all delineations of the entire Hilbert space, the sum over $x'|y'$ considers all possible unordered bipartitions over $\mathcal{H'}$ \cite{linowski2020entangling, Eva2023}. For the three-spin case, $x'|y' \in \{.|0'1'2', 0'|1'2', 0'1'|2', 0'1'2'|., 1'|0'2', 2'|0'1', 0'2'|1', 1'2'|0'\},$ where ``." represents an empty subset of spins. 

Previous works \cite{Eva2023, dakis2024high} have demonstrated the tractability of the one-tangling power as an entanglement measure for electron-nuclear central-spin systems in color centers in diamond and SiC, respectively. Crucially, the Hamiltonians of these systems possess the general structure 
\begin{equation}
    H = \sigma_{00} \otimes H_0 + \sigma_{11}\otimes H_{1},
    \label{eq:Hgen}
\end{equation}
\noindent where $\sigma_{jj} = |j\rangle\langle j|$ ($j\in 0,1$) projects onto the electronic spin states. This prior work showed that for block-diagonal unitary evolutions that arise from Eq.\ (\ref{eq:Hgen}) with and without additional $\delta$-function pulse sequences applied to the central spin \cite{Eva2023}, it follows that the one-tangling power is linearly related to the first Makhlin invariant $G_1$, e.g., $\epsilon_{p|q} = \frac{2}{9}(1 - G_1)$ \cite{Eva2023, dakis2024high, balakrishnan2010entangling}. These unitaries can be written in the form of $U = \sigma_{00} \otimes R_{n_0} + \sigma_{11} \otimes R_{n_1}$, where $R_{n_i}$ represents a conditional nuclear spin rotation described by axis $\mathbf{n_i}$ \cite{Eva2023}. This previous result applies to systems containing spin-$1/2$ nuclei, in which case $G_1$ is the usual Makhlin invariant defined for two \emph{qubits} \cite{kraus2001optimal,makhlin2002nonlocal}. However, the existence of important higher-spin systems calls for a more general entanglement metric. In what follows, we will introduce a generalized version of the Makhlin invariant that applies to \emph{qudits}. We define this generalization via the one-tangling power, and we also denote it by the symbol ``$G_1$." Our derivation for the higher-spin one-tangling power similarly applies to driven and undriven systems described by block-diagonal unitary evolutions as in Eq.\ (\ref{eq:Hgen}), with the distinction that $U_i$ can now be higher-dimensional, unlike previously studied cases.

\section{Generalized one-tangling power}
\label{Sec:generalizedexprs}

Starting with Eq.\ (\ref{eq:epq}), we find that the one-tangling power for an electron-nuclear central-spin system of arbitrary size and total spin is given by one of two different expressions. These expressions differ based upon which subsystem is bipartitioned from the rest. In the case where a single nuclear spin is isolated from the electron and remaining ensemble, we obtain the nuclear one-tangling power (see Appendix \ref{appendix:nuclearonetangles}):
\begin{equation}
\epsilon^{\mathrm{nuclear},i}_{p|q} = \frac{1}{3}\Big(\frac{d_i}{1+d_i}\Big)\big(1 - G_1^{(i)}\big),
    \label{eq:nuclearepanalytical}
\end{equation} 
\noindent where $d_i$ represents the dimension of the isolated nuclear spin labeled by $i$, and its associated generalized Makhlin invariant is given by $G_1^{(i)} = \frac{1}{d_i^2}|\text{Tr}[R_{n_0}^{\dagger}R_{n_1}]|^2$. This quantity is similar to a traditional Makhlin invariant \cite{makhlin2002nonlocal} in the way that it contributes to the classification of the entangling capabilities of a unitary, but it is not restricted to two-qubit evolutions. Notice that the nuclear one-tangling power only depends on the properties of the nuclear spin that has been isolated from the rest of the system, and on the central electron. In the derivation for this expression, we make the assumption that $H_0$ and $H_1$ are separable into tensor products of operators that act on individual nuclei, which implies that there are no direct inter-nuclear interactions. Even though the systems described by Eq.\ (\ref{eq:nuclearepanalytical}) do not include appreciable inter-nuclear interactions, we cannot assume that the amount of entanglement quantified by this expression is entirely attributed to the central electron. The dynamics of the system include one-to-all interactions between the electron and surrounding nuclei which can also lead to indirect all-to-all correlations. 

We also present the one-tangling power with respect to the central electron spin. In this case, we consider a single electron bipartitioned from an ensemble of $n$ nuclear spins and the one-tangling power reads (see Appendix \ref{appendix:electroniconetangles})
\begin{equation}
    \epsilon^{\mathrm{electronic}}_{p|q} = \frac{1}{3} -  \frac{1}{3}\prod_{i=1}^n\Big(\frac{1}{d_i + 1}\Big) \Big(1 + d_i G_1^{(i)} \Big).
    \label{eq:electronicepanalytical}
\end{equation}
\noindent We can see that the amount of entanglement between the electron and surrounding nuclei depends on the product of generalized Makhlin invariants associated with every nuclear spin interacting appreciably with the electron. 

The one-tangling power expressions, Eqs.\ (\ref{eq:nuclearepanalytical}) and (\ref{eq:electronicepanalytical}), readily apply to many different central-spin systems. We only need knowledge of the physical parameters in the Hamiltonian and the two-spin evolution operators generated by the interaction of the electron with each individual nucleus to convert the general formulas, Eqs.\ (\ref{eq:nuclearepanalytical}) and (\ref{eq:electronicepanalytical}), into explicit expressions. Each two-spin evolution operator determines one of the $G_1^{(i)}$. We focus on the example of an (In)GaAs QD to demonstrate this real-world application, offering a procedure that can be followed for other platforms. We also explore the growth of entanglement in the QD with two different non-collinear terms in the Hamiltonian (Eq.\ \ref{eq:Hgen}) to illustrate the scope of this application to other systems subject to these types of interactions. 

\section{One-tangling power of a quantum dot}
\label{Sec:QDexample}

We first consider a Hamiltonian containing the always-on hyperfine contact interaction and strain-induced quadrupolar effects present in self-assembled QD systems, as well as a higher-order quadrupolar-induced non-collinear hyperfine term \cite{gangloff2019quantum}: 
\begin{equation}
    H = \omega_e S_z + \sum_i \omega_i I_i^z + H_{\text{Q}} + H_{\text{c}} + H_{\text{nc}}.
    \label{eq:H}
\end{equation}
\noindent Here, $I_i$ represents the spin operator corresponding to a given nucleus and $S_z$ corresponds to the electron ($S_z = \frac{1}{2}\sigma_z$). The first and second terms are the electronic and nuclear Zeeman terms, respectively, which are typically on the order of $\omega_e \sim$ GHz and $\omega_i \sim$ MHz \cite{stockill2016quantum, Zaporskithesis}. Relevant experiments are typically in the few-Tesla regime, for the purpose of suppressing quadrupolar-based dephasing \cite{stockill2016quantum} and initializing the electron spin when the g-factor is small \cite{shofer2024tuning, appel2025many}. Note that the electronic Zeeman term commutes with the rest of the Hamiltonian, so we can transform it away and do not need to consider it in the analysis that follows.

The third term accounts for quadrupolar effects that are present because we are dealing with a strained system containing nuclei of spin $I_i > 1/2$ \cite{urbaszek2013nuclear}. The following form retains only the diagonal elements of the full quadrupolar Hamiltonian given in \cite{urbaszek2013nuclear, gangloff2019quantum}. It is given by
\begin{equation}
    H_{\mathrm{Q}} = \sum_i \Delta_{\mathrm{Q},i}(I_i^z)^2.
\end{equation}
\noindent This term introduces an anharmonicity to the nuclear energy levels and is proportional to the quadrupolar coupling parameter, $\omega_{\mathrm{Q},i}$: $\Delta_{\mathrm{Q},i} = \omega_{\mathrm{Q},i}(\sin^2(\theta_i)- \frac{1}{2}\cos^2(\theta_i))$ \cite{gangloff2019quantum}. Here, $\theta_i$ is defined by the offset of the nuclear quantization axis due to strain caused by couplings between nuclear quadrupole moments and nearby electric field gradients \cite{urbaszek2013nuclear}. Although it is typically on the order of hundreds of kHz for GaAs QDs, $\Delta_{\mathrm{Q},i}$ can be tuned via the quadrupolar strain \cite{appel2025many, shofer2024tuning, zaporski2023ideal} and is on average a few MHz in systems with higher strain such as InGaAs \cite{stockill2016quantum, Zaporskithesis}.

The fourth term, known as the Overhauser term, quantifies the shift on the electron spin precession due to the effective magnetization of the nuclei \cite{urbaszek2013nuclear}. This collinear hyperfine contribution arises from the interaction between the electronic wavefunction and surrounding nuclear lattice sites \cite{urbaszek2013nuclear}. This term is expressed as
\begin{equation}
    H_{\mathrm{c}} = S_z \sum_i a_i I_i^z,
\end{equation}
\noindent where $a_i$ denotes the hyperfine coupling strength between the electron and a given nucleus. This coupling is proportional to the magnitude of the electronic wavefunction at the location of the nucleus. Because this wavefunction decays with distance from the center of the QD, so do the hyperfine couplings. Most of the results we present below are valid for any distribution of these couplings.

The last term is known as the non-collinear hyperfine term, and is written as
\begin{equation}
\begin{split}
    H_{\mathrm{nc}} = -S_z \sum_i a_i^{\mathrm{nc}} &[\cos^2\theta_i((I_i^x)^2 - (I_i^y)^2) \\&+ \sin2\theta_i(I_i^zI_i^x + I_i^xI_i^z)],
    \label{eq:Hnc}
\end{split}
\end{equation}
\noindent where $a_{i}^{\mathrm{nc}} = \frac{a_i \omega_{\mathrm{Q},i}}{2 \omega_i}$ \cite{gangloff2019quantum}. This term arises from strain and enables $\Delta m = \pm 1$ and $\Delta m = \pm 2$ electron spin-dependent transitions between the levels of individual nuclei \cite{denning2019collective}. Note that we treat $\Delta_{\text{Q},i}$ and $a_{i}^{\mathrm{nc}}$ independently to study the relative importance and individual effects of these terms. It is experimentally possible to perform this independent tuning via e.g., magnetic field angle relative to strain axes, regardless of whether $a^{\text{nc}}_i$ arises from quadrupolar interactions or g-factor anisotropy \cite{gangloff2021witnessing}. Overall, we consider relative parameter strengths according to the following hierarchy: $\omega_e \gg \omega_i \gg a_i\gg\Delta_{\mathrm{Q},i},a_{i}^{\mathrm{nc}}$.

Each of these terms in the Hamiltonian plays an important role in the entanglement dynamics that take place in the system. Note that this entire Hamiltonian, Eq.\ (\ref{eq:H}), is block-diagonal and can be expressed in the form of Eq.\ (\ref{eq:Hgen}), where
\begin{flalign}
\begin{aligned}
H_{0/1} &= \pm \frac{1}{2}\omega_e\mathbb{I} + \sum_i \omega_i I_i^z + \sum_i \Delta_{\mathrm{Q},i}(I_i^z)^2  \\
&\pm \sum_i \frac{a_i}{2} I_i^z \mp \sum_i  \frac{a_i^{\mathrm{nc}}}{2}[\cos^2\theta_i ((I_i^x)^2 - (I_i^y)^2) \\ 
&+ \sin2\theta_i(I_i^zI_i^x + I_i^xI_i^z)].
\label{eq:Hi}
\end{aligned}&&&
\end{flalign}
\noindent It is this block-diagonal structure that lends Eq.\ (\ref{eq:H}) conveniently to the one-tangles formalism. We will next apply the one-tangling power expressions, Eqs.\ (\ref{eq:nuclearepanalytical}) and (\ref{eq:electronicepanalytical}), to study the effects of each term in the Hamiltonian on the growth of entanglement in the system. 

In this case, we form the explicit one-tangling power expressions by assuming that all nuclei are spin-$3/2$, corresponding to either Ga or As. Note these expressions still hold with the inclusion of nuclear species of different total spin quantum numbers. Appendix \ref{Appendix:Spinninehalfex} illustrates the additional consideration of spin-$9/2$ $^{115}$In nuclei in the QD. For now, we assign $d_0 = 2$ (corresponding to the central electron spin) and $d_1,...,d_n = 4$, so Eq.\ (\ref{eq:nuclearepanalytical}) becomes
\begin{equation}
    \epsilon^{\mathrm{nuclear},i}_{p|q} = \frac{4}{15} \big(1 - G_1^{(i)}\big).
    \label{eq:nuclearep3/2}
\end{equation}
\noindent This expression holds for any system that possesses the block-diagonal structure of Eq.\ (\ref{eq:Hgen}) where the singled-out nucleus is spin-$3/2$. Notice that in the case of a maximally entangling evolution, when $G_1^{(i)} = 0$, our expression is upper-bounded by $\epsilon_{p|q}(U) = \frac{4}{15}$. Contrarily, for a nonentangling class of unitaries, $G_1^{(i)} = 1$ and the one-tangling power is zero. We obtain the analogous expression for the electronic one-tangling power, in which Eq.\ (\ref{eq:electronicepanalytical}) becomes
\begin{equation}
\epsilon^{\mathrm{electronic}}_{p|q} = \frac{1}{3} -  \frac{1}{3}\Big(\frac{1}{5}\Big)^n \prod_{i=1}^n\Big(1 + 4 G_1^{(i)} \Big).
    \label{eq:electronicepa3/2}
\end{equation}
\noindent This applies to systems made up of arbitrarily many nuclei all of dimension $d_i = d = 4$. Again, adjustments can be made if we choose to consider nuclei of different total spin (see Appendix \ref{Appendix:Spinninehalfex}). We can see that this expression is upper-bounded differently than Eq.\ (\ref{eq:nuclearep3/2}). This indicates an entirely different amount of entanglement that can be generated by the one-to-all interaction between the electron and surrounding nuclei. In what follows, we will use Eqs.\ (\ref{eq:nuclearep3/2}) and (\ref{eq:electronicepa3/2}) to compute the entanglement dynamics of our example (In)GaAs electron-nuclear QD system for both free evolution (Sec.\ \ref{Sec:freeevolution}) and in the presence of dynamical decoupling pulses (Sec.\ \ref{Sec:ddsequences}).

\section{Entanglement growth during free evolution}
\label{Sec:freeevolution}

Under free evolution, the QD spins obey dynamics governed by
\begin{equation}
    U_{\mathrm{free}} = \sigma_{00} \otimes R_{n_0} + \sigma_{11} \otimes R_{n_1},
    \label{eq:Ufree}
\end{equation}
\noindent where $R_{n_0} = e^{-i H_0 t}$ and $R_{n_1} = e^{-i H_1 t}$. We  employ the one-tangling power expressions, Eqs.\ (\ref{eq:nuclearep3/2}) and (\ref{eq:electronicepa3/2}), to examine entanglement growth between freely evolving spins in the QD system. First we present an analysis of the case when the central electron is coupled to a single nuclear spin. Note that because the nuclear one-tangling power only depends on the isolated nuclear spin in question and its interaction with the electronic spin, Eq.\ (\ref{eq:nuclearep3/2}) holds regardless of how many additional nuclei are present. These additional nuclei could even have total spin other than 3/2, so long as the nucleus of interest is spin-$3/2$, and we can gain insight into the entangling capabilities of $U_{\text{free}}$ pertaining to a large ensemble by first studying this simpler case. 

\subsection{Single nuclear spin}
\begin{figure}
    \centering
    \includegraphics[width=\linewidth]{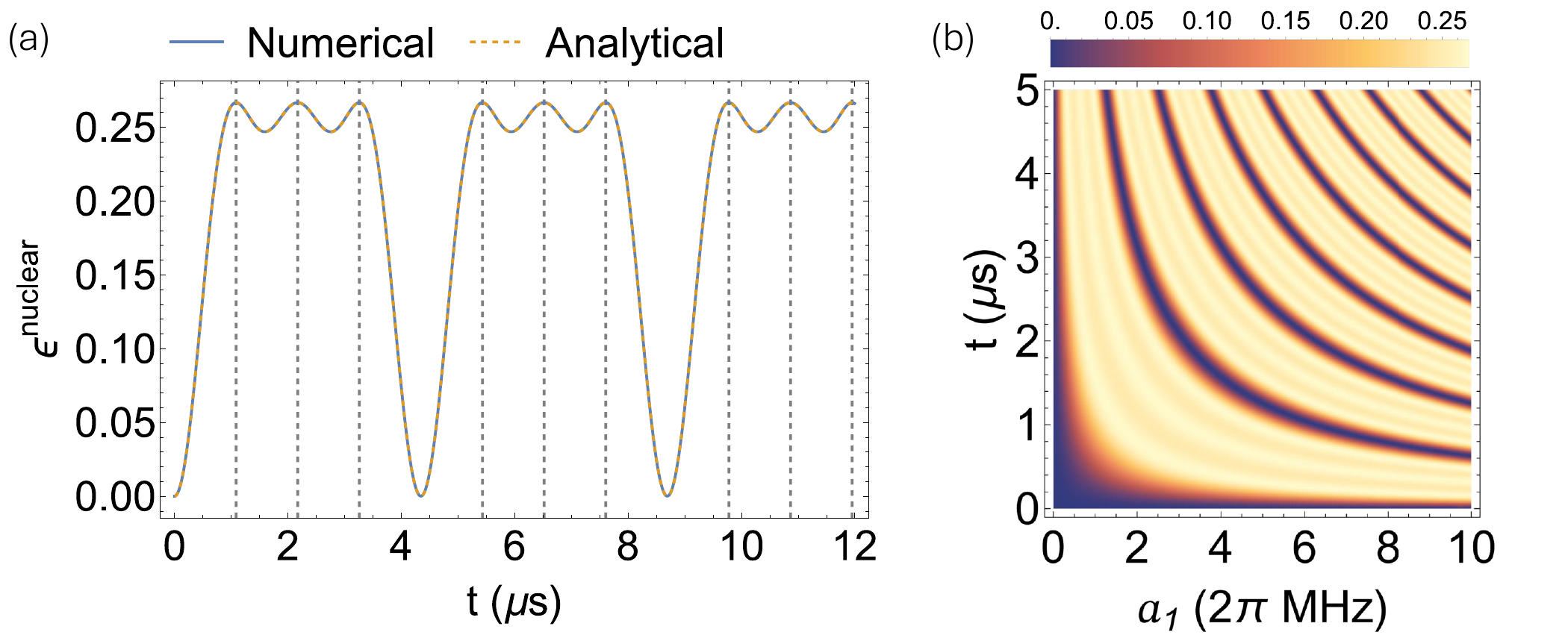}
    \caption{(a) Numerical (blue, solid) and analytical (orange, dashed) one-tangling power with respect to a single nucleus coupled to the electron. The distinction between the numerical and analytical results lies in which expression for $G_1^{(1)}$ is used to plot Eq.\ (\ref{eq:e12}). The numerics refer to the use of $G_1^{(1)} = \frac{1}{d_1^2}|\text{Tr}[R_{n_0}^{\dagger}R_{n_1}]|^2$ and the analytical result is formed using Eq.\ (\ref{eq:G1analytical}). The vertical, gray dashed lines correspond to the resonances up to $k = 10$ and for $a_1/2\pi = 0.23$ MHz. The peaks in this figure represent times at which the nucleus and electron are maximally entangled, while the zeros show when these spins are decoupled. (b) Numerical one-tangling power as a function of time and hyperfine coupling strength. This illustrates how the period of $G_1$ inversely depends on $a_1$. The parameter values used for both plots include $\omega_1/2\pi = 12.98$ MHz (corresponding to the Larmor frequency of $^{71}$Ga when the external magnetic field is $1$ T), $a^{\mathrm{nc}}_1/2\pi = 0.056$ MHz, $\Delta_{\mathrm{Q}, 1}/2\pi = 0.034$ MHz, and $\theta_1 = \frac{\pi}{3}$.}
    \label{fig:analytics}
\end{figure}
The amount of entanglement that $U_{\mathrm{free}}$ can generate between the central electron and a single nuclear spin is expressed as
\begin{equation}
    \epsilon_{0|1} = \frac{4}{15}(1 - G_1^{(1)}).
    \label{eq:e12}
\end{equation}
\noindent Note that $\epsilon^{\mathrm{nuclear}}_{0|1} =\epsilon^{\mathrm{electronic}}_{0|1} = \epsilon_{0|1}$, where $0$ denotes the electron spin and $1$ refers to the nucleus. In other words, Eqs.\ (\ref{eq:nuclearep3/2}) and (\ref{eq:electronicepa3/2}) both simplify to Eq.\ (\ref{eq:e12}) when we consider a single nucleus. The block-diagonal form of $U$ leads to the generalized Makhlin invariant given by $G_1^{(1)} = \frac{1}{16}|\text{Tr}[R_{n_0}^{\dagger}R_{n_1}]|^2$. An exact, analytical expression for this generalized $G_1$, that holds for any parameter values when $a^{\mathrm{nc}}_i = 0$, is given in Eq.\ (\ref{eq:analyticG1}) of Appendix \ref{appendix:G1analytical}. We demonstrate the use of this expression by choosing $\omega_1 = \frac{1}{2}a_1$, and $\Delta_{\mathrm{Q}, 1} = a^{\mathrm{nc}}_1 = 0$. Under these conditions, $G_1^{(1)}$ simplifies to (see Appendix \ref{appendix:G1analytical})
\begin{equation}
    G_1^{(1)} = \frac{1}{2} \big[\cos(a_1 t)^2(1 + \cos(a_1 t)) \big], 
    \label{eq:G1analytical}
\end{equation}
\noindent where $a_1$ corresponds to the collinear hyperfine coupling strength associated with the single nucleus. This analytical $G_1^{(1)}$ holds in the regime of small $a^{\mathrm{nc}}_1$, lending itself as an accessible tool for identifying instances in which we can achieve maximal or minimal entanglement. Also, because the nuclear one-tangling power, Eq.\ (\ref{eq:nuclearepanalytical}), only depends on the central electron and the nuclear spin of interest, this analytical expression holds for larger ensembles as well. In the analysis that follows, we will compare the numerical and analytical results for the free evolution one-tangling power to illustrate the use of this expression for examining nuclear entanglement as well as electron spin dephasing.

We can also use the analytical $G_1^{(1)}$ to directly solve for times at which a given spin is maximally entangled with the rest of the system, i.e., when $G_1^{(1)} = 0$. We can see from Eq.\ (\ref{eq:G1analytical}) and from  Fig.\ \ref{fig:analytics}(a) that this happens on a periodic basis at values known as resonance times. The resonance times are given by
\begin{equation}
    t_k = \frac{(2k + 1)\pi}{a_1}  \text{ and }  
    \frac{(2k + 1)\pi}{2a_1},
    \label{eq:tk}
\end{equation}
\noindent where $k$ is an integer. Note the way the resonance times depend on the collinear hyperfine coupling, $a_1$. Fig.\ \ref{fig:analytics}(b) demonstrates this relationship, showing how the period of the one-tangling power increases when $a_1$ decreases. This indicates that nuclei at different locations in the dot experience varied levels of entanglement with the electron, also shown in Fig.\ \ref{fig:analytics}(b). This is discussed further in Sec.\ \ref{subsec:spatialdep}.  

The resonance times demonstrate the important role that the hyperfine coupling plays in the growth of entanglement in the system. Previous work \cite{Eva2023, dakis2024high} has already demonstrated the use of resonance times for the purpose of designing single-shot entangling gates for defects in diamond and SiC that target desired nuclei while decoupling unwanted spins. We can apply a similar idea here, except instead of treating individual spins, we address groups of nuclei that are estimated to have the same hyperfine coupling strength based on their radial positions relative to the center of the QD. We next consider a model that introduces realistic inhomogeneities in the hyperfine and quadrupolar coupling strengths across the nuclear ensemble to uncover detailed insight into system dynamics and compute electron spin dephasing exactly. 

\subsection{Ensemble of nuclei}
\label{subsec:spatialdep}

\begin{figure*}
\centering
\includegraphics[width=\linewidth]{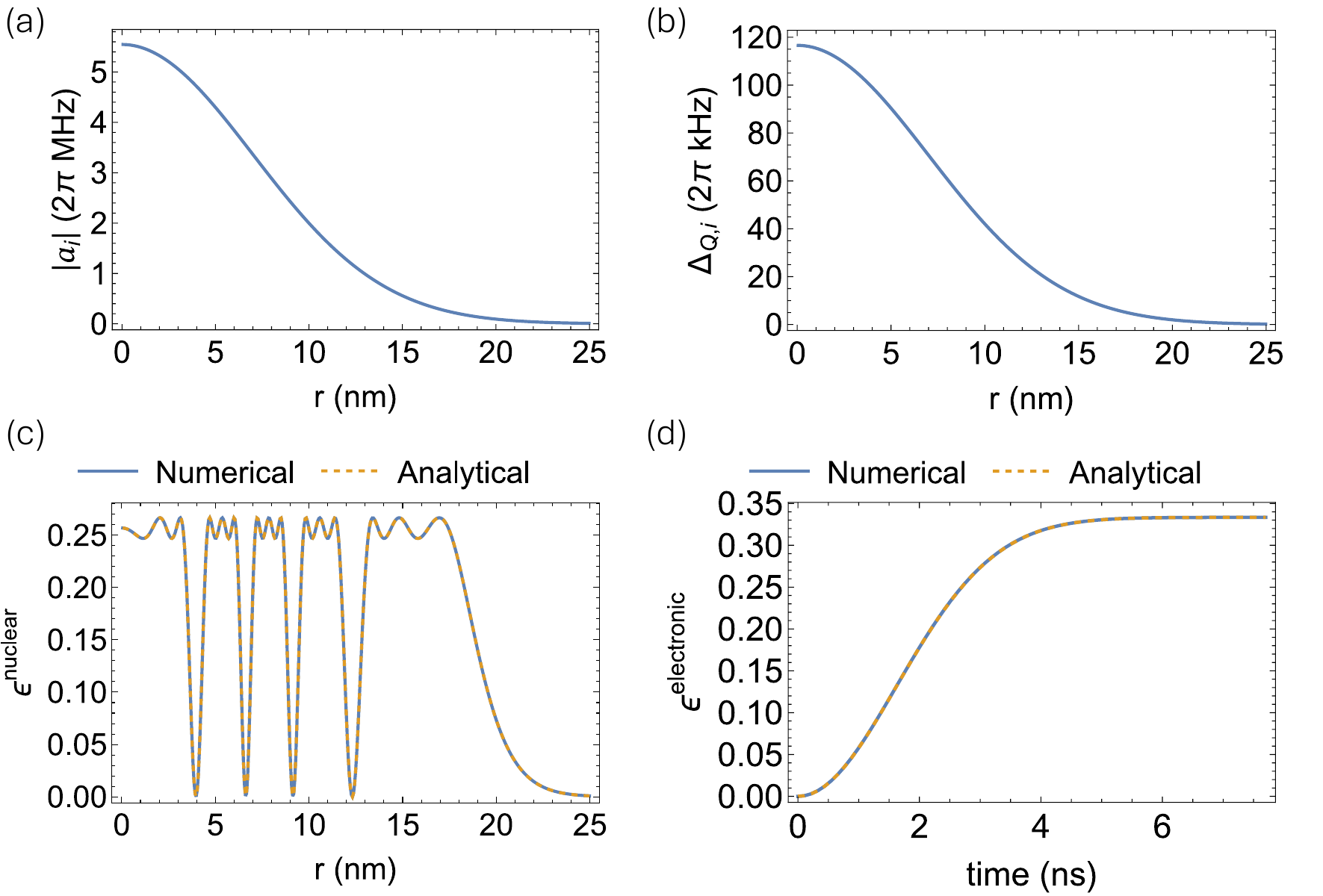}
\caption{(a) Gaussian distribution of hyperfine coupling parameters representing nuclei spread across a QD of radius $25$ nm, given by the distribution $a(r_i)/2\pi = -0.88 e^{-r_i^2/98}$ (in MHz). This is the normal distribution generated using the mean and standard deviation $\mu = 0$ nm and $\sigma = 7$ nm, multiplied by a scaling constant of $-15.49 * 2\pi$. The average value of $|a_i|$ is $0.87$ MHz and we sampled over $80,247$ $r$ values that represent the various radial positions of the QD nuclei, ranging from $r_0 = 0$ nm to $r_{80,247} = 25$ nm with a step size of $\Delta r = 0.056$ nm. The total of all hyperfine coupling values in the ensemble is $A/2\pi = -11.12$ GHz. (b) Distribution for quadrupolar coupling values, given by $\omega_{\text{Q},i}/2\pi = 0.030 e^{-r_i^2/98}$ MHz, which has the same mean, standard deviation and was sampled over the same $r$ values as those given for (a), except now the scaling constant is $0.521 * 2\pi$. We compute the average value of ${\omega}_{\text{Q},i}/2\pi$ to be $4.7$ kHz. This leads to a mean $\Delta_{\text{Q},i}/2\pi$ value of $2.91$ kHz when $\theta_i = \frac{\pi}{3}$. (c) Free evolution nuclear one-tangling power for a single nucleus bipartitioned from the ensemble of $80,247$ nuclear spins and the electron, plotted both numerically (blue, solid) and analytically (orange, dashed) at time $t = 5.32$ $\upmu$s. Several other random times were tested, yielding similar results. This figure shows the spatial dependence of the one-tangling power, providing insight into the amount entanglement between a given nuclear spin (based on its position in the dot) and the rest of the ensemble. (d) Amount of entanglement generated between the electron and surrounding $80,247$ nuclei during free evolution, computed both numerically (blue, solid) and analytically (orange, dashed). As the amount of entanglement between the electron and surrounding nuclei saturates in this plot, we can interpret this in the context of dephasing. We arrive at the electron spin dephasing time of $T_2 \approx 5$ ns using $\epsilon^{\text{electronic}}$. Both (c) and (d) were plotted using the Gaussian distributions representing $a_i$ and $\Delta_{\text{Q},i}$, given in (a) and (b). Other parameters used in both (c) and (d) include $a^{\mathrm{nc}}_i/2\pi = 0.0051$ MHz, $\theta_i = \frac{\pi}{3}$, $\omega_i/2\pi = 12.98$ MHz. The analytical plots are generated using Eq.\ (\ref{eq:G1analytical}). }
\label{fig:freeensembles}
\end{figure*}

When taking into account an ensemble of QD spins, the nuclear one-tangling power assumes the same form as in the single nuclear spin case (Eq.\ (\ref{eq:e12})) because it only depends on the isolated spin and the central electron. The Makhlin invariant corresponding to each nucleus is in general distinct as it depends on the hyperfine and quadrupolar coupling values associated with that particular spin. Therefore, we can simulate a realistic ensemble of nuclei by generating a distribution of values for $a_i$ and $\Delta_{\mathrm{Q},i}$.  

Due to its dependence on the $3$D-confined electronic wavefunction \cite{urbaszek2013nuclear}, the hyperfine coupling strength is greatest near the center of the dot and decreases as the distance from the center grows. We simulate this using a Gaussian distribution that is shown in Fig.\ \ref{fig:freeensembles}(a). To obtain this distribution of values, we assume a uniform density of nuclei that, assuming a disk-like geometry for the QD, leads to an approximately linear relationship between the number of nuclei and their radial distance from the center. We assign the same hyperfine coupling strength to nuclei at a given radius from the electron and sample the corresponding number of values from the Gaussian distribution. Therefore, there are concentric annuli of nuclear spins that all have the same hyperfine coupling strength. This allows us to distinguish between subsets of nuclei of the same species based on which annulus they reside at within the dot. In this case, we simulate the ensemble for a single nuclear spin species, estimating that for $^{71}$Ga nuclei the average value of $|a_i|$ should be about $1$ MHz, and the total of all coupling strengths reach the known constant value of $A/2\pi = \sum_i^n a_i/2\pi = -11.12$ GHz \cite{Zaporskithesis, stockill2016quantum, cywinski2009pure}. We explore the addition of a second species, $^{115}$In, in Appendix \ref{Appendix:Spinninehalfex}. The sum of all nuclei at each radial position, along with the chosen step size between annuli is what determines the simulated number of nuclear spins in the dot. We aim for a simulated number of spins of around $80,000$, in accordance with what is known about these systems \cite{urbaszek2013nuclear}. Our choice of a step size of $\Delta r = 0.056$ nm leads to an ensemble made up of $n = 80,247$ nuclei. This step size is required to produce smooth plots in Fig.\ \ref{fig:freeensembles}. 

There is a $\Delta_{\mathrm{Q},i}$ value assigned to nuclear spins located at given radial positions in the QD as well. This quadrupolar interaction strength arises from coupling between the nuclear quadrupole moments and nearby, strain-induced electric field gradients \cite{urbaszek2013nuclear}. We can study any distribution of quadrupolar parameter values using the one-tangling power. In the case of InGaAs QDs, this distribution follows the strain profile of the dot \cite{bulutay2011quadrupolar}. For the purpose of our generic analysis, we assume the Gaussian distribution shown in Fig.\ \ref{fig:freeensembles}(b). Recall that $\Delta_{\text{Q},i} = \omega_{\text{Q},i}[\sin^2(\theta_i) - \frac{1}{2}\cos^2(\theta_i)]$. 

When these distributions are applied, the free evolution nuclear one-tangling power offers an exact look at the evolution of entanglement in the system for a large ensemble, made simpler by the analytical expression for the generalized Makhlin invariant (see Eqs.\ (\ref{eq:G1analytical}) and (\ref{eq:analyticG1})). We examine the amount of entanglement generated in the QD between a single nuclear spin and the rest as a function of its distance from the center of the QD at a particular instant in time in Fig.\ \ref{fig:freeensembles}(c). We see that the entanglement oscillates at a frequency that varies nonmonotonically with distance. The period of the one-tangling power depends on how quickly the hyperfine coupling values are changing. For instance, the spacings between the peaks of entanglement are smallest towards the middle of the dot, where the slope of the distribution is steepest. Fig.\ \ref{fig:freeensembles}(c) also demonstrates the applicability of the analytical nuclear one-tangling power by verifying that it matches the numerical expression in the regime of low $a^{\mathrm{nc}}_i$.

We saw above that the electronic one-tangling power, given by Eq.\ (\ref{eq:electronicepa3/2}), depends on the product of all generalized Makhlin invariants corresponding to each nucleus interacting appreciably with the electron. We can draw meaningful conclusions about hyperfine-induced decoherence and offer a straightforward method for computing $T_2^*$ through the lens of the one-tangling power. Considering the electron to be surrounded by a lattice of $80,247$ nuclei allows us to reframe the problem as one of a central-spin coupled to its environment. In general, decoherence can be interpreted as the growth of entanglement between a system and its surroundings. Since this information is captured in the one-tangling power, we can also interpret it as the ``decohering power" of an evolution operator \cite{zanardi2000entangling}. We find that for $n = 80,247$, the timescale at which the decohering power saturates is consistent with the known electron spin dephasing times of a few ns for strained QDs \cite{stockill2016quantum}, shown in Fig.\ \ref{fig:freeensembles}(d). This figure also shows that the analytical result agrees with the numerics, promoting this simpler function, given by Eqs.\ (\ref{eq:G1analytical}) and (\ref{eq:analyticG1}), as an accessible method for computing the dephasing time. In the case of the electronic one-tangling power, we have applied the product of $80,247$ copies of Eq.\ (\ref{eq:G1analytical}) with different hyperfine and quadrupolar couplings. 

During free evolution, the QD electron spin decoherence is mainly attributed to the effective magnetic field fluctuations felt by the electron due to random nuclear spin dynamics. Spin echo pulses are well-documented as a treatment for hyperfine-induced noise in these QD systems \cite{petta2005coherent, bluhm2011dephasing, malinowski2017notch, Taminiau, medford2012scaling, zaporski2023ideal}. We next simulate this protocol in the form of the Carr-Purcell-Meiboom-Gill (CPMG) sequence \cite{carr1954effects, meiboom1958modified} to examine and verify electron spin dephasing times, as well as study the growth and minimization of entanglement in the QD when it is subject to dynamical decoupling. 

\section{Entanglement growth during dynamical decoupling}
\label{Sec:ddsequences}

Here we illustrate the use of the CPMG sequence, a single unit of which is composed of three free evolution intervals separated by two $\pi$-pulses, in conjunction with our generalized expressions for the one-tangling power. We describe the CPMG protocol in terms of $(t/4-\pi-t/2-\pi-t/4)^N$, where $N$ represents the number of times the sequence in parenthesis is repeated and a single iteration takes place over the time interval $t$. The first part of our analysis explores the single-unit evolution $(N=1)$ until Sec.\ \ref{Sec:Ng2CPMG}, at which point we explore the addition of $N>1$ iterations of the CPMG sequence. In the case of $N>1$ pulses, the total duration of the evolution is given as $\tau = Nt$. The unitary describing the QD system subject to this single-unit decoupling protocol is given as (see Appendix (\ref{appendix:CPMGU}))
\begin{equation}
\label{eq:UCPMG}
\begin{split}
    U_{\mathrm{CPMG}} &= \sigma_{00} \otimes R_{n_0}(t_3) R_{n_1}(t_2) R_{n_0}(t_1) \\
    &+ \sigma_{11} \otimes R_{n_1}(t_3) R_{n_0}(t_2) R_{n_1}(t_1) \\
    &= \sigma_{00} \otimes R_{n_0}^{\mathrm{CPMG}} + \sigma_{11} \otimes R_{n_1}^{\mathrm{CPMG}},
\end{split}
\end{equation}
\noindent where $t_3 = t_1 = \frac{t}{4}$, $t_2 = \frac{t}{2}$, and the $R_i(t)$ are the free evolution rotations given in Eq.\ (\ref{eq:Ufree}). We also assume that the system is subject to an external magnetic field that enables optical control of the electron, e.g., the Voigt geometry \cite{gangloff2019quantum}. We again employ Eqs.\ (\ref{eq:nuclearep3/2}) and (\ref{eq:electronicepa3/2}) to study electron-nuclear entanglement as well as electron spin dephasing generated by this evolution.

\begin{figure*}
    \centering
    \includegraphics[width=\linewidth]{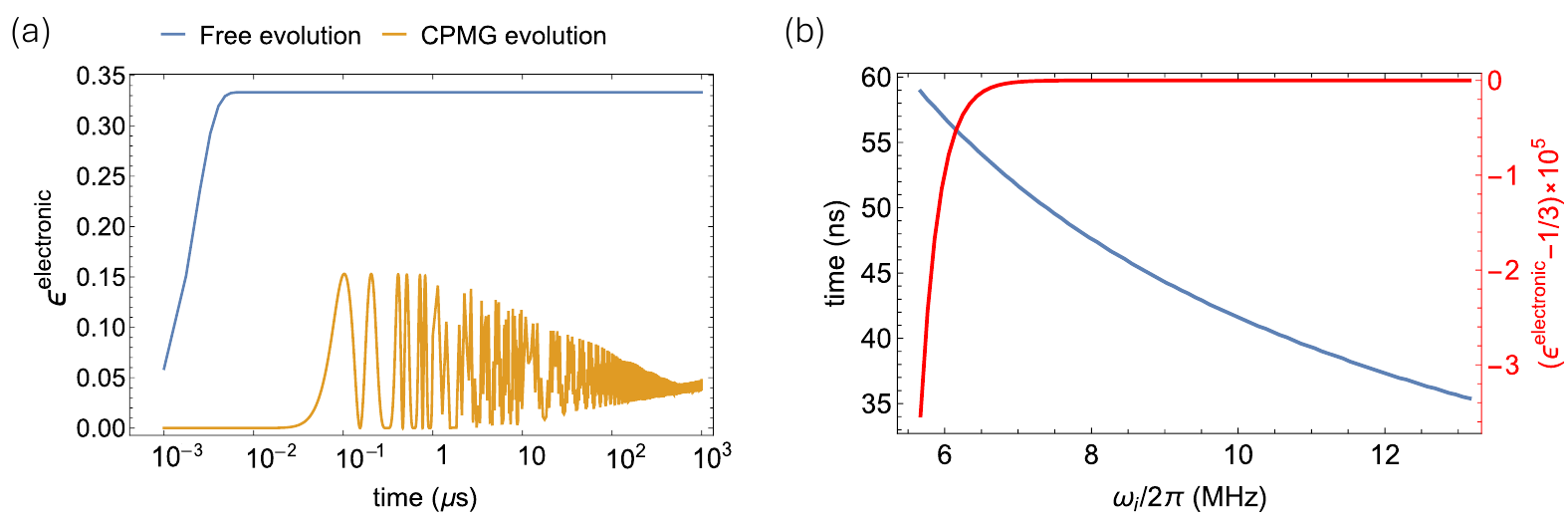}
    \caption{(a) Decohering power as a function of log-scaled time for an electron coupled to $80,247$ nuclei when it is freely evolving (blue) and subject to a single unit of the CPMG sequence (orange), assuming Gaussian distributions from Sec.\ \ref{subsec:spatialdep} for hyperfine and quadrupolar coupling strengths and $\omega_i/2\pi = 12.98 $ MHz. We use the analytical free evolution expression given by Eq.\ (\ref{eq:G1analytical}). The $t$ axis represents the duration of both the free and CPMG evolutions, so the entire CPMG sequence is simulated over $t = 250$ ns. Other parameters include $a^{\mathrm{nc}}_i/2\pi = 0.0051$ MHz and $\theta_i = \frac{\pi}{3}$. This figure shows the improvement in electron spin decoherence when just a single iteration of the CPMG sequence is applied. (b) Dependence of electron spin dephasing time on the Larmor frequency ($\omega_i/2\pi$) of an ensemble of nuclei. We simulate this ensemble using the same distributions given in Sec.\ \ref{subsec:spatialdep}, except now we sample $n = 272$ values. The dephasing times are recorded when the electronic one-tangling power reaches half its maximum value. We sample over a range of $\omega_i/2\pi$ values from $5.67$ MHz to $13.2$ MHz with a step size of $\Delta \omega_i/2\pi = 0.096$ MHz. We also use time steps of $\Delta t = 0.06$ ns, as well as $a^{\mathrm{nc}}_i/2\pi = 0.58$ MHz and $\theta_i = \frac{\pi}{3}$. This figure illustrates the effects of tuning the Larmor frequencies of the QD nuclei (presumably by tuning the external magnetic field) on the electron spin decoherence.}
    \label{fig:ensemble}
    \label{fig:omegavst}
    \label{fig:ensemblefordelta}
    \label{fig:ensembleforanc}
    \label{fig:ensemblefortheta}
\end{figure*}

\subsection{Effects of CPMG sequence on electron spin dephasing}

In this section, we demonstrate the use of the one-tangling power for characterizing and controlling electron spin dephasing under a single unit of the CPMG sequence. The decohering power given by Eq.\ (\ref{eq:electronicepa3/2}) can readily be plotted for the simulated nuclear ensemble by adjusting the form of the generalized Makhlin invariant to be $G_1^{(i)} = \frac{1}{d_i^2}|\text{Tr}\big(R_{n_0}^{\mathrm{CPMG},(i)}\big)^{\dagger}R_{n_1}^{\mathrm{CPMG},(i)}|^2$. Figure \ref{fig:ensemble}(a) reveals that the electronic one-tangling power of $U_{\text{CPMG}}$ is generally smaller than during free evolution, saturating to the maximum value on a much slower timescale depending on the value of $a^{\text{nc}}_i$. For $a^{\text{nc}}_i/2\pi = 0.0051$ MHz, we checked that the one-tangling power saturates at the value attained at $t = 10^3$ $\upmu$s (see Fig.\ \ref{fig:ensemble}(a)) at least for timescales of up to $10$ ms. This result demonstrates how a single unit of the CPMG sequence reduces hyperfine-induced noise in the QD by working to effectively cancel out the Overhauser field \cite{petta2005coherent}. In the next section, we show how to suppress decoherence further by tailoring our driving scheme appropriately. 

We explore the crucial role of $\omega_i$ and therefore, the external magnetic field, on electron spin dephasing by examining the coherence times of the electron coupled to the simulated ensemble of $80,247$ nuclei for a range of nuclear Larmor frequencies, shown in Fig.\ \ref{fig:omegavst}(b). We assume the same distributions of hyperfine and quadrupolar coupling parameters as in Sec.\ \ref{subsec:spatialdep}, but now compute the electron spin dephasing times for different values of $\omega_i$. We define the dephasing times to be the instances in which the decohering power reaches half its maximum value within a given evolution. Note that the total amount of entanglement can vary with each $\omega_i$. 

We have shown how implementing the CPMG sequence on the QD electron cancels out entanglement with unwanted nuclei. Next we explore ways to maximally entangle desired subsets of nuclei using realistic parameter values informed by the nuclear one-tangling power and supported by analytics that capture when degeneracies of the QD Hamiltonian occur. 

\subsection{Mechanisms for entanglement growth between a QD nucleus and electron that is subject to dynamical decoupling}
\begin{figure*}
    \centering
    \includegraphics[width=\linewidth]{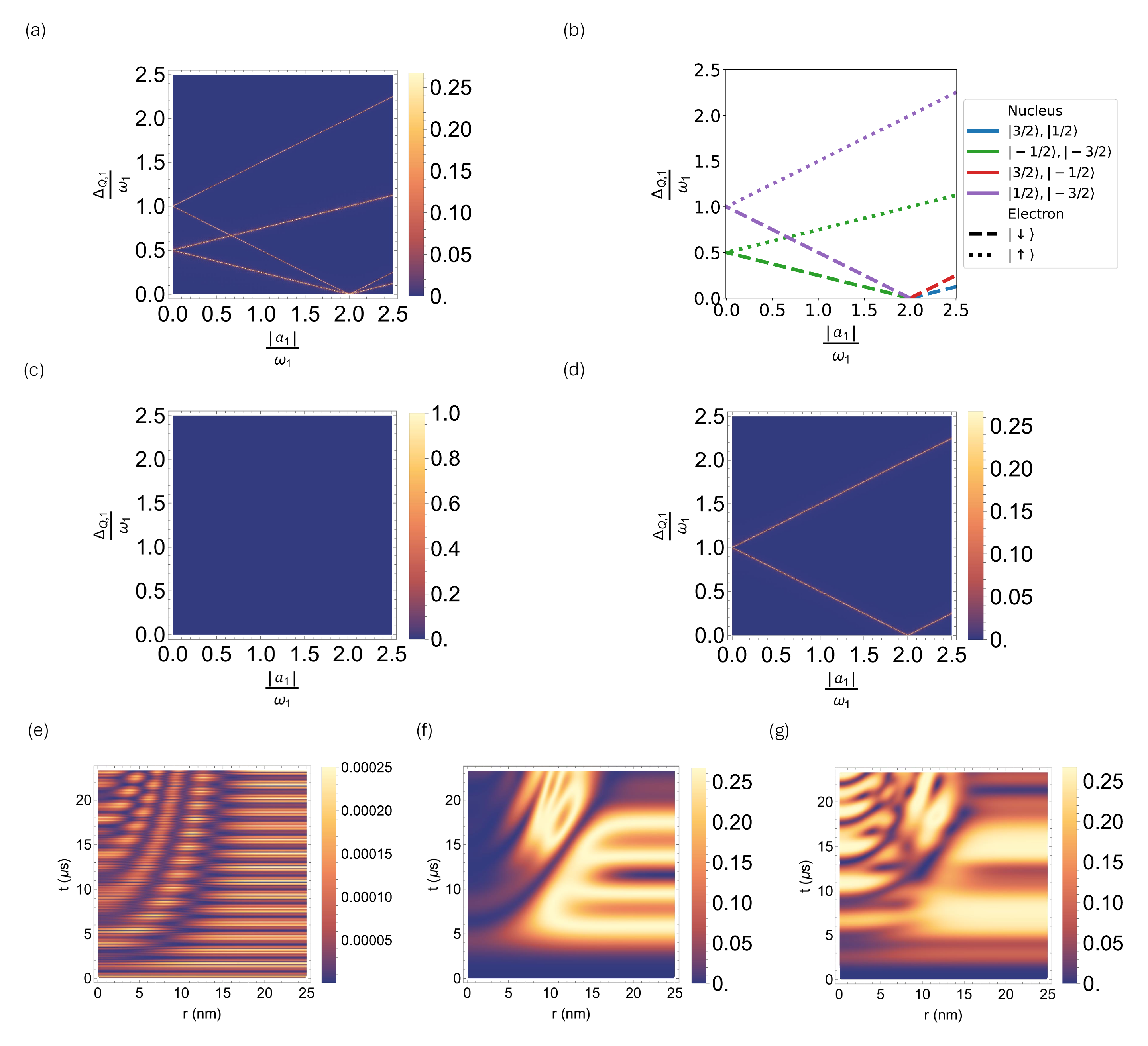}
    \caption{(a) One-tangling power of $U_{\text{CMPG}}$ with respect to a single nucleus paired to the QD electron as a function of $\Delta_{\text{Q},1}/\omega_1$ and $|a_1|/\omega_1$ for $\omega_1/2\pi = 12.98$ MHz, $a^{\mathrm{nc}}_1/2\pi = 0.058$ MHz, and $\theta_1= \frac{\pi}{4}$. (b) Occurrences of degeneracies in the QD Hamiltonian for the same parameter values as in (a). (c) CPMG nuclear one-tangling power with all parameters used in (a) except $\theta_1 = \frac{\pi}{2}$. (d) CPMG nuclear one-tangling power with all parameters used in (a) except $\theta_1 = \pi$. Panels\ (a)-(d) all reveal the parameter values for which the entanglement between the electron and nuclear spin is maximal: along lines corresponding to system degeneracies. (e) CPMG one-tangling power between a single nucleus and the rest of the QD nuclear bath plus the electron. The QD ensemble includes $n = 446$ nuclei that are assigned values from the hyperfine and quadrupolar coupling distributions in Sec. \ref{subsec:spatialdep}, as well as $\theta_i = \frac{\pi}{3}$, $\omega_i/2\pi = 12.98$ MHz, and $a_i^{\mathrm{nc}}/2\pi = 0.051$ MHz. This shows the entangling capabilities of a single iteration of the CPMG evolution for arbitrary parameter values. (f) CPMG nuclear one-tangling power with all parameters used in (e) except $\omega_i = \frac{1}{2} \bar{a}$, where $|\bar{a}|/2\pi = 0.14$ MHz. This shows the increase in entanglement achieved when the value of $\omega_i$ satisfies a condition that yields system degeneracies. (g) CPMG nuclear one-tangling power with parameters used in (e) except $\omega_i = \frac{1}{2} \bar{a}$ and rather than using the distribution of quadrupolar coupling values, $\Delta_{\mathrm{Q},i} = \omega_i$. This illustrates the further increase of entanglement across the nuclear ensemble when an additional parameter condition yielding degenerate eigenstates is applied. A single unit of the CPMG sequence of total duration $t = 23.4$ $\upmu$s is applied for all plots except for (b) which is time independent.}
    \label{fig:diffthetas}
    \label{fig:degeneracies}
    \label{fig:nuclearensemble}
\end{figure*}

Before discussing ways to purposely generate and control electron-nuclear entanglement in the following subsections, we first take a closer look at the role of the various terms in the QD Hamiltonian in creating entanglement. We revisit the case of the QD electron paired with a single nucleus for simplicity, now evolving according to $U_{\mathrm{CPMG}}$. This leads to the same general expression for the single-nuclear free evolution one-tangling power (Eq.\ (\ref{eq:e12})), except now $G_1^{(i)}$ is formed by applying the rotation operators in $U_{\mathrm{CPMG}}$, as given in the previous section. We solve this numerically to find that for arbitrary parameter values, the magnitude of electron-nuclear entanglement is much less in the CPMG case than during free evolution. As previously mentioned, the CPMG pulses effectively cancel out the Overhauser field, so entanglement growth now arises from higher-order processes. Correspondingly, Fig.\ \ref{fig:degeneracies}(a) shows that there is a smaller range of Hamiltonian parameter values that yields maximal entanglement under a single unit of the CPMG sequence. The narrow bands of high one-tangling power indicate that the QD parameters must be carefully tuned (unlike in the free evolution case) in order for spins to become maximally entangled. This density plot in Fig.\ \ref{fig:degeneracies}(a) reveals parameter values that enable $U_{\mathrm{CPMG}}$ to generate correlations between the spins in the QD system. 

Since it is less tractable to analytically compute the generalized Makhlin invariant in the CPMG case, we leverage the fact that the appearance of maximal entanglement coincides with occurrences of energy level degeneracies when $a^{\mathrm{nc}}_i$ is small, as shown in Fig.\ \ref{fig:degeneracies}(b). Working in the physically realistic regime of small $a^{\mathrm{nc}}_i$ ensures that the entanglement-generating nuclear spin transitions induced by the non-collinear term (Eq.\ (\ref{eq:Hnc})) only happen when eigenstates of the Hamiltonian are very close together. This allows us to defer to knowledge of degeneracies of the Hamiltonian for predicting and controlling the appearance of entanglement in the system. This also grants us insight into which nuclear spin levels are being populated at a given time, because we can map the one-tangling power to the corresponding degeneracy for a given set of parameter values of the system. 

The relationship between the non-collinear term and one-tangling power is further illustrated in Figs.\  \ref{fig:diffthetas}(c) and (d). These plots contain the lines of maximal one-tangling power that appear for various quadrupolar strain angles. In Fig.\ \ref{fig:diffthetas}(c), we find that for $\theta_1 = \frac{\pi}{2}$, when the non-collinear term is zero, there is virtually no entanglement between spins. This confirms that the nuclear transitions induced by the non-collinear term are essential for the formation of entanglement in the system when it is subject to the CPMG sequence. For values of $\theta_1$ that yield a nonzero contribution from the non-collinear term, the lines of maximal one-tangling power appear depending on which transitions are allowed. For instance, Fig.\ \ref{fig:diffthetas}(d) shows that when $\theta_1 = \pi$, entanglement only arises along lines that correspond to degeneracies between eigenstates that differ by $\Delta m = \pm 2$. This reflects the transitions that are allowed by the first part of the non-collinear term, $(I^x_i)^2 - (I^y_i)^2$. For angles that lead to nonzero contributions from both parts of the non-collinear term, we see that both $\Delta m = \pm 1$ and $\pm 2$ transitions yield maximal one-tangling power, as illustrated in Fig.\ \ref{fig:degeneracies}(a). For a discussion on higher-order transitions and the appearance of entanglement for a wider range of parameter regimes, see Appendix \ref{Appendix:Degeneracies}. This connection between maximal entanglement and degeneracies within the QD system offers a general approach for cultivating entanglement between spins that are subject to dynamical decoupling and controlled via a non-collinear interaction. This analysis is readily applicable to other systems with a similar non-collinear term, and in Appendix \ref{Appendix:Degeneracies}, we discuss the degeneracies that arise via a different $H_{\mathrm{nc}}$ to further demonstrate this method for analyzing other systems. 

We have confirmed that the activation of the non-collinear term is crucial for the generation of entanglement in the system, and for small $a^{\mathrm{nc}}_i$, the non-collinear term is only able to induce entanglement-generating nuclear spin transitions when eigenstates of the system are degenerate. Next we harness knowledge of these degeneracies and the controllability of the system to deliberately maximize and minimize entanglement.  

\subsection{Tuning QD parameters to control electron-nuclear entanglement}

We compute the nuclear one-tangling power of $U_{\mathrm{CPMG}}$ to study the interactions contributing to the growth of entanglement between QD spins and utilize them to build up or eliminate such correlations. We again utilize the coupling parameter distributions from Sec.\ \ref{subsec:spatialdep}, now to investigate ways to tune the Hamiltonian in order to achieve maximal entanglement under a single unit of the CPMG sequence.

\begin{figure}
    \centering
    \includegraphics[width=\linewidth]{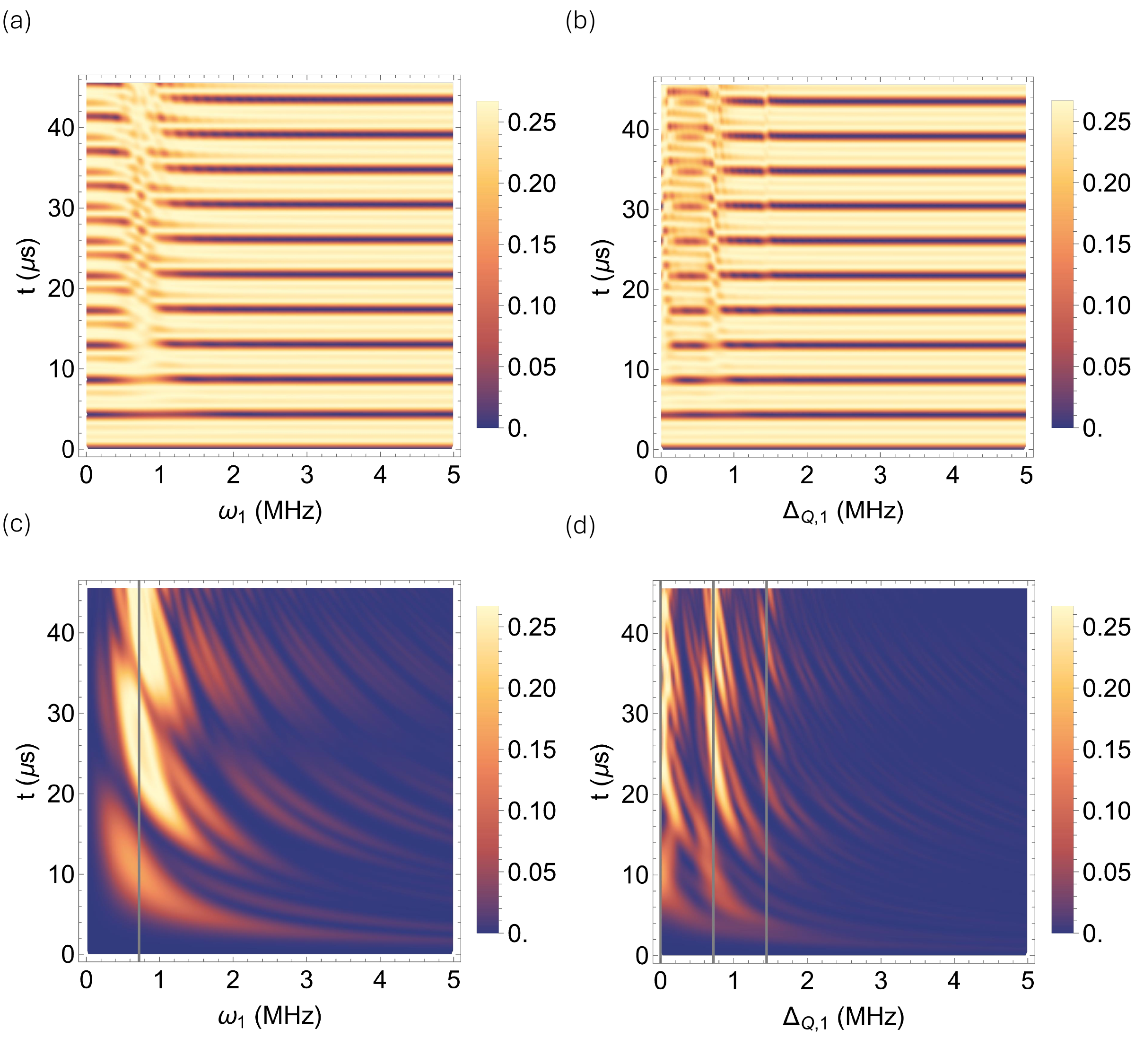}
    \caption{(a) Free evolution nuclear one-tangling power as a function of time and nuclear Larmor frequency when $\Delta_{\text{Q},1}/2\pi = 0.0058$ MHz. (b) Free evolution nuclear one-tangling power as a function of time and quadrupolar coupling strength when $\omega_1 = \frac{1}{2}a_1$. (c) Nuclear one-tangling power when a single iteration of the CPMG sequence is applied over the duration $t = 45.6$ $\upmu$s, when $\Delta_{\text{Q},1}/2\pi = 0.0058$ MHz. This plot, in comparison to (a), shows how maximal entanglement arises when $\omega_1$ at or near the degeneracy condition $\omega_1 = \frac{1}{2} a_1$. (d) Nuclear one-tangling power when a single unit of the CPMG sequence is applied over the duration $t = 45.6$ $\upmu$s, and $\omega_1 = \frac{1}{2}a_1$. This illustrates how maximal entanglement arises for values of $\Delta_{\text{Q},1}$ that coincide with Hamiltonian degeneracies. Other parameters common among the plots include $a_1/2\pi = 0.23$ MHz, $a_1^{\mathrm{nc}}/2\pi = 0.021$ MHz, $\theta_1 = \frac{\pi}{3}$. The vertical gray lines in (c) and (d) indicate the values of $\omega_1$ and $\Delta_{\text{Q},1}$ that yield maximal one-tangling power in Fig.\ \ref{fig:degeneracies}(a). In (c), this is at $\omega_1 = \frac{1}{2}a_1$ and in (d), $\Delta_{\text{Q},1} = 0$ (when $\omega_1 = \frac{1}{2}a_1$), $\frac{1}{2}\omega_1$, and $\omega_1$. }

    %Free evolution nuclear one-tangling power as a function of time and (a) nuclear Larmor frequency and (b) quadrupolar coupling strength when $\omega_1 = \frac{1}{2} a_1$. Nuclear one-tangling power when a single unit of the CPMG sequence is applied over the duration of $t = 45.6$ $\upmu$s, as a function of time and (c) nuclear Larmor frequency and (d) quadrupolar coupling strength when $\omega_1 = \frac{1}{2} a_1$. Other parameters common among the plots include $a_1/2\pi = 0.23$ MHz, $a_1^{\mathrm{nc}}/2\pi = 0.021$ MHz, $\theta_1 = \frac{\pi}{3}$. In (a) and (c), $\Delta_{\text{Q},1}/2\pi = 0.0058$ MHz. The vertical gray lines in (c) and (d) indicate the values of $\omega_1$ and $\Delta_{\text{Q},1}$ that yield maximal one-tangling power in Fig.\ \ref{fig:degeneracies}(a). In (c), this is at $\omega_1 = \frac{1}{2}a_1$ and in (d), $\Delta_{\text{Q},1} = 0$ (when $\omega_1 = \frac{1}{2}a_1$), $\frac{1}{2}\omega_1$, and $\omega_1$. }
    \label{fig:omegaanddelta}
\end{figure}

\begin{figure*}
    \centering
    \includegraphics[width=\linewidth]{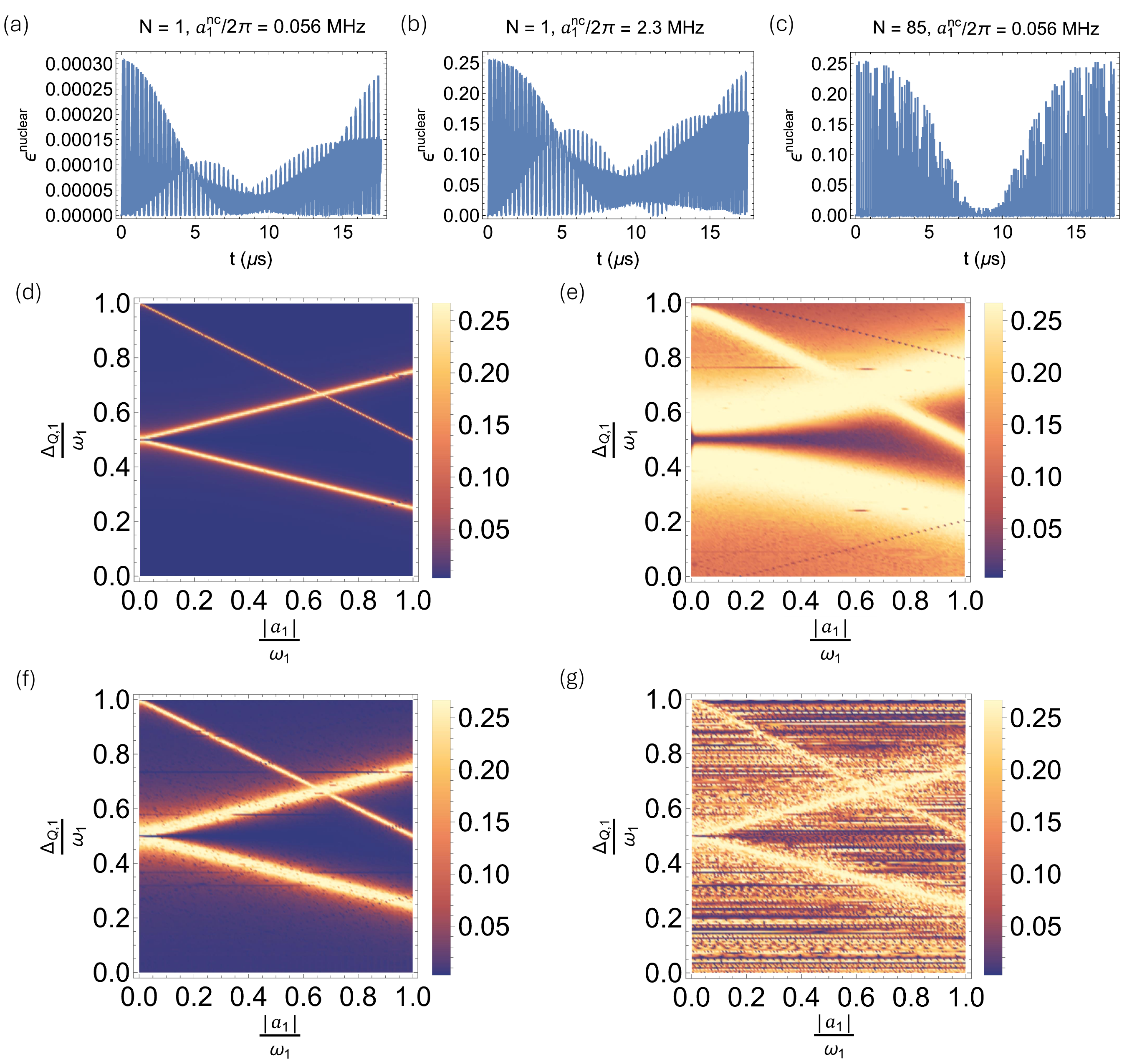}
    \caption{(a) Nuclear one-tangling power of $U_{\text{CPMG}}$ applied over a duration of $17.6$ $\upmu$s for $N=1$ iteration and $a^{\mathrm{nc}}_1/2\pi = 0.056$ MHz. (b) Nuclear one-tangling power of $U_{\text{CPMG}}$ applied over a duration of $17.6$ $\upmu$s for $N=1$ iteration and $a^{\mathrm{nc}}_1/2\pi = 2.3$ MHz. This plot, in comparison to (a), shows how the electron-nuclear entanglement can be maximized by tuning $a^{\mathrm{nc}}_1$. (c) Nuclear one-tangling power of $U_{\text{CPMG}}$ applied over a duration of $17.6$ $\upmu$s for $N=85$ iteration and $a^{\mathrm{nc}}_1/2\pi = 0.056$ MHz. This exemplifies how entanglement between a given nuclear spin and the electron can be maximized when additional iterations of the CPMG sequence are applied while the arbitrary parameter values from (a) are used. The common parameters used for (a)-(c) are $\omega_1/2\pi = 12.98$ MHz, $\Delta_{\text{Q},1}/2\pi = 0.034$ MHz, $\theta_1 = \frac{\pi}{3}$, and $a_1/2\pi = 0.23$ MHz. (d) Dependence of the single-nuclear CPMG one-tangling power on $\Delta_{\text{Q},1}/\omega_1$ and $|a_1|/\omega_1$ for $N=1$ and $a^{\mathrm{nc}}_1/2\pi = 0.063$ MHz. (e) Single-nuclear CPMG one-tangling power arising when $N=1$ and $a^{\mathrm{nc}}_1/2\pi = 1.3$ MHz. This illustrates how entanglement is maximized for parameter values beyond those that yield system degeneracies when the non-colinear coupling strength is increased. (f) Single-nuclear CPMG one-tangling power for $N = 5$ and $a^{\mathrm{nc}}_1/2\pi = 0.063$ MHz. This shows how entanglement can be grown out from values that strictly yield degenerate eigenstates when the number of pulses is increased slightly. (g) Single-nuclear CPMG one-tangling power for $N = 85$ and $a^{\mathrm{nc}}_1/2\pi = 0.063$ MHz. This demonstrates how entanglement can further be maximized for a wider range of parameter values when the number of pulses is increased beyond (f). Figures (d)-(g) are all optimized over the time interval $t = 36.7$ $\upmu$s and have parameter values $\omega_1/2\pi = 12.98$ MHz and $\theta_1 = \frac{\pi}{3}$. Note that the horizontal lines in (g) are most likely numerical artifacts.} 
    %Nuclear one-tangling power of $U_{\text{CPMG}}$ applied over a duration of $17.6$ $\upmu$s for (a) $N=1$ iteration and $a^{\mathrm{nc}}_1/2\pi = 0.056$ MHz, (b) $N = 1$ and $a^{\mathrm{nc}}_1/2\pi = 2.3$ MHz, and (c) $N = 85$ and $a^{\mathrm{nc}}_1/2\pi = 0.056$ MHz. Note that for $N=85$, all pulses are still implemented over total duration $\tau = 17.6$ $\upmu$s. The unit time for a single iteration of the sequence in this case is given as the total duration divided by the number of pulses, $t = 0.207$ $\upmu$s. The common parameters used for (a)-(c) are $\omega_1/2\pi = 12.98$ MHz, $\Delta_{\text{Q},1}/2\pi = 0.034$ MHz, $\theta_1 = \frac{\pi}{3}$, and $a_1/2\pi = 0.23$ MHz. Dependence of the single-nuclear one-tangling power on $\Delta_{\text{Q},1}/\omega_1$ and $|a_1|/\omega_1$ for (d) $N=1$ and $a^{\mathrm{nc}}_1/2\pi = 0.063$ MHz, (e) $N=1$ and $a^{\mathrm{nc}}_1/2\pi = 1.3$ MHz, (f) $N = 5$ and $a^{\mathrm{nc}}_1/2\pi = 0.063$ MHz, and (g) $N = 85$ and $a^{\mathrm{nc}}_1/2\pi = 0.063$ MHz. (d)-(g) are all optimized over the time interval $t = 36.7$ $\upmu$s and have parameter values $\omega_1/2\pi = 12.98$ MHz and $\theta_1 = \frac{\pi}{3}$. Note that the horizontal lines in (g) are most likely numerical artifacts.
    \label{fig:nucCPMG}
    \label{fig:NiterationsCPMG}
    \label{fig:nucdeponanc}
\end{figure*}

We have already found that, for arbitrary parameters, a single unit of the CPMG sequence typically yields low amounts of entanglement between a nucleus and the remaining QD spins. This is further supported by Fig.\ \ref{fig:nuclearensemble}(e), which shows the nuclear one-tangling power as a function of position and time for an ensemble of $n = 446$ spins. However, if we implement parameter conditions that yield degeneracies within the system, we can maximize this entanglement between the electron and a subset of nuclei. We first focus on tuning the nuclear Larmor frequency. Now that the Overhauser field has been effectively removed by the spin echo sequence, there is a greater dependence on $\omega_i$ for the formation of entanglement between a nucleus and the rest of the system. We see in Figs.\  \ref{fig:omegaanddelta}(a) and (c) that the dependence of the nuclear one-tangling power on the value of $\omega_1$ is vastly different in the CPMG versus free evolution case. Fig.\ \ref{fig:omegaanddelta}(c) reveals that maximal entanglement only arises between a single nuclear spin and the electron at $\omega_1 = \frac{1}{2}a_1$ after ambient entanglement generated by the collinear hyperfine interaction has been stripped away. This is consistent with Fig.\ \ref{fig:degeneracies}(a), which shows the maximal one-tangling power occurring at this value of $\omega_1$. We find in Fig.\ \ref{fig:nuclearensemble}(f) that when we set all values of the nuclear Larmor frequencies in the ensemble to $\omega_i = \omega = \frac{1}{2}\bar{a}$, where $\bar{a}$ is the average of all hyperfine coupling values, the dispersion and amount of entanglement changes, increasing at times to the maximum value that was not reached during the arbitrary parameter case. Note that we obtain the same result when $\omega = \frac{1}{2}|\bar{a}|$. Appendix \ref{Appendix:Degeneracies} contains the explicit, analytical conditions that yield degeneracies between particular eigenstates. We find that for transitions present in the degeneracy plot in Fig.\ \ref{fig:degeneracies}(b), the corresponding relations between Hamiltonian parameters outlined in Appendix \ref{Appendix:Degeneracies} reduce to the condition $\omega_i = \pm \frac{1}{2} a_i$ when $\Delta_{\mathrm{Q},i} = 0$. 

Because the CPMG sequence effectively eliminates terms in the Hamiltonian that are proportional to $S_z$, the quadrupolar coupling strength now plays a larger role in the buildup of entanglement as well. The dependence of the one-tangling power on $\Delta_{\mathrm{Q},i}$ indicates the influence of the energy level spacings of the nuclear spins on the entanglement dynamics of the system. The result in Fig.\ \ \ref{fig:omegaanddelta}(d) shows the particular quadrupolar coupling values for which we obtain maximal entanglement under the previously imposed condition of $\omega_i = \frac{1}{2}\bar{a}$. These values occur at integer multiples of $\omega_i$, specifically $\Delta_{\mathrm{Q},i} = 0, \omega_i, \text{ and } 2\omega_i$. We further increase the amount of entanglement in the system in Fig.\ \ \ref{fig:nuclearensemble}(g) by applying the condition $\Delta_{\mathrm{Q},i} = \omega_i$. 

We have demonstrated our approach to exploring tunable parameter regimes that yield maximal entanglement between target spins in the QD. However, not all physical parameters are tunable, and this flexibility generally depends on the platform. With this in mind, we point out that there are ways in which we can generate entanglement within the system without having to tune more than one (if any) parameters. In what follows, we consider methods for maximizing electron-nuclear entanglement by tuning only the non-collinear coupling strength as well as increasing the number of iterations of the applied CPMG sequence. 

\subsection{Selectivity of dynamical decoupling when generating electron-nuclear entanglement}
\label{Sec:Ng2CPMG}

In the previous section, we used the nuclear one-tangling power to inform ways to tune the QD parameters in order to grow electron-nuclear entanglement. We also offer methods for building up electron-nuclear entanglement without changing any system parameters. The two methods we describe in the following section include increasing the strength of the non-collinear term and implementing additional iterations of the applied CPMG sequence. 

We have offered a guide for tuning system parameters in the form of analytical conditions that yield degeneracies between eigenstates of the QD Hamiltonian (Eq.\ (\ref{eq:Hgen})) when the non-collinear coupling is weak. It is in this regime that the entanglement-generating transitions only occur between degenerate eigenstates of the system. While the small $a^{\mathrm{nc}}_i$ regime is physically realistic, we can consider higher values to increase entanglement in cases when other system parameters are fixed. When $a^{\mathrm{nc}}_i$ is unbounded, we find that the non-collinear term is able to induce entanglement for arbitrary parameters, independent of the appearance of degeneracies within the system. Figures \ref{fig:nucdeponanc}(b), (d), and (e) illustrate the increase in maximal entanglement for different parameter values as $a^{\mathrm{nc}}_i$ is increased.

As mentioned in Sec.\ \ref{Sec:freeevolution}, we can in principle design pulse sequences to entangle a subset of nuclei with common hyperfine coupling values. Nuclei with the same $a_i$ are considered in our model to have the same radial distance from the electron. We can think about adjusting our pulse sequences to entangle ``shells" of nuclei some distance from the center at once while decoupling others. Figure \ref{fig:NiterationsCPMG}(a) shows low entanglement for arbitrary parameters when a single unit of the CPMG sequence is applied. In Fig.\ \ \ref{fig:NiterationsCPMG}(c) we demonstrate how applying additional iterations of the CPMG sequence, within the same duration as in Fig.\ \ \ref{fig:NiterationsCPMG}(a), can actually build up entanglement between this subset of nuclei. The same principle applies for Figs.\ \ \ref{fig:NiterationsCPMG}(f) and (g), where $N$ is increased from $5$ to $85$, resulting in an increase in the amount of entanglement between the nucleus and electron during a given time interval. 

\section{Conclusions and Outlook}
\label{Sec:conclusion}
We derive expressions for the one-tangling power to study entanglement dynamics arising in central-spin systems of arbitrary size and total spin. These expressions can be harnessed for a variety of physical systems, and offer a general means for uncovering information about spin-spin correlations using only knowledge of system parameters and the evolution of interest. We demonstrate how the one-tangling power can be used to control the growth of entanglement in a central-spin system via the example of a strained QD. We study this system with and without the application of dynamical decoupling, fleshing out numerical results that are also supported analytically in both cases. We demonstrate the one-tangling power as an exact method for identifying subtle dependencies of the growth of entanglement on the radial position, and therefore hyperfine coupling strength, of spins in an ensemble. We verify that hyperfine interactions are the primary source of entanglement in the system during free evolution, as well as the culprit of electron spin dephasing. Applying dynamical decoupling to the system allows us to gain insight into the underlying mechanisms responsible for generating entanglement in the system in a controlled manner. We find that naturally-occurring degeneracies in the energy levels of our system coincide with observed instances of maximal one-tangling power under the CPMG evolution for low non-collinear coupling strengths, and harness these results to control entanglement among subsets of spins. Knowledge obtained about the QD system through the one-tangling power offers a generalizable scheme for leveraging natural features of various physical platforms for the purpose of controlling entanglement dynamics between spins. 

While we apply the generalized one-tangling power to probe for parameter conditions and driving schemes that yield maximal or minimal entanglement between spins in QDs, an interesting future direction is to apply a similar analysis to other central-spin systems. In the case of QDs, a natural continuation of this work is the formal design of many-body entangling gates capable of selectively entangling target nuclei while decoupling spins assigned to the bath. Resolving between subsets of nuclei in such a dense ensemble could enable the harnessing of memory and computation qudits, or even qubits within the levels of higher-spin nuclei. 

\section{Acknowledgments}

The authors thank Evangelia Takou for useful discussions. This work was supported by the National Science Foundation (grant nos. 1847078, 2137953, and 2137645). KS acknowledges funding from the EPSRC Doctoral Training Programme. DG acknowledges funding from a Royal Society University Research Fellowship. \\

The data that support the findings of this article are openly available \cite{isabela_gnasso_2026_18226303}, embargo periods may apply.

\newpage

\appendix
\begin{widetext}
\section{Generalized one-tangling power expressions}
    \subsection{Nuclear one-tangling power}
\label{appendix:nuclearonetangles}
This section outlines the original derivation of the generalized nuclear one-tangling power expression given as Eq.\ (\ref{eq:nuclearepanalytical}). Note that there is also a shorter, alternative derivation in Appendix \ref{appendix:simplerderivationnuc}. For this derivation, we begin with (\ref{eq:epq}), the general expression for the one-tangling power \cite{linowski2020entangling}, 
\begin{equation}
    \epsilon_{p|q}(U) = 1 - \big(\prod_{i = 0}^n \frac{d_i}{d_i + 1}\big)\sum_{x'|y'}\text{Tr}\{(\text{Tr}_{px'}[|U\rangle \langle U|])^2\}, \\
    \label{eq:epqappendix}
\end{equation}
\noindent we will compute the one-tangling power with respect to a single nuclear spin, labeled with the subscript $q$, bipartitioned from the rest of the system denoted by $p$. In total, the system contains $n+1$ spins. This includes $n$ nuclei and one extra electron. In the above expression, $x'|y'$ represents the possible unordered bipartitions of the copy system. We proceed by computing the trace of the partial trace of the outer product of our unitary squared, $\text{Tr}[\text{Tr}_{px'}(|U\rangle \langle U|)^2]$. We first write this expression in a general way (given by Eq.\ (\ref{eq:simplefiedTr})) that streamlines further calculations by treating the matrix form of $U$. To compute this simplification, we first note that the vectorized unitary acting on $n$ nuclei and one electron, $|U\rangle$, can in general be written as
\begin{equation}
    |U\rangle = \frac{1}{\sqrt{d_0 d_1...d_n}}U^{j_0...j_n}_{j_{0'}...j_{n'}}|j_0...j_n j_{0'}...j_{n'}\rangle 
    = \frac{1}{\sqrt{d_0 d_1...d_n}}U^{j_{\{n\}}}_{j_{\{n'\}}}|j_{\{n\}} j_{\{n'\}}\rangle,
    \label{eq:Uvec}
\end{equation}
\noindent where the set $\{n\}$ is comprised of all of the indices representing spins in the real and copy ensemble, denoted by $i(j)$ and $i'(j')$, respectively. The electron is denoted with $0$ in each corresponding system. Also note that we follow Einstein summation notation for the majority of this derivation unless explicitly stated otherwise. Taking the outer product of Eq.\ (\ref{eq:Uvec}) leads to 
\begin{equation}
    |U\rangle \langle U| = \frac{1}{d_0...d_n} (U^*)^{i_{\{n\}}}_{i_{\{n'\}}}U^{j_{\{n\}}}_{j_{\{n'\}}}|j_{\{n\}} j_{\{n'\}}\rangle\langle i_{\{n\}} i_{\{n'\}}|. 
    %= \frac{1}{d_0...d_n} (U^*)^{i_0...i_n}_{i_{0'}...i_{n'}}U^{j_0...j_n}_{j_{0'}...j_{n'}}|j_0...j_n j_{0'}...j_{n'}\rangle\langle i_0...i_n i_{0'}...i_{n'}| //
    \label{eq:outer}
\end{equation}
\noindent We can then take the partial trace with respect to the set of spins in $p$ and $x'$, indicated here by the inner product of $|k\rangle$ indexed by the label of all spins in $p$ and $x'$:
\begin{equation}
    \begin{split}
        \text{Tr}_{px'}[|U\rangle \langle U|] &= \frac{1}{{d_0...d_n}} (U^*)^{i_{\{n\}}}_{i_{\{n'\}}}U^{j_{\{n\}}}_{j'_{\{n\}}} \langle k_{\{p\}} k_{\{x'\}}|j_{\{n\}} j_{\{n'\}}\rangle \langle i_{\{n\}} i_{\{n'\}}|k_{\{p\}} k_{\{x'\}}\rangle. \\
    \label{eq:Trpx}
    \end{split}
\end{equation}
\noindent At this point, we can relabel the result of the inner product of $\langle k_{\{p\}}k_{\{x'\}}|j_{\{n\}} j_{\{n'\}}\rangle$ in terms of the indices corresponding to the bipartitioned spin $q$, and the spins in $\{y'\}$. Mathematically, this is represented as $\langle k_{\{p\}}|j_{\{n\}}\rangle = \delta^{k_{\{p\}}}_{j_{\{p\}}}|j_q\rangle$. Note that we choose to label the nuclei in the set of $\{y'\}$ as $j_{\{y'\}}$. Similarly, $\langle k_{\{x'\}}|j_{\{n'\}}\rangle = \delta^{k_{\{x'\}}}_{j_{\{x'\}}} |j_{\{y'\}}\rangle$ because any spin not in $\{y'\}$ is in $\{x'\}$. Simplifying (\ref{eq:Trpx}), we have
\begin{equation}
    \begin{split}
        \text{Tr}_{px'}[|U\rangle \langle U|] &= \frac{1}{{d_0...d_n}} (U^*)^{k_{\{p\}} i_q}_{k_{\{x'\}} i_{\{y'\}}} U^{k_{\{p\}} j_q}_{{k_{\{x'\}} j_{\{y'\}}}} |j_q j_{\{y'\}}\rangle \langle i_q i_{\{y'\}}|. \\
    \end{split}
\end{equation}
\noindent We choose to relabel the unitary amplitudes as $\lambda_{i_{\{y'\}}j_{\{y'\}}}^{i_q j_q}$, which leads to 
\begin{equation}
    \begin{split}
        \text{Tr}_{px'}[|U\rangle \langle U|] &= \frac{1}{{d_0...d_n}} \lambda_{i_{\{y'\}}j_{\{y'\}}}^{i_q j_q} |j_q j_{\{y'\}}\rangle \langle i_q i_{\{y'\}}|. \\
    \end{split}
\end{equation}
\noindent Squaring this gives
\begin{equation}
    \begin{split}
        (\text{Tr}_{p x'}[|U\rangle \langle U|])^2 &= \frac{1}{(d_0...d_n)^2} \lambda_{i_{\{y'\}}j_{\{y'\}}}^{i_q j_q}\lambda_{l_{\{y'\}}m_{\{y'\}}}^{l_q m_q} |j_q j_{\{y'\}}\rangle \langle i_q i_{\{y'\}}|m_q m_{\{y'\}}\rangle \langle l_q l_{\{y'\}}| \\
        &= \frac{1}{(d_0...d_n)^2} \lambda_{i_{\{y'\}}j_{\{y'\}}}^{i_q j_q}\lambda_{l_{\{y'\}} i_{\{y'\}}}^{l_q i_q} |j_q j_{\{y'\}} \rangle \langle l_q l_{\{y'\}}|. \\
    \end{split}
\end{equation}
\noindent Taking the full trace leaves
\begin{equation}
    \begin{split}
        \text{Tr}[(\text{Tr}_{p x'}[|U\rangle \langle U|])^2] &= \frac{1}{(d_0...d_n)^2} \lambda_{i_{\{y'\}}j_{\{y'\}}}^{i_q j_q}\lambda_{l_{\{y'\}} i_{\{y'\}}}^{l_q i_q} \delta_{l_q}^{j_q}\delta_{l_{\{y'\}}}^{j_{\{y'\}}}\\
        &= \frac{1}{(d_0...d_n)^2} \lambda_{i_{\{y'\}}j_{\{y'\}}}^{i_q j_q}\lambda_{j_{\{y'\}}i_{\{y'\}}}^{j_q i_q}. \\
    \label{eq:genTr}
    \end{split}
\end{equation}
\noindent To accommodate the matrix form of our evolution $U$, we can rewrite (\ref{eq:genTr}) in terms of the general amplitudes of $U$ and $U^{\dagger}$. We first define the matrices
\begin{equation*}
    \begin{split}
        & U = U^{a_{\{n\}}}_{a_{\{n'\}}}|a_{\{n\}}\rangle \langle a_{\{n'\}}| \text{  and  } U^{\dagger} = (U^*)^{b_{\{n\}}}_{b_{\{n'\}}}|b_{\{n'\}} \rangle \langle b_{\{n\}}|. \\
    \end{split}
\end{equation*}
\noindent We consider the amplitudes of these unitaries with respect to our isolated nuclear spin, denoted by $n$ and a subset of spins in our bipartition, $i_{\{y'\}}$:
\begin{equation*}
    \begin{split}
        \langle i_{\{y'\}} | U^{\dagger}| i_q \rangle &=  (U^*)^{b_{\{n\}}}_{b_{\{n'\}}} \langle i_{\{y'\}}|b_{\{n'\}} \rangle \langle b_{\{n\}}| i_q \rangle \\
        &= (U^*)_{b_{\{x'\}}i_{\{y'\}}}^{b_{\{p\}}i_q} |b_{\{x'\}}\rangle \langle b_{\{p\}}|, 
    \end{split}
\end{equation*}
\noindent and 
\begin{equation*}
    \begin{split}
        \langle j_q | U |j_{\{y'\}}\rangle &= U^{a_{\{n\}}}_{a_{\{n'\}}} \langle j_q | a_{\{n\}} \rangle \langle a_{\{n'\}}|j_{\{y'\}}\rangle \\&= U_{a_{\{x'\}} j_{\{y'\}}}^{a_{\{p\}} j_q} |a_{\{p\}} \rangle \langle a_{\{x'\}}|. \\
    \end{split}
\end{equation*}
\noindent Finally, taking the trace, we find that we recover the resulting form of (\ref{eq:genTr})
\begin{equation}
    \begin{split}
        \text{Tr}[\langle i_{\{y'\}} | U^{\dagger}| i_q \rangle \langle j_q | U |j_{\{y'\}}\rangle]&= \text{Tr}[(U^*)^{i_q b_{\{p\}}}_{i_{\{y'\}} b_{\{x'\}}} U_{a_{\{x'\}} j_{\{y'\}}}^{a_{\{p\}} j_q} |b_{\{x'\}}\rangle \langle b_{\{p\}}|a_{\{p\}} \rangle \langle a_{\{x'\}}| ] \\
        &= (U^*)^{i_q b_{\{p\}}}_{i_{\{y'\}}b_{\{x'\}}} U_{a_{\{x'\}} j_{\{y'\}}}^{a_{\{p\}} j_q} \delta^{b_{\{p\}}}_{a_{\{p\}}}\delta^{b_{\{x'\}}}_{a_{\{x'\}}} \\
        &= (U^*)^{i_q a_{\{p\}}}_{i_{\{y'\}}a_{\{x'\}}} U_{a_{\{x'\}} j_{\{y'\}}}^{a_{\{p\}} j_q} \\
        &= \lambda^{i_q j_q}_{i_{\{y'\}}j_{\{y'\}}}. \\
    \end{split} 
\end{equation} 
\noindent With this equivalent expression for $\lambda^{i_q j_q}_{i_{\{y'\}}j_{\{y'\}}}$, we can simplify the RHS of (\ref{eq:genTr}) as (note that there is no implicit sum in this expression):
\begin{equation}
    \begin{split}
        \lambda^{i_q j_q}_{i_{\{y'\}}j_{\{y'\}}} \lambda^{j_q i_q}_{j_{\{y'\}} i_{\{y'\}}} = \Big|\text{Tr}[\langle i_{\{y'\}}|U^{\dagger}|i_q \rangle \langle j_q|U|j_{\{y'\}} \rangle] \Big|^2. 
        \label{eq:finalformnuc}
    \end{split}
\end{equation}
\noindent This leaves the convenient simplification for Eq.\ (\ref{eq:genTr}) (now including the implicit sums): 
\begin{equation}
    \text{Tr}[(\text{Tr}_{px'}[|U\rangle \langle U|])^2] = \prod_{m=0}^n\frac{1}{d_m^2}\sum_{i_q,j_q,i_{\{y'\}}j_{\{y'\}}}|\text{Tr}[\langle i_{\{y'\}}|U^{\dagger}|i_{q}\rangle \langle j_{q} |U|j_{\{y'\}} \rangle]|^2. 
    \label{eq:simplefiedTr}
\end{equation}
\noindent In order to compute the average entangling capabilities of a given unitary, we need to consider all possible configurations of the copy system, represented above by the indices contained in $i_{\{y'\}}$ and $j_{\{y'\}}$. There are $2^{n+1}$ possible unordered bipartitions of the system. These possibilities can be roughly listed as (choosing to denote the electron as $0'$):
\begin{equation}
\begin{split}
    & \{0'1'2'...(n-2)'(n-1)'n'|., \\
    & 0'1'2'...(n-2)'(n-1)'|n', \\
    & 0'1'2'...(n-2)'|(n-1)'n', \\
    &\vdots \\
    & 0'1'|2'...(n-2)'(n-1)'n', \\
    & 0'|1'2'...(n-2)'(n-1)'n' \\
    & .|0'1'2'...(n-2)'(n-1)'n' \\
    & \vdots \\
    & 2'|0'1'...(n-2)'(n-1)'n' \\ 
    &\vdots \\
    & (n-1)'|0'1'2'...(n-2)'n' \\
    & n'|0'1'2'...(n-2)'(n-1)'\}.
\end{split}
\end{equation}
\noindent We group the contributions of each of these bipartitions into four different types. These categories are distinguished from each other based on the locations of the indices of the electron and isolated nuclear spin. The possibilities are: both the electron and nuclear spin of interest are in $\{y'\}$, neither of them are, or one of them is while the other is not. Depending on which category we are dealing with, the contribution to the one-tangling power expression will be different.

We begin with a case that yields a trivial (constant value) contribution, when both the bipartitioned nucleus and the electron are included in $\{y'\}$ (e.g., $1'2'...(n-1)'|0'n'$). This accounts for $\frac{2^{n+1}}{4}$ contributions that take on the form given below. We assume an evolution of the form 
\begin{equation*}
    U = \sigma_{00} \otimes R_{n_0} + \sigma_{11} \otimes R_{n_1}. \\
\end{equation*}
\noindent To solve for the corresponding contribution to the one-tangling power expression, we insert $U$ into the simplified expression, Eq.\  (\ref{eq:simplefiedTr}), setting $q = n$:
\begin{equation*}
    \begin{split}
        n' \in y', 0' \in y':\\
        \text{Tr}[(\text{Tr}_{0...(n-1) x'}[|U\rangle \langle U|])^2] &= \prod_{m=0}^n \frac{1}{d_m^2}\sum_{i_n, j_n, i_{\{y'\}},j_{\{y'\}}}\Big|\text{Tr}[\langle i_{\{y' \}}|(\sigma_{00} \otimes R_{n_0} + \sigma_{11}\otimes R_{n_1}))^{\dagger}|i_n\rangle \langle j_n|(\sigma_{00} \otimes R_{n_0} + \sigma_{11}\otimes R_{n_1})|j_{\{y' \}}\rangle ]\Big|^2 \\
        &= \prod_{m=0}^n \frac{1}{d_m^2}\sum_{i_n, j_n, i_{\{y'\}},j_{\{y'\}}}\Big|\text{Tr}[\langle i_{0'}| \sigma_{00}|j_{0'}\rangle \otimes \langle i_{\{y'\neq 0' \}}|R_{n_0}^{\dagger}|i_n\rangle \langle j_n|R_{n_0}|j_{\{y'\neq 0' \}}\rangle]\\
        &+ \text{Tr}[\langle i_{0'}|\sigma_{11}|j_{0'}\rangle \otimes \langle i_{\{y'\neq 0' \}}|R_{n_1}^{\dagger}|i_n\rangle \langle j_n|R_{n_1}|j_{\{y'\neq 0' \}}\rangle] \Big|^2.  \\
    \end{split}
\end{equation*}
\noindent We can write the nuclear rotation operators as the tensor product of the rotation acting on the bipartitioned spin and the rotation acting on the remaining nuclei: $R_{n_{j}} = R_{n_j}^n \otimes R_{n_j}^{\{n-1\}}$. This allows us to simplify further and obtain:
\begin{equation*}
    \begin{split}
        \text{Tr}[(\text{Tr}_{0...(n-1) x'}[|U\rangle \langle U|])^2] &= \prod_{m=0}^n \frac{1}{d_m^2}\sum_{i_n, j_n, i_{\{y'\}},j_{\{y'\}}}\Big|\delta_{0}^{i_{0'}}\delta_{j_{0'}}^0\text{Tr}[\langle i_{\{y'\neq 0' \}}|(R_{n_0}^{\{n-1\}}\otimes R_{n_0}^n)^{\dagger}|i_n\rangle \langle j_n|R_{n_0}^{\{n-1\}}\otimes R_{n_0}^n|j_{\{y'\neq 0' \}}\rangle] \\
        &+ \delta_{1}^{i_{0'}}\delta_{j_{0'}}^1 \text{Tr}[\langle i_{\{y'\neq 0' \}}|(R_{n_1}^{\{n-1\}}\otimes R_{n_1}^n)^{\dagger}|i_n\rangle \langle j_n|R_{n_1}^{\{n-1\}}\otimes R_{n_1}^n|j_{\{y'\neq 0' \}}\rangle] \Big|^2 \\
        &= \prod_{m=0}^n \frac{1}{d_m^2}\sum_{i_n, j_n, i_{\{y'\}},j_{\{y'\}}}\Big|\delta_{0}^{i_{0'}}\delta_{j_{0'}}^0\text{Tr}[\langle i_{\{y'\neq 0', n' \}}|\big(R_{n_0}^{\{n-1\}}\big)^{\dagger}R_{n_0}^{\{n-1\}}|j_{\{y'\neq 0', n' \}} \rangle]\\
        &\cdot \langle i_{n'} |(R_{n_0}^n)^{\dagger}|i_{n}\rangle\langle j_n|R_{n_0}^n|j_{n'}\rangle + \delta_{1}^{i_{0'}}\delta_{j_{0'}}^1 \text{Tr}[\langle i_{\{y'\neq 0', n' \}}|(R_{n_1}^{\{n-1\}})^{\dagger}R_{n_1}^{\{n-1\}}|j_{\{y'\neq 0', n' \}}\rangle] \\
        &\cdot \langle i_{n'}|(R_{n_1}^n)^{\dagger}|i_{n}\rangle\langle j_n|R_{n_1}^n|j_{n'}\rangle \Big|^2.  \\
    \end{split}
\end{equation*}
\noindent Throughout this derivation, we assume that $R_{n_j}^{\{n-1\}}$ can be written as the tensor product of all rotations acting on each nuclear spin in $\{p\}$. So, $R_{n_j}^{\{n-1\}} = R_{n_j}^1\otimes R_{n_j}^2\otimes ... \otimes R_{n_j}^{n-1}$. We recognize $\big(R_{n_j}^{\{n-1\}}\big)^{\dagger}R_{n_{j}}^{\{n-1\}}$ as the identity matrix of dimension defined by all of the nuclei in the system (there are $n+1$ of them in total), excluding the isolated nucleus. The identity matrix formed by the conjugation of this rotation operator has dimensions $d_1 \cdot d_2\cdot ...\cdot d_{n-1}$, where $d_i$ corresponds to the dimension of the $i^{\text{th}}$ nucleus. After taking the inner product with all of the nuclei in $\{y'\}$ not including the isolated spin, $\langle i_{\{y'\neq 0',n'\}}|\big(R_{n_j}^{\{n-1\}}\big)^{\dagger}R_{n_j}^{\{n-1\}}|j_{\{y'\neq 0',n'\}}\rangle$, we obtain an identity matrix of dimension corresponding to the spins in ${\{x'\}}$. We label this as $\mathbb{I}_{\{x'\}} = \mathbb{I}_{x'_1}\otimes \mathbb{I}_{x'_2} \otimes...\otimes \mathbb{I}_{x'_{n + 1 -k}}$ (where $k$ is the number of spins in $\{y'\}$). This identity matrix therefore has dimensions corresponding to the product of dimensions of each nuclear spin in $\{x'\}$: $\text{dim}(\mathbb{I}_{\{x'\}}) = d_{x'_1}\cdot d_{x'_2}\cdot...\cdot d_{x'_{n+1-k}} = d_{\{x'\}} = \text{Tr}[\mathbb{I}_{\{x'\}}]$. Thus, the explicit dimension of this matrix will depend on the dimension of each spin in $\{x'\}$. There are many possible permutations of different nuclei in $\{x'\}$, and we will account for this later on in the calculation when we compute $\sum_{x'|y'} \text{Tr}[(\text{Tr}_{0...(n-1)x'}[|U\rangle \langle U|])^2]$. For now, we can write the trace of this identity matrix squared generally as $(d_{\{x'\}})^2$ to obtain:
\begin{equation*}
    \begin{split}
        \text{Tr}[(\text{Tr}_{0...(n-1) x'}[|U\rangle \langle U|])^2] &= \prod_{m=0}^n \frac{1}{d_m^2}\sum_{i_n, j_n, i_{\{y'\}},j_{\{y'\}}}\Big|\delta_{0}^{i_{0'}}\delta_{j_{0'}}^0\text{Tr}[\mathbb{I}_{\{x'\}}]\delta^{i_{\{y'{\neq 0',n'}\}}}_{j_{\{y'{\neq 0',n'}\}}}\langle i_{n'} |(R_{n_0}^n)^{\dagger}|i_{n}\rangle\langle j_n|R_{n_0}^n|j_{n'}\rangle \\
        &+ \delta_{1}^{i_{0'}}\delta_{j_{0'}}^1 \text{Tr}[\mathbb{I}_{\{x'\}}]\delta^{i_{\{y'{\neq 0',n'}\}}}_{j_{\{y'{\neq 0',n'}\}}}\langle i_{n'}|(R_{n_1}^n)^{\dagger}|i_{n}\rangle\langle j_n|R_{n_1}^n|j_{n'}\rangle \Big|^2 \\
        &=\prod_{m=0}^n \frac{1}{d_m^2} (d_{\{x'\}})^2 \sum_{i_n, j_n, i_{\{y'\}},j_{\{y'\}}} \delta^{i_{\{y'{\neq 0',n'}\}}}_{j_{\{y'{\neq 0',n'}\}}} \Big(\Big|\delta_{0}^{i_{0'}}\delta_{j_{0'}}^0 \langle i_{n'} |(R_{n_0}^n)^{\dagger}|i_{n}\rangle\langle j_n|R_{n_0}^n|j_{n'}\rangle \Big|^2 \\
        &+ \Big|\delta_{1}^{i_{0'}}\delta_{j_{0'}}^1 \langle i_{n'}|(R_{n_1}^n)^{\dagger}|i_{n}\rangle\langle j_n|R_{n_1}^n|j_{n'}\rangle \Big|^2  \Big). \\
    \end{split}
\end{equation*}
\noindent We can also simplify $\sum_{i_{\{y'\}},j_{\{y'\}}} \delta^{i_{\{y'{\neq 0',n'}\}}}_{j_{\{y'{\neq 0',n'}\}}}$ in terms of the dimension of each spin in $\{y'\neq 0',n'\}$. This leaves $d_{\{y'\neq 0',n'\}}$, which depends on the dimension of each spin in this set and this is simplified when we compute $\sum_{x'|y'}\text{Tr}[(\text{Tr}_{0...(n-1)x'}[|U\rangle \langle U|])^2]$. Continuing, we have
\begin{equation}
    \begin{split}
        \text{Tr}[(\text{Tr}_{0...(n-1) x'}[|U\rangle \langle U|])^2] &= \prod_{m=0}^n \frac{1}{d_m^2}(d_{\{x'\}})^2 \sum_{i_{\{y'\neq 0',n'\}}, j_{\{y'\neq 0',n'\}}} \delta^{i_{\{y'{\neq 0',n'}\}}}_{j_{\{y'{\neq 0',n'}\}}} \sum_{i_{0' }, j_{0'}} (\delta_{0}^{i_{0'}}\delta_{j_{0'}}^0 + \delta_{1}^{i_{0'}}\delta_{j_{0'}}^1)d_n^2 \\
        &= \prod_{m=0}^n \frac{1}{d_m^2}(d_{\{x'\}})^2 d_{\{y'\neq 0',n'\}} 2d_n^2. \\
    \end{split}
    \label{eq:Trniny0iny}
\end{equation}
\noindent As previously mentioned, to perform the calculation of the one-tangling power, we will eventually consider the sum over all possible bipartitions of the copy system, $\sum_{x'|y'}$. We will collect all four types of contributions to $\text{Tr}[(\text{Tr}_{0...(n-1) x'})^2]$ before computing $\sum_{x'|y'}\text{Tr}[(\text{Tr}_{0...(n-1) x'})^2]$. For now, we will leave Eq.\ (\ref{eq:Trniny0iny}) as is and revisit it after computing the other three contributions. Next we consider the case when $n' \notin y'$ and $0' \in y'$:
\begin{equation*}
    \begin{split}
        n' \notin y', 0' \in y':\\
        \text{Tr}[(\text{Tr}_{0...(n-1) x'}[|U\rangle \langle U|])^2] &= \prod_{m=0}^n\frac{1}{d_m^2}\sum_{i_{\{y'\}}, j_{\{y' \}},i_n, j_n} \Big|\text{Tr}[\langle i_{\{y'\}}|(\sigma_{00} \otimes R_{n_0} + \sigma_{11}\otimes R_{n_1}))^{\dagger}|i_{n}\rangle \langle j_n|(\sigma_{00} \otimes R_{n_0} + \sigma_{11}\otimes R_{n_1})|j_{\{y'\}}\rangle ]\Big|^2 \\
        &=\prod_{m=0}^n\frac{1}{d_m^2}\sum_{i_{\{y'\}}, j_{\{y' \}},i_n, j_n} \Big|\delta_{0}^{i_{0'}}\delta_{j_{0'}}^0\text{Tr}[\langle i_{\{y'\neq 0'\}}|(R_{n_0}^{\{n-1\}})^{\dagger}R_{n_0}^{\{n-1\}}|j_{\{y'\neq 0'\}}\rangle]\text{Tr}[(R_{n_0}^n)^{\dagger}|i_n\rangle\langle j_n|R_{n_0}^n] \\
        &+ \delta_{1}^{i_{0'}}\delta_{j_{0'}}^1 \text{Tr}[\langle i_{\{y'\neq 0'\}}|(R_{n_1}^{\{n-1\}})^{\dagger}R_{n_1}^{\{n-1\}}|j_{\{y'\neq 0'\}}\rangle ]\text{Tr}[(R_{n_1}^n)^{\dagger}|i_n\rangle\langle j_n|R_{n_1}^n] \Big|^2. \\
    \end{split}
\end{equation*}
\noindent Once again, the dimension of the identity operator resulting from $\langle i_{\{y'\neq 0'\}}|(R_{n_j}^{\{n-1\}})^{\dagger}R_{n_j}^{\{n-1\}}|j_{\{y' \neq 0'\}}\rangle $ is dictated by the nuclei in the ensemble that are in $\{x'\}$. Since $n'$ is now in $\{x'\}$, we represent this matrix with $\mathbb{I}_{\{x'\neq n'\}}$ below. The square of the trace of this matrix can be written as $(\text{Tr}[\mathbb{I}_{\{x'\neq n'\}}])^2 = (d_{\{x'\neq n'\}})^2$. Similarly, $\sum_{i_{\{y'\neq 0'\}}, j_{\{y' \neq 0'\}}} \delta^{i_{\{y'{\neq 0'}\}}}_{j_{\{y'{\neq 0'}\}}} = d_{\{y'\neq 0'\}}$. At this point, we can simplify the rest of the expression to obtain:
\begin{equation}
    \begin{split}
        \text{Tr}[(\text{Tr}_{0...(n-1) x'}[|U\rangle \langle U|])^2] &= \prod_{m=0}^n\frac{1}{d_m^2}\sum_{i_{\{y'\}}, j_{\{y' \}},i_n, j_n} \Big|\delta_{0}^{i_{0'}}\delta_{j_{0'}}^0\text{Tr}[\mathbb{I}_{\{x'\neq n'\}}] \delta^{i_{\{y'{\neq 0'}\}}}_{j_{\{y'{\neq 0'}\}}}\delta_{j_n}^{i_n} + \delta_{1}^{i_{0'}}\delta_{j_{0'}}^1\text{Tr}[\mathbb{I}_{\{x'\neq n'\}}] \delta^{i_{\{y'{\neq 0'}\}}}_{j_{\{y'{\neq 0'}\}}} \delta_{j_n}^{i_n} \Big|^2 \\
        &= \prod_{m=0}^n\frac{1}{d_m^2}\Big(\text{Tr}[\mathbb{I}_{\{x'\neq n'\}}] \Big)^2 \sum_{i_{\{y'{\neq 0'}\}}, j_{\{y'{\neq 0'}\}}} \delta^{i_{\{y'{\neq 0'}\}}}_{j_{\{y'{\neq 0'}\}}} \sum_{i_0', j_0', i_n, j_n} \Big|\delta_{0}^{i_{0'}}\delta_{j_{0'}}^0 \delta_{j_n}^{i_n} + \delta_{1}^{i_{0'}}\delta_{j_{0'}}^1 \delta_{j_n}^{i_n} \Big|^2 \\
        &= \prod_{m=0}^n\frac{1}{d_m^2}(d_{\{x'\neq n'\}})^2 d_{\{y'\neq 0'\}} (2 d_n). \\
    \end{split}
    \label{eq:Trnnotiny0iny}
\end{equation}
\noindent Next, we consider a case that yields nontrivial contributions, i.e., contributions that contain the generalized Makhlin invariant, $G_1$. This happens when the bipartitioned nucleus is in $\{y'\}$ but the electron is not:
\begin{equation*}
    \begin{split}
        n' \in y', 0' \notin y': \\
        \text{Tr}[(\text{Tr}_{0...(n-1) x'}[|U\rangle \langle U|])^2] &= \prod_{m=0}^n\frac{1}{d_m^2} \sum_{i_{\{y'\}}, j_{\{y' \}},i_n, j_n} \Big|\text{Tr}[\langle i_{\{y' \}}|(\sigma_{00} \otimes R_{n_0} + \sigma_{11}\otimes R_{n_1}))^{\dagger}|i_n \rangle \langle j_n|(\sigma_{00} \otimes R_{n_0} + \sigma_{11}\otimes R_{n_1})|j_{\{y' \}} \rangle ]\Big|^2 \\
        &= \prod_{m=0}^n\frac{1}{d_m^2} \sum_{i_{\{y'\}}, j_{\{y' \}},i_n, j_n} \Big|\text{Tr}[\langle i_{\{y' \}}|R_{n_0}^{\dagger}|i_n\rangle \langle j_n|R_{n_0}|j_{\{y' \}}\rangle  + \langle i_{\{y' \}}|R_{n_1}^{\dagger}|i_n \rangle \langle j_n|R_{n_1}|j_{\{y' \}} \rangle]\Big|^2 \\
        &= \prod_{m=0}^n\frac{1}{d_m^2} \sum_{i_{\{y'\}}, j_{\{y' \}},i_n, j_n} \Big|\text{Tr}[\langle i_{\{y' \}}|(R_{n_0}^{\{n-1\}}\otimes R_{n_0}^n)^{\dagger}|i_n\rangle \langle j_n|R_{n_0}^{\{n-1\}}\otimes R_{n_0}^n|j_{\{y' \}} \rangle  \\
        &+ \langle i_{\{y' \}}|(R_{n_1}^{\{n-1\}}\otimes R_{n_1}^n)^{\dagger}|i_n\rangle \langle j_n|R_{n_1}^{\{n-1\}}\otimes R_{n_1}^n|j_{\{y' \}} \rangle ]\Big|^2 \\
        &= \prod_{m=0}^n\frac{1}{d_m^2}\sum_{i_{\{y'\}}, j_{\{y' \}},i_n, j_n} \delta^{i_{\{y'\neq n'\}}}_{j_{\{y' \neq n'\}}} \Big|\text{Tr}[\mathbb{I}_{\{x'\neq 0'\}}]\langle i_{n'}|( R_{n_0}^n)^{\dagger}|i_n\rangle \langle j_n| R_{n_0}^n|j_{n'}\rangle \\
        &+ \text{Tr}[\mathbb{I}_{\{x'\neq 0'\}}]|j_{\{y' \neq n'\}}]\langle i_{n'}| R_{n_1}^n)^{\dagger}|i_n\rangle \langle j_n| R_{n_1}^n|j_{n'}\rangle\Big|^2. \\
    \end{split}
\end{equation*}
\noindent Similar to the previous case, the identity matrix resulting from $\langle i_{\{y'{\neq n'}\}}|(R_{n_1}^{\{n-1\}})^{\dagger}R_{n_1}^{\{n-1\}}|j_{\{y' \neq n'\}}\rangle$ depends on the dimensions of $n - k$ nuclear spins, and we denote it as $\mathbb{I}_{\{x'\neq 0'\}}$. It follows that $(\text{Tr}[\mathbb{I}_{\{x'\neq 0'\}}])^2 = (d_{\{x'\neq 0'\}})^2$. We also note that $\sum_{i_{\{y'\neq n'\}}, j_{\{y'\neq n'\}}} \delta^{i_{\{y'\neq n'\}}}_{j_{\{y'\neq n'\}}} = d_{\{y'\neq n'\}}$. continuing, we have: 
\begin{equation*}
    \begin{split}
        \text{Tr}[(\text{Tr}_{0...(n-1) x'}[|U\rangle \langle U|])^2] &= \prod_{m=0}^n\frac{1}{d_m^2}\Big(\text{Tr}[\mathbb{I}_{\{x'\neq 0'\}}]\Big)^2\sum_{i_{\{y'\}}, j_{\{y' \}},i_n, j_n} \delta^{i_{\{y'\neq n'\}}}_{j_{\{y'\neq n'\}}}  \Big|\langle i_{n'}|( R_{n_0}^n)^{\dagger}|i_n\rangle \langle j_n| R_{n_0}^n|j_{n'}\rangle \\
        &+ \langle i_{n'}|( R_{n_1}^n)^{\dagger}|i_n\rangle \langle j_n| R_{n_1}^n|j_{n'}\rangle\Big|^2\\
        &= \prod_{m=0}^n\frac{1}{d_m^2} (d_{\{x'\neq 0'\}})^2 d_{\{y'\neq n'\}} \sum_{i_n, j_n, i_{n'}, j_{n'}} \Big(\Big|\langle i_{n'}|(R_{n_0}^n)^{\dagger}|i_n\rangle \langle j_{n}| R_{n_0}^n|j_{n'}\rangle \Big|^2 \\
        &+ \Big|\langle i_{n'}|( R_{n_1}^n)^{\dagger}|i_n\rangle \langle j_n| R_{n_1}^n|j_{n'}\rangle\Big|^2 + 2 \text{Re}[\langle i_{n'}|( R_{n_0}^n)^{\dagger}|i_n\rangle \langle j_n| R_{n_0}^n|j_{n'}\rangle\langle i_{n'}|( R_{n_1}^n)^{\dagger}|i_n\rangle \langle j_{n}| R_{n_1}^n|j_{n'}\rangle]\Big). \\
    \end{split}
\end{equation*}
\noindent Above, the RHS expression includes three terms that result from the norm squared of the trace. The first two terms in this sum each contribute $d_n^2$ to the one-tangling power. The third term yields:
\begin{equation}
    \begin{split}
        \text{Tr}[(\text{Tr}_{0...(n-1) x'}[|U\rangle \langle U|])^2] &= \prod_{m=0}^n\frac{1}{d_m^2} (d_{\{x'\neq 0'\}})^2 d_{\{y'\neq n'\}} \Big(2d_n^2 + 2 \sum_{i_n,i_{n'},j_n,j_{n'}}\text{Re}[\langle i_{n'}|( R_{n_0}^n)^{\dagger}|i_n\rangle \\
        &\cdot \langle j_n| R_{n_0}^n|j_{n'}\rangle\langle i_{n'}|( R_{n_1}^n)^{\dagger}|i_n\rangle \langle j_n| R_{n_1}^n|j_{n'}\rangle]\Big)\\
        &= \prod_{m=0}^n\frac{1}{d_m^2} (d_{\{x'\neq 0'\}})^2 d_{\{y'\neq n'\}} \Big(2d_n^2 + 2 \sum_{i_{n'},j_n} \text{Re}[\langle i_{n'}|( R_{n_0}^n)^{\dagger}R_{n_1}^n|i_{n'}\rangle \langle j_n| R_{n_0}^n (R_{n_1}^n)^{\dagger}|j_n\rangle]\Big)\\
        &= \prod_{m=0}^n\frac{1}{d_m^2} (d_{\{x'\neq 0'\}})^2 d_{\{y'\neq n'\}} 2(d_n^2 + |\text{Tr}[(R_{n_0}^n)^{\dagger}R_{n_1}^n]|^2). \\
    \end{split}
    \label{eq:Trniny0notiny}
\end{equation}
\noindent The nontrivial term, $|\text{Tr}[(R_{n_0}^n)^{\dagger}R_{n_1}^n]|^2$, is interpreted as the generalized Makhlin invariant corresponding to the bipartitioned nucleus multiplied by its normalization: $d_n^2G_1^{(n)}$. Moving forward, we will drop the superscript $n$ on the nuclear rotation operators, $R_{n_j}^n$, for simplicity, but note that these operators act only on the spin of interest. Thus, the generalized Makhlin invariant, $G_1^{(n)}$, pertains only to the correlation between the isolated nucleus and the central electron. This brings us to the last contribution, which is given by the case when neither the electron nor the isolated nucleus is in $\{y'\}$. This leads to another trivial contribution:
\begin{equation*}
    \begin{split}
        n' \notin y', 0' \notin y':\\
        \text{Tr}[(\text{Tr}_{0...(n-1) x'}[|U\rangle \langle U|])^2] &= \prod_{m=0}^n\frac{1}{d_m^2} \sum_{i_{\{y'\}}, j_{\{y'\}},i_n,j_n} \Big|\text{Tr}[\langle i_{\{y'\}}|(\sigma_{00} \otimes R_{n_0} + \sigma_{11}\otimes R_{n_1}))^{\dagger}|i_n\rangle \langle j_n|(\sigma_{00} \otimes R_{n_0} + \sigma_{11}\otimes R_{n_1})|j_{\{y'\}}\rangle ]\Big|^2 \\
        &= \prod_{m=0}^n\frac{1}{d_m^2} \sum_{i_{\{y'\}}, j_{\{y'\}},i_n,j_n} \Big|\text{Tr}[\langle i_{\{y'\}}|(R_{n_0}^{\{n-1\}})^{\dagger}R_{n_0}^{\{n-1\}}|j_{\{ y'\}}\rangle]\text{Tr}[(R_{n_0}^n)^{\dagger}|i_n\rangle\langle j_n|R_{n_0}^n] \\
        &+ \text{Tr}[\langle i_{\{y'\}}|(R_{n_1}^{\{n-1\}})^{\dagger}R_{n_1}^{\{n-1\}}|j_{\{y'\}}\rangle]\text{Tr}[(R_{n_1}^n)^{\dagger}|i_{n'}\rangle\langle j_n|R_{n_1}^n] \Big|^2. \\
    \end{split}
\end{equation*}
\noindent In this case, the identity matrix arising from $\langle i_{\{y' \}}|(R_{n_{0/1}}^{n-1})^{\dagger}R_{n_{0/1}}^{n-1}|j_{\{ y'\}}\rangle$ is represented by $\mathbb{I}_{\{x'\neq 0',n'\}}$ and it depends on the dimensions of $n - k - 1$ number of spins. Thus, $(\text{Tr}[\mathbb{I}_{\{x'\neq 0',n'\}}])^2 = (d_{\{x'\neq 0',n'\}})^2$. Also, $\sum_{i_{\{y' \}}, j_{\{y'\}}} \delta_{j_{\{y'\}}}^{i_{\{y'\}}} = d_{\{y'\}}$. So, we end up with
\begin{equation}
    \begin{split}
        \text{Tr}[(\text{Tr}_{0...(n-1) x'}[|U\rangle \langle U|])^2] &= \prod_{m=0}^n\frac{1}{d_m^2}\sum_{i_{\{y' \}}, j_{\{y'\}},i_n,j_n} \delta_{j_{\{y'\}}}^{i_{\{y'\}}} \Big|\text{Tr}[\mathbb{I}_{\{x'\neq 0',n'\}}](\text{Tr}[(R_{n_0}^n)^{\dagger}|i_n\rangle\langle j_n|R_{n_0}^n] + \text{Tr}[(R_{n_1}^n)^{\dagger}|i_n\rangle\langle j_n|R_{n_1}^n]) \Big|^2 \\
        &= \prod_{m=0}^n\frac{1}{d_m^2} (d_{\{x'\neq 0',n'\}})^2 d_{\{y'\}} 4\sum_{i_n,j_n}\delta_{i_n}^{j_n}\\
        &= \prod_{m=0}^n\frac{1}{d_m^2} (d_{\{x'\neq 0',n'\}})^2 d_{\{y'\}} 4 d_n. \\
    \end{split}
    \label{eq:Trnnotiny0notiny}
\end{equation}
\noindent Now that we have obtained simplified, but still somewhat general expressions for the different contributions making up $\text{Tr}[(\text{Tr}_{0...(n-1)x'}[|U\rangle \langle U|])^2]$, we can collect the terms we have computed and derive the one-tangling power as it depends on $\sum_{x'|y'}\text{Tr}[(\text{Tr}_{0...(n-1)x'}[|U\rangle \langle U|])^2]$. In other words, we now have to sum over all possible bipartitions of our fictitious system in order to compute the generalized nuclear one-tangling power. We now can collect the terms from Eqs.\ (\ref{eq:Trniny0iny}), (\ref{eq:Trnnotiny0iny}), (\ref{eq:Trniny0notiny}), and (\ref{eq:Trnnotiny0notiny}), and plug them into Eq.\ (\ref{eq:epqappendix}):
\begin{equation}
    \begin{split}
        \epsilon_{0...(n-1)|n}(U) &= 1 - \Big(\prod_{m = 0}^n \frac{d_m}{d_m + 1}\Big)\sum_{x'|y'} \text{Tr}[(\text{Tr}_{0...(n-1)x'}[|U\rangle \langle U|)^2]] \\
        &= 1 - \frac{1}{6}\Big(\prod_{m = 1}^n\frac{1}{d_m(d_m + 1)}\Big)\Big[\sum_{x'|y', 0'\in \{y'\},n'\in \{y'\}}(d_{\{x'\}})^2 d_{\{y'\neq 0',n'\}} 2d_n^2 \\
        &+ \sum_{x'\neq n'|y', 0'\in \{y'\},n'\notin \{y'\}} (d_{\{x'\neq n'\}})^2 d_{\{y'\neq 0'\}} 2d_n \\
        &+ \sum_{x'\neq 0'|y', 0'\notin \{y'\},n'\in \{y'\}} (d_{\{x'\neq 0'\}})^2 d_{\{y'\neq n'\}} 2(d_n^2 + |\text{Tr}[R_{n_0}^\dagger R_{n_1}]|^2) \\
        &+ \sum_{x'\neq 0',n'|y', 0'\notin \{y'\},n'\notin \{y'\}}(d_{\{x'\neq 0',n'\}})^2 d_{\{y' \}} 4 d_n \Big]. \\
    \end{split}
    \label{eq:collectedterms}
\end{equation}
\noindent The next step of this calculation is to compute the sum over all unordered bipartitions over $x'|y'$ for all four terms in the square brackets above. We will begin by outlining the solution to the first term in the square brackets of Eq.\ (\ref{eq:collectedterms}):
\begin{equation}
    \sum_{x'|y', 0'\in \{y'\},n'\in \{y'\}}(d_{\{x'\}})^2 d_{\{y'\}} 2d_n^2. 
    \label{eq:exfirstterm}
\end{equation}
\noindent Notice first that $2d_n d_{\{x'\}}d_{\{y'\neq 0',n'\}}$ is the dimension of the entire Hilbert space. We can rewrite this as $\prod_{m = 0}^n d_m$. Then it follows that
\begin{equation*}
\begin{split}
    \sum_{x'|y', 0'\in \{y'\},n'\in \{y'\}}(d_{\{x'\}})^2 d_{\{y'\}} 2d_n^2 &= d_n \prod_{m = 0}^n d_{m}\sum_{x'|y'}d_{\{x'\}} \\
    &= 2 d_n^2 \prod_{m = 1}^{n-1} d_m \sum_{x'|y'}d_{\{x'\}}.\\
\end{split} 
\end{equation*}
\noindent To illustrate the consideration of the sum over all possible permutations of the bipartition $x'|y'$, we begin with the explicit example of an ensemble of four nuclei. Recall that this term above arises when $n'\in y'$ and $0'\in y'$. Correspondingly, the possible biparitions of this example system include: ${\{.|0'1'2'3'4', 1'|0'2'3'4', 2'|0'1'3'4', 3'|0'1'2'4', 1'2'|0'3'4', 1'3'|0'2'4', 2'3'|0'1'4', 1'2'3'|0'4'\}}$ (recall that $``."$ denotes an empty bipartition and $``0'"$ pertains to the electronic spin in this copy Hilbert space). We choose $n' = 4'$ corresponding to the spin we are computing the one-tangling power with respect to. We do not assume explicit values for the dimensions of any of these nuclei. We first compute this sum just over the trace of the identity matrix to obtain:
\begin{equation}
    \begin{split}
    \sum_{x'|y'}d_{\{x'\}} &= 1 + d_{1'} + d_{2'} + d_{3'} + d_{1'}d_{2'} + d_{1'}d_{3'} +  d_{2'}d_{3'} + d_{1'}d_{2'}d_{3'} \\
    &= \prod_{m = 1}^{3}(1 + d_m). \\
    %\text{Tr}[\mathbb{I}_{.}] + \text{Tr}[\mathbb{I}_{1'}] + \text{Tr}[\mathbb{I}_{2'}] + \text{Tr}[\mathbb{I}_{3'}] + \text{Tr}[\mathbb{I}_{1'2'}] + \text{Tr}[\mathbb{I}_{1'3'}] + \text{Tr}[\mathbb{I}_{2'3'}] + \text{Tr}[\mathbb{I}_{1'2'3'}] \\
    \end{split}
\end{equation}
Plugging this result into Eq.\ (\ref{eq:exfirstterm}) gives
\begin{equation*}
    \sum_{x'|y', 0'\in \{y'\},n'\in \{y'\}}(d_{\{x'\}})^2 d_{\{y'\}} 2d_n^2 = 2 d_n^2 \prod_{m = 1}^{n-1} d_m(d_m + 1). \\
\end{equation*}
\noindent We can apply a similar procedure to the three other contributing terms in Eq.\ (\ref{eq:collectedterms}) to find that the identity matrices and sums over kronecker deltas for the various cases lead to the same overall factors as shown above. We next adjust these results to the general case and return to Eq.\ (\ref{eq:collectedterms}): 
\begin{equation}
    \begin{split}
        \epsilon_{0...(n-1)|n}(U) &= 1 - \frac{1}{6}\Big(\prod_{m = 1}^n\frac{1}{d_m(d_m + 1)}\Big)\Big[2d_n^2\prod_{m = 1}^{n - 1}d_m(d_m+1) + 2d_n \prod_{m = 1}^{n - 1}d_m(d_m+1)\\
        &+ 2(d_n^2 + d_n^2G_1^{(n)})\prod_{m = 1}^{n - 1}d_m(d_m+1) + 4 d_n \prod_{m = 1}^{n - 1}d_m(d_m+1) \Big] \\
        &= 1 - \frac{1}{6}\cdot \Big(\frac{1}{d_n(d_n + 1)}\Big)(4d_n^2 + 6d_n + 2d_n^2 G_1^{(n)}) \\
        &= \frac{1}{3}\frac{d_n}{d_n + 1}(1 - G_1^{(n)}), \\
        \label{eq:nuclearappendix}
    \end{split}
\end{equation}
\noindent which is the final expression as given in Eq.\ (\ref{eq:nuclearepanalytical}).

\subsection{Electronic one-tangling power}
\label{appendix:electroniconetangles}
This section outlines the original derivation for the electronic one-tangling power, which applies when the central electron is bipartitioned from surrounding nuclear spins, Eq.\ (\ref{eq:electronicepanalytical}). The process is similar to that of the previous section, but again note that there exists a shorter derivation in Appendix \ref{appendix:simplerderivationelec}. Here, we again aim to solve Eq.\ (\ref{eq:epq}) by first computing $\text{Tr}[(\text{Tr}_{1...nx'}[|U\rangle \langle U|])^2]$. As shown previously, we compute this quantity in parts, but unlike the nuclear one-tangling power case, we only consider two categories: one in which the primed index corresponding to the electron ($0'$) is in $\{y'\}$, and one in which it is not. We will first examine the case where $0' \notin {\{y'\}}$. Starting from Eq.\ (\ref{eq:simplefiedTr}) we have:
\begin{equation}
\begin{split}
0' \notin \{y'\}:\\
        \text{Tr}[(\text{Tr}_{1...nx'}[|U\rangle \langle U|])^2] &= \prod_{m = 0}^n\frac{1}{d_m^2} \sum_{i_0,j_0, i_{\{y'\}},j_{\{y'\}}} \Big|\text{Tr}[\langle i_{\{y' \}}|U^{\dagger}|i_{0}\rangle \langle j_0|U|j_{\{y'\}}\rangle ] \Big|^2. \\
\end{split}
\end{equation}
\noindent Notice that now $q = 0$ because we are bipartitioning the electron from the nuclear ensemble. This also means that $\{p\} = \{1...n\}$. Just as we did before, we can again plug in the form of the evolution, $U = \sigma_{00}\otimes R_{n_0} + \sigma_{11}\otimes R_{n_1}$ and simplify further for this case:
\begin{equation*}
    \begin{split}
        \text{Tr}[(\text{Tr}_{1...nx'}[|U\rangle \langle U|])^2] &= \prod_{m = 0}^n\frac{1}{d_m^2} \sum_{i_0,j_0, i_{\{y'\}},j_{\{y'\}}} \Big|\text{Tr}[\langle i_{\{y' \}}|\sigma_{00}\otimes R_{n_0}^{\dagger} + \sigma_{11}\otimes R_{n_1}^{\dagger}|i_0\rangle \langle j_0|\sigma_{00}\otimes R_{n_0} + \sigma_{11}\otimes R_{n_1}|j_{\{y'\}} \rangle ] \Big|^2 \\
        &= \prod_{m = 0}^n\frac{1}{d_m^2} \sum_{i_0,j_0, i_{\{y'\}},j_{\{y'\}}} \Big|\text{Tr}[(\sigma_{00}|i_0\rangle \otimes \langle i_{\{y' \}}|R_{n_0}^{\dagger} + \sigma_{11}|i_0\rangle \otimes \langle i_{\{y' \}}|R_{n_1}^{\dagger})\\
        &\cdot ( \langle j_0|\sigma_{00} \otimes R_{n_0}|j_{\{y'\}}\rangle  + \langle j_0|\sigma_{11}\otimes R_{n_1}|j_{\{y'\}}\rangle)] \Big|^2 \\
        &= \prod_{m = 0}^n\frac{1}{d_m^2} \sum_{i_0,j_0, i_{\{y'\}},j_{\{y'\}}} \Big|\text{Tr}[\langle j_0|\sigma_{00}|i_0\rangle \otimes \langle i_{\{y'\}}|R_{n_0}^{\dagger}R_{n_0}|j_{\{y'\}}\rangle  + \langle j_0|\sigma_{11}|i_0\rangle \otimes \langle i_{\{y' \}}|R_{n_1}^{\dagger}R_{n_1}|j_{\{y'\}}\rangle] \Big|^2 \\
        &= \prod_{m = 0}^n\frac{1}{d_m^2} \sum_{i_0,j_0, i_{\{y'\}},j_{\{y'\}}} \Big|\delta_{0}^{j_0}\delta_{i_0}^0 \text{Tr}[\langle i_{\{y'\}}|R_{n_0}^{\dagger}R_{n_0}|j_{\{y'\}}\rangle]  + \delta_{1}^{j_0}\delta_{i_0}^1\text{Tr}[\langle i_{\{y' \}}|R_{n_1}^{\dagger}R_{n_1}|j_{\{y'\}}\rangle] \Big|^2. \\
    \end{split}
\end{equation*}
\noindent At this point, we see again the amplitudes that contribute an identity matrix to the expression: $\langle i_{\{y' \}}|R_{n_j}^{\dagger}R_{n_j}|j_{\{y'\}}\rangle$. In this case, we can write this generally as $\mathbb{I}_{\{x'\neq 0'\}}$ to describe the identity matrix of dimension determined by the spins in $\{x'\}$ that are not the bipartitioned electron. We again make no assumptions about the dimensions of these nuclei until later in the calculation, so for now we write this generally. The squared trace of this amplitude amounts to $(\text{Tr}[\langle i_{\{y' \}}|R_{n_j}^{\dagger}R_{n_j}|j_{\{y'\}}\rangle])^2 = (\text{Tr}[\mathbb{I}_{\{x'\neq 0'\}}])^2 = (d_{\{x'\neq 0'\}})^2$. The remaining spins that are included in $\{y'\}$ but are not acted on by $R_{n_j}$ lead to an additional factor of $\sum_{i_{\{y'\}},j_{\{y'\}}} \delta_{j_{\{y'\}}}^{i_{\{y'\}}} = d_{\{y'\}}$. Continuing, the previous expression simplifies further: 
\begin{equation*}
    \begin{split}
        \text{Tr}[(\text{Tr}_{1...nx'}[|U\rangle \langle U|])^2] &= \prod_{m = 0}^n\frac{1}{d_m^2} (d_{\{x'\neq 0'\}})^2 \sum_{i_0,j_0, i_{\{y'\}},j_{\{y'\}}} \delta_{j_{\{y'\}}}^{i_{\{y'\}}} \Big|\delta_{0}^{j_0}\delta_{i_0}^0  + \delta_{1}^{j_0}\delta_{i_0}^1 \Big|^2. \\
        &= \prod_{m = 0}^n\frac{1}{d_m^2} (d_{\{x'\neq 0'\}})^2 d_{\{y'\}} \sum_{i_0,j_0} \Big|\delta_{0}^{j_0}\delta_{i_0}^0  + \delta_{1}^{j_0}\delta_{i_0}^1\Big|^2 \\
    \end{split}
\end{equation*}
\noindent Finally, because the squared sum over the kronecker delta of the primed indices with the electronic spin states amounts to the squared dimension of the electron: $\Big|\delta_{0}^{j_0}\delta_{i_0}^0  + \delta_{1}^{j_0}\delta_{i_0}^1\Big|^2 = \delta_{0}^{j_0}\delta_{i_0}^0  + \delta_{1}^{j_0}\delta_{i_0}^1 = 2$, we obtain:
\begin{equation}
    \text{Tr}[(\text{Tr}_{1...nx'}[|U\rangle \langle U|])^2] = \prod_{m = 0}^n\frac{1}{d_m^2} \cdot 2(d_{\{x'\neq 0'\}})^2 d_{\{y'\}}. \\
    \label{eq:firstcontribution}
\end{equation}
\noindent We now can account for the other half of the contributions to the electronic one-tangling power, which arise when $0' \in \{y'\}$. This case yields a nontrivial term that depends on the generalized Makhlin invariants corresponding to each nuclear spin interacting with the electron. To start, we have:
\begin{equation*}
    \begin{split}
    0 \in \{y'\}:\\
        \text{Tr}[(\text{Tr}_{1...nx'}[|U\rangle \langle U|])^2] &= \prod_{m = 0}^n\frac{1}{d_m^2} \sum_{i_0,j_0, i_{\{y'\}},j_{\{y'\}}} \Big|\text{Tr}[\langle i_{\{y' \}}|U^{\dagger}|i_0\rangle \langle j_0|U|j_{\{y'\}}\rangle ] \Big|^2 \\
        &= \prod_{m = 0}^n\frac{1}{d_m^2} \sum_{i_0,j_0, i_{\{y'\}},j_{\{y'\}}} \Big|\text{Tr}[\langle i_{\{y'\}}|\sigma_{00}\otimes R_{n_0}^{\dagger} + \sigma_{11}\otimes R_{n_1}^{\dagger}|i_0 \rangle \langle j_0|\sigma_{00}\otimes R_{n_0} + \sigma_{11}\otimes R_{n_1}|j_{\{y'\}}\rangle ] \Big|^2 \\
        &= \prod_{m = 0}^n\frac{1}{d_m^2} \sum_{i_0,j_0, i_{\{y'\}},j_{\{y'\}}} \Big|\text{Tr}[(\langle i_{0'}|\sigma_{00}|i_0\rangle \otimes \langle i_{\{y'{\neq 0' \}}}|R_{n_0}^{\dagger} + \langle y_{0'}|\sigma_{11}|i_0\rangle \otimes \langle i_{\{y'{\neq 0'}\}}|R_{n_1}^{\dagger})\\
        &( \langle j_0|\sigma_{00}|j_{0'}\rangle \otimes R_{n_0}|j_{\{y'{\neq 0'}\}}\rangle  + \langle j_0|\sigma_{11}|j_{\{y'\neq 0'\}}\rangle\otimes R_{n_1}|j_{\{y'{\neq 0'}\}} \rangle)] \Big|^2 \\
        &= \prod_{m = 0}^n\frac{1}{d_m^2} \sum_{i_0,j_0, i_{\{y'\}},j_{\{y'\}}} \Big|\text{Tr}[(\langle i_{0'}|\sigma_{00}|i_0 \rangle \langle j_0|\sigma_{00}|j_{0'}\rangle] \text{Tr}[\langle i_{\{y' \neq 0' \}}|R_{n_0}^{\dagger}R_{n_0}|j_{\{y'{\neq 0'}\}}\rangle] \\
        &+ \text{Tr}[(\langle i_{0'}|\sigma_{00}|i_0 \rangle \langle j_0|\sigma_{11}|j_{0'}\rangle] \text{Tr}[\langle i_{\{y'{\neq 0'} \}}|R_{n_0}^{\dagger}R_{n_1}|j_{\{y'{\neq 0'}\}}\rangle] \\
        &+ \text{Tr}[\langle i_{0'}|\sigma_{11}|i_0\rangle\langle j_0|\sigma_{00}|j_{0'} \rangle] \text{Tr}[\langle i_{\{y'{\neq 0'}\}}|R_{n_1}^{\dagger}R_{n_0}|j_{\{y'{\neq 0'}\}}\rangle] \\
        &+ \text{Tr}[\langle i_{0'}|\sigma_{11}|i_0 \rangle\langle j_0|\sigma_{11}|j_{0'}\rangle] \text{Tr}[\langle i_{\{y'{\neq 0'}\}}|R_{n_1}^{\dagger}R_{n_1}|j_{\{y'{\neq 0'}\}}\rangle]) \Big|^2. \\
        \label{eq:electrontangle}
    \end{split}
\end{equation*}
\noindent Of the four terms above, the nuclear parts of the first and fourth will yield $\sum_{i_{\{y'\neq 0'\}},j_{\{y'\neq 0'\}}}\text{Tr}[\langle i_{\{y' \neq 0' \}}|R_{n_j}^{\dagger}R_{n_j}|j_{\{y'{\neq 0'}\}}\rangle] = \text{Tr}[\mathbb{I}_{\{x'\}}]\sum_{i_{\{y'\neq 0'\}},j_{\{y'\neq 0'\}}}\delta_{j_{\{y'\neq 0'\}}}^{i_{\{y'\neq 0'\}}}$. We will see in forthcoming steps that the second and third terms lead to a nontrivial result amounting to the generalized Makhlin invariants:
\begin{equation*}
    \begin{split}
        \text{Tr}[(\text{Tr}_{1...nx'}[|U\rangle \langle U|])^2] &= \prod_{m = 0}^n\frac{1}{d_m^2} \sum_{i_0,j_0, i_{\{y'\}},j_{\{y'\}}} \Big|\big(\delta_{0}^{i_{0'}}\delta_{i_{0}}^{0} \delta_{0}^{j_0}\delta_{j_{0'}}^{0} + \delta_{1}^{i_{0'}}\delta_{i_{0'}}^{1} \delta_{1}^{j_0}\delta_{j_{0'}}^{1}\big)\text{Tr}[\mathbb{I}_{\{x'\}}] \delta_{j_{\{y'{\neq 0'}\}}}^{i_{\{y' {\neq 0'}\}}} \\
        &+ \delta_{0}^{i_{0'}}\delta_{i_0}^{0} \delta_{1}^{j_0}\delta_{j_{0'}}^{1} \text{Tr}[\langle i_{\{y'{\neq 0'}} \}|R_{n_0}^{\dagger}R_{n_1}|j_{\{y'{\neq 0'}\}}\rangle] + \big(\delta_{1}^{i_{0'}}\delta_{i_0}^{1} \delta_{0}^{j_0}\delta_{j_{0'}}^{0} \text{Tr}[\langle i_{\{y'{\neq 0'}\}}|R_{n_1}^{\dagger}R_{n_0}|j_{\{y'{\neq 0'}\}} \rangle]\big) \Big|^2 \\
        &= \prod_{m = 0}^n\frac{1}{d_m^2} \sum_{i_0,j_0, i_{\{y'\}},j_{\{y'\}}} \Big[ \Big|\big(\delta_{0}^{i_{0'}}\delta_{i_0}^{0} \delta_{0}^{j_0}\delta_{j_{0'}}^{0} + \delta_{1}^{i_{0'}}\delta_{i_0}^{1} \delta_{1}^{j_0}\delta_{j_{0'}}^{1}\big)\text{Tr}[\mathbb{I}_{\{x'\}}] \delta_{j_{{\{y'{\neq 0'}\}}}}^{i_{\{y'{\neq 0'}\}}}\Big|^2 \\
        &+ \Big|\delta_{0}^{i_{0'}}\delta_{i_0}^{0} \delta_{1}^{j_0}\delta_{j_{0'}}^{1} \text{Tr}[\langle i_{\{y'{\neq 0'}\}}|R_{n_1}^{\dagger}R_{n_0}|j_{\{y'{\neq 0'}\}}\rangle] + \delta_{1}^{i_{0'}}\delta_{i_0}^{1} \delta_{0}^{j_0}\delta_{j_{0'}}^{0}  \text{Tr}[\langle i_{\{y' {\neq 0'} \}}|R_{n_0}^{\dagger}R_{n_1}|\{j_{{y' \neq 0'}\}}\rangle] \Big|^2 \\
        &+ 2 \text{Re}\Big[  \big(\delta_{0}^{i_{0'}}\delta_{i_0}^{0} \delta_{0}^{j_0}\delta_{j_{0'}}^{0} + \delta_{1}^{i_{0'}}\delta_{i_0}^{1} \delta_{1}^{j_0}\delta_{j_{0'}}^{1}\big)\text{Tr}[\mathbb{I}_{\{x'\}}]\delta_{j_{{\{y'{\neq 0'}\}}}}^{i_{\{y'{\neq 0'}\}}} \\
        &\cdot \big( \delta_{0}^{i_{0'}}\delta_{i_0}^{0} \delta_{1}^{j_0}\delta_{j_{0'}}^{1} \text{Tr}[\langle j_{\{y'{\neq 0'}\}}|R_{n_0}^{\dagger}R_{n_1}|i_{\{y'{\neq 0'}\}}\rangle] + \delta_{1}^{i_{0'}}\delta_{i_0}^{1} \delta_{0}^{j_0}\delta_{j_{0'}}^{0}  \text{Tr}[\langle j_{\{y'{\neq 0'} \}}|R_{n_1}^{\dagger}R_{n_0}|i_{\{y'{\neq 0'}\}} \rangle] \big)\Big]. \\
    \end{split}
\end{equation*}
\noindent We recognize that in the first term, $(\text{Tr}[\mathbb{I}_{\{x'\}}])^2\sum_{i_{\{y'\neq 0'\}},j_{\{y'\neq 0'\}}}\delta_{j_{\{y'\neq 0'\}}}^{i_{\{y'\neq 0'\}}} = (d_{\{x'\}})^2 d_{\{y'\neq 0'\}}$. At this point, we can also note that the third term (beginning with $2\text{Re}[...]$) is zero when we sum over $i_0,i_{0'},j_0, j_{0'}$. This leaves:
\begin{equation}
    \begin{split}
        \text{Tr}[(\text{Tr}_{1...nx'}[|U\rangle \langle U|])^2] &= \prod_{m = 0}^n\frac{1}{d_m^2} \Big[2 (d_{\{x'\}})^2 d_{\{y'\neq 0'\}} + \sum_{i_0,j_0,i_{\{y'\}},j_{\{y'\}}}\Big|\delta_{0}^{i_{0'}}\delta_{i_0}^{0} \delta_{1}^{j_0}\delta_{j_{0'}}^{1} \text{Tr}[\langle i_{\{y'{\neq 0'}\}}|R_{n_1}^{\dagger}R_{n_0}|j_{\{y'{\neq 0'}\}}\rangle] \\
        &+ \delta_{1}^{i_{0'}}\delta_{i_0}^{1} \delta_{0}^{j_0}\delta_{j_{0'}}^{0}  \text{Tr}[\langle i_{\{y' {\neq 0'} \}}|R_{n_0}^{\dagger}R_{n_1}|\{j_{{y' \neq 0'}\}}\rangle] \Big|^2 \Big]. \\
    \end{split}
    \label{eq:fullmodsq}
\end{equation}
\noindent Next, we can compute the second term (the modulus squared of the traces of inner products of $R_{n_i}^{\dagger}R_{n_j}$). Isolating this term, we have:
\begin{equation*}
    \begin{split}
        &\sum_{i_0, j_0, i_{\{y'\}},j_{\{y'\}}} \Big|\delta_{0}^{i_{0'}}\delta_{i_0}^{0} \delta_{1}^{j_0}\delta_{j_{0'}}^{1} \text{Tr}[\langle i_{\{y'\neq 0'\}}|R_{n_1}^{\dagger}R_{n_0}|j_{\{y' \neq 0'\}}\rangle] + \delta_{1}^{i_{0'}}\delta_{i_0}^{1} \delta_{0}^{j_0}\delta_{j_{0'}}^{0}  \text{Tr}[\langle i_{\{y'\neq 0'\}}|R_{n_0}^{\dagger}R_{n_1}|j_{\{y'\neq 0'\}}\rangle] \Big|^2 \\ 
        &= \sum_{i_0, j_0, i_{\{y'\}},j_{\{y'\}}} \Big[ \Big|\delta_{0}^{i_{0'}}\delta_{i_0}^{0} \delta_{1}^{j_0}\delta_{j_{0'}}^{1} \text{Tr}[\langle i_{\{y'\neq 0'\}}|R_{n_1}^{\dagger}R_{n_0}|j_{\{y'\neq 0'\}}\rangle]\Big|^2 + \Big|\delta_{1}^{i_{0'}}\delta_{i_{0}}^{1} \delta_{0}^{j_0}\delta_{j_{0'}}^{0}  \text{Tr}[\langle i_{\{y'\neq 0'\}}|R_{n_0}^{\dagger}R_{n_1}|j_{\{y'\neq 0'\}} \rangle] \Big|^2  \\
        &+ 2 \text{Re}\big[\delta_{0}^{i_{0'}}\delta_{i_0}^{0} \delta_{1}^{j_0}\delta_{j_{0'}}^{1} \text{Tr}[\langle i_{\{y'\neq 0'\}}|R_{n_1}^{\dagger}R_{n_0}|j_{\{y'\neq 0'\}}\rangle] \cdot \delta_{1}^{i_{0'}}\delta_{i_{0}}^{1} \delta_{0}^{j_0}\delta_{j_{0'}}^{0}  \text{Tr}[\langle j_{\{y'\neq 0'\}}|R_{n_1}^{\dagger}R_{n_0}|i_{\{y'\neq 0'\}} \rangle] \big] \Big]. \\
\end{split}
\end{equation*}
\noindent The last term on the RHS above is equal to zero under the sum over $i_0,j_0, i_{0'}, j_{0'}$, leaving:
\begin{equation*}
    \begin{split}
    &\sum_{i_0, j_0, i_{\{y'\}},j_{\{y'\}}} \Big|\delta_{0}^{i_{0'}}\delta_{i_0}^{0} \delta_{1}^{j_0}\delta_{j_{0'}}^{1} \text{Tr}[\langle i_{\{y'\neq 0'\}}|R_{n_1}^{\dagger}R_{n_0}|j_{\{y' \neq 0'\}}\rangle] + \delta_{1}^{i_{0'}}\delta_{i_0}^{1} \delta_{0}^{j_0}\delta_{j_{0'}}^{0}  \text{Tr}[\langle i_{\{y'\neq 0'\}}|R_{n_0}^{\dagger}R_{n_1}|j_{\{y'\neq 0'\}}\rangle] \Big|^2 \\ 
    &= \sum_{i_{\{y'\neq 0'\}},j_{\{y' \neq 0'\}}} \Big|\text{Tr}[\langle i_{\{y'\neq 0'\}}|R_{n_1}^{\dagger}R_{n_0}|j_{\{y'\neq 0'\}}\rangle]\Big|^2 + \Big|\text{Tr}[\langle i_{\{y'\neq 0'\}}|R_{n_0}^{\dagger}R_{n_1}|j_{\{y'\neq 0'\}}\rangle] \Big|^2. \\
    \end{split}
\end{equation*}
\noindent Note that we can split up the nuclear spin operators in terms of the the parts that act on spins in $\{x'\}$ and spins in $\{y'\}$. So if we write them as $R_{n_i} = R_{n_i}^{\{y'\}} \otimes R_{n_i}^{\{x'\}}$, we obtain: 
\begin{equation*}
    \begin{split}
    &\sum_{i_0, j_0, i_{\{y'\}},j_{\{y'\}}} \Big|\delta_{0}^{i_{0'}}\delta_{i_0}^{0} \delta_{1}^{j_0}\delta_{j_{0'}}^{1} \text{Tr}[\langle i_{\{y'\neq 0'\}}|R_{n_1}^{\dagger}R_{n_0}|j_{\{y' \neq 0'\}}\rangle] + \delta_{1}^{i_{0'}}\delta_{i_0}^{1} \delta_{0}^{j_0}\delta_{j_{0'}}^{0}  \text{Tr}[\langle i_{\{y'\neq 0'\}}|R_{n_0}^{\dagger}R_{n_1}|j_{\{y'\neq 0'\}}\rangle] \Big|^2 \\ 
    &= \sum_{i_{\{y'\neq 0'\}},j_{\{y'\neq 0'\}}} \Big[ \Big|\text{Tr}[\langle i_{\{y'\neq 0'\}}|(R_{n_0}^{\{y'\}}\otimes R_{n_0}^{\{x'\}})^{\dagger}(R_{n_1}^{\{y'\}}\otimes R_{n_1}^{\{x'\}})|j_{\{y'\neq 0'\}}\rangle]\Big|^2 \\
    &+ \Big|\text{Tr}[\langle i_{\{y'\neq 0'\}} \}|(R_{n_1}^{\{y'\}}\otimes R_{n_1}^{\{x'\}})^{\dagger}(R_{n_0}^{\{y'\}}\otimes R_{n_0}^{\{x'\}})|j_{\{y'\neq 0'\}}\rangle] \Big|^2 \Big]\\
    &= \sum_{i_{\{y'\neq 0'\}},j_{\{y'\neq 0'\}}} \Big|\text{Tr}[(R_{n_0}^{\{x'\}})^{\dagger}R_{n_1}^{\{x'\}}] \text{Tr}[\langle i_{\{y'\neq 0'\}}|(R_{n_0}^{\{y'\}})^{\dagger}R_{n_1}^{\{y'\}}|j_{\{y'\neq 0'\}}\rangle ] \Big|^2 \\ 
    &+ \Big|\text{Tr}[(R_{n_1}^{\{x'\}})^{\dagger}R_{n_0}^{\{x'\}}] \text{Tr}[\langle i_{\{y'\neq 0'\}}|(R_{n_1}^{\{y'\}})^{\dagger}R_{n_0}^{\{y'\}}|j_{\{y'\neq 0'\}}\rangle] \Big|^2. \\
    \end{split}
\end{equation*}
\noindent If we expand the first part of the RHS above, we find that we can use the completeness relation:
\begin{equation*}
\begin{split}
    &\sum_{i_0, j_0, i_{\{y'\}},j_{\{y'\}}} \Big|\delta_{0}^{i_{0'}}\delta_{i_0}^{0} \delta_{1}^{j_0}\delta_{j_{0'}}^{1} \text{Tr}[\langle i_{\{y'\neq 0'\}}|R_{n_1}^{\dagger}R_{n_0}|j_{\{y' \neq 0'\}}\rangle] + \delta_{1}^{i_{0'}}\delta_{i_0}^{1} \delta_{0}^{j_0}\delta_{j_{0'}}^{0}  \text{Tr}[\langle i_{\{y'\neq 0'\}}|R_{n_0}^{\dagger}R_{n_1}|j_{\{y'\neq 0'\}}\rangle] \Big|^2 \\
    &= \sum_{i_{\{y'\neq 0'\}},j_{\{y'\neq 0'\}}} \big(\text{Tr}[(R_{n_0}^{\{x'\}})^{\dagger}R_{n_1}^{\{x'\}}] \text{Tr}[\langle i_{\{y'\neq 0'\}}|(R_{n_0}^{\{y'\}})^{\dagger}R_{n_1}^{\{y'\}}|j_{\{y'\neq 0'\}}\rangle ] \big) \cdot \big(\text{Tr}[(R_{n_1}^{\{x'\}})^{\dagger}R_{n_0}^{\{x'\}}] \text{Tr}[\langle j_{\{y'\neq 0'\}}|(R_{n_1}^{\{y'\}})^{\dagger}R_{n_0}^{\{y'\}}|i_{\{y'\neq 0'\}}\rangle] \big) \\ 
    &+ \Big|\text{Tr}[(R_{n_1}^{\{x'\}})^{\dagger}R_{n_0}^{\{x'\}}]\text{Tr}[\langle i_{\{y'\neq 0'\}}|(R_{n_1}^{\{y'\}})^{\dagger}R_{n_0}^{\{y'\}}|j_{\{y'\neq 0'\}}\rangle] \Big|^2 \Big]\\
    &= \big|\text{Tr}[(R_{n_0}^{\{x'\}})^{\dagger}R_{n_1}^{\{x'\}}]\big|^2 \sum_{i_{\{y'\neq 0'\}},j_{\{y'\neq 0'\}}} \Big[ \langle i_{\{y'\neq 0'\}}|(R_{n_0}^{\{y'\}})^{\dagger}R_{n_1}^{\{y'\}}\underbrace{|j_{\{y'\neq 0'\}}\rangle \langle j_{\{y'\neq 0'\}}|}_{=\mathbb{I}}(R_{n_1}^{\{y'\}})^{\dagger}R_{n_0}^{\{y'\}}|i_{\{y'\neq 0'\}}\rangle \\ 
    &+ \Big|\text{Tr}[(R_{n_1}^{\{x'\}})^{\dagger}R_{n_0}^{\{x'\}}]\text{Tr}[\langle i_{\{y'\neq 0'\}}|(R_{n_1}^{\{y'\}})^{\dagger}R_{n_0}^{\{y'\}}|j_{\{y'\neq 0'\}}\rangle] \Big|^2 \Big] \\
    &= \big|\text{Tr}[(R_{n_0}^{\{x'\}})^{\dagger}R_{n_1}^{\{x'\}}]\big|^2 \sum_{i_{\{y'\neq 0'\}},j_{\{y'\neq 0'\}}} \Big[\underbrace{\text{Tr}[\langle i_{\{y'\neq 0'\}}|(R_{n_0}^{\{y'\}})^{\dagger}R_{n_1}^{\{y'\}}(R_{n_1}^{\{y'\}})^{\dagger}R_{n_0}^{\{y'\}}|i_{\{y'\neq 0'\}}\rangle]}_{{=1}} \\
    &+ \Big|\text{Tr}[(R_{n_1}^{\{x'\}})^{\dagger}R_{n_0}^{\{x'\}}]\text{Tr}[\langle i_{\{y'\neq 0'\}}|(R_{n_1}^{\{y'\}})^{\dagger}R_{n_0}^{\{y'\}}|j_{\{y'\neq 0'\}}\rangle] \Big|^2 \Big]. \\
    \end{split}
\end{equation*}
\noindent We can apply the same result to the second part of the expression on the RHS above, leaving:
\begin{equation}
    \begin{split}
        &\sum_{i_0, j_0, i_{\{y'\}},j_{\{y'\}}} \Big|\delta_{0}^{i_{0'}}\delta_{i_0}^{0} \delta_{1}^{j_0}\delta_{j_{0'}}^{1} \text{Tr}[\langle i_{\{y'\neq 0'\}}|R_{n_1}^{\dagger}R_{n_0}|j_{\{y' \neq 0'\}}\rangle] + \delta_{1}^{i_{0'}}\delta_{i_0}^{1} \delta_{0}^{j_0}\delta_{j_{0'}}^{0}  \text{Tr}[\langle i_{\{y'\neq 0'\}}|R_{n_0}^{\dagger}R_{n_1}|j_{\{y'\neq 0'\}}\rangle] \Big|^2 \\
        &= \sum_{i_{\{y'\neq 0'\}},j_{\{y'\neq 0'\}}} \Big(\big|\text{Tr}[(R_{n_0}^{\{x'\}})^{\dagger}R_{n_1}^{\{x'\}}]\big|^2
        + \big|\text{Tr}[(R_{n_1}^{\{x'\}})^{\dagger}R_{n_0}^{\{x'\}}]\big|^2 \Big) \\
        &= 2 \big|\text{Tr}[(R_{n_0}^{\{x'\}})^{\dagger}R_{n_1}^{\{x'\}}]\big|^2 \sum_{i_{\{y'\}},j_{\{y'\}}} 1 \\
        &= 2 d_{\{y'\neq 0'\}} \big|\text{Tr}[(R_{n_0}^{\{x'\}})^{\dagger}R_{n_1}^{\{x'\}}]\big|^2. \\
    \end{split}
    \label{eq:secondterm}
\end{equation}
\noindent Observe that $\big|\text{Tr}[(R_{n_0}^{\{x'\}})^{\dagger}R_{n_1}^{\{x'\}}]\big|^2$ is composed of a tensor product of rotations acting on each nucleus in $\{x'\}$. Therefore, we can generally write $R_{n_j}^{\{x'\}} = R_{n_j}^{x'_1}\otimes R_{n_j}^{x'_2}\otimes ...\otimes R_{n_j}^{x'_{n+1-k}}$, and our tensor product turns into a product of generalized Makhlin invariants. Therefore, we can write $\big|\text{Tr}[(R_{n_0}^{\{x'\}})^{\dagger}R_{n_1}^{\{x'\}}]\big|^2 = (d_{\{x'\}})^2 G_{1}^{(\{x'\})}$ and plug the result, Eq.\ (\ref{eq:secondterm}), into Eq.\ (\ref{eq:fullmodsq}) to find:
\begin{equation}
    \begin{split}
        \text{Tr}[(\text{Tr}_{1...nx'}[|U\rangle \langle U|])^2] &= \prod_{m = 0}^n\frac{1}{d_m^2} \Big[2 d_{\{y'\neq 0'\}}(d_{\{x'\}})^2 + 2 d_{\{y'\neq 0'\}} (d_{\{x'\}})^2 G_{1}^{(\{x'\})}\Big] \\
        &= \prod_{m = 0}^n\frac{1}{d_m^2} 2 d_{\{y'\neq 0'\}} (d_{\{x'\}})^2 \Big[1 + G_{1}^{(\{x'\})}\Big]. 
    \end{split}
    \label{eq:secondcontribution}
\end{equation}
\noindent We now have the terms contributing to the electronic one-tangling power, Eqs.\ (\ref{eq:firstcontribution}) and (\ref{eq:secondcontribution}). We next plug these into Eq.\ (\ref{eq:epqappendix}) and acquire the expression:
\begin{equation}
    \begin{split}
        \epsilon_{1...n|0}(U) &= 1 - \prod_{m = 0}^n \Big(\frac{d_m}{d_m + 1}\Big)\sum_{x'|y'}\text{Tr}[(\text{Tr}_{1...nx'}[|U\rangle \langle U|])^2]\\
        &= 1 - \prod_{m = 0}^n \Big(\frac{1}{d_m(d_m + 1)}\Big)\Big[2 \sum_{x'|y', 0'\notin \{y'\}} (d_{\{x'\neq 0'\}})^2 d_{\{y'\}} + 2\sum_{x'|y', 0'\in \{y'\}} (d_{\{x'\}})^2 d_{\{y'\neq 0'\}}(1 + G_1^{(\{x'\})})\Big]\\
        &= 1 - \prod_{m = 0}^n \Big(\frac{1}{d_m + 1}\Big)\Big[\sum_{x'|y', 0'\notin \{y'\}} d_{\{x'\neq 0'\}} + \sum_{x'|y', 0'\in \{y'\}} d_{\{x'\}} (1 + G_1^{(\{x'\})})\Big].\\
    \end{split}
    \label{eq:electronicepqcollectedterms}
\end{equation}
\noindent Observe that once again, each term contains the dimension of the entire Hilbert space, $\prod_{m = 0}^n d_m$. Next we consider the sums over the permutations of possible spins in the bipartition $x'|y'$. We again consult the familiar four-nuclear example from the previous section and begin by focusing on the first term above. This first term arises from the case when $0'\in \{y'\}$, which corresponds to the $16$ possible configurations:
\begin{equation*}
    \begin{matrix}
        0'|1'2'3'4' & 0'1'|2'3'4' & 0'1'2|'3'4' & 0'1'2'3'|4' \\
        0'1'2'3'4'|. & 0'2'|1'3'4' & 0'3'|1'2'4' & 0'4'|1'2'3' \\
        0'1'3'|2'4' & 0'1'4'|2'3' & 0'2'3'|1'4' & 0'2'4'|1'3' \\
        0'3'4'|1'2' & 0'1'2'4'|3' & 0'1'3'4'|2' & 0'2'3'4'|1'. \\
    \end{matrix}
\end{equation*}
\noindent Plugging these possibilities into the first term, we obtain:
\begin{equation*}
    \begin{split}
        \sum_{x'|y', 0'\notin \{y'\}} d_{\{x'\neq 0'\}} &= 1 + d_{1'} + d_{1'}d_{2'} + d_{1'}d_{2'}d_{3'} +...+ d_{1'}d_{3'}d_{4'} + d_{2'}d_{3'}d_{4'}\\
        &= \prod_{m = 1}^4 (1 + d_m).
    \end{split}
\end{equation*}
\noindent Note that unlike in the derivation for the nuclear one-tangling power, this product is now bounded by the fourth nuclear spin, rather than the third. This generalizes to
\begin{equation*}
    \sum_{x'|y', 0'\notin \{y'\}} d_{\{x'\neq 0'\}} = \prod_{m = 1}^n (1 + d_m).
\end{equation*}
\noindent Now we can consider the second term in Eq.\ (\ref{eq:electronicepqcollectedterms}). This term arose from the case when $0' \in \{y'\}$, which is illustrated by the list of possible configurations above, if we flip the indexed spins on each side of the bipartition. We find that $\sum_{x'|y', 0'\in \{y'\}} d_{\{x'\}} + d_{\{x'\}}G_1^{(\{x'\})} = \prod_{m = 1}^4 (1 + d_m) + \prod_{m = 1}^4 (1 + d_m G_1^{(m)}) $ for this example. Generally, this is expressed as:
\begin{equation*}
    \begin{split}
        \sum_{x'|y', 0'\in \{y'\}} d_{\{x'\}} (1 + G_1^{(x')}) &= \prod_{m = 1}^n (1 + d_m) + \prod_{m = 1}^n (1 + d_m G_1^{(m)}). \\
    \end{split}
\end{equation*}
\noindent Plugging this result back into Eq.\ (\ref{eq:electronicepqcollectedterms}) yields the final expression for the electronic one-tangling power (Eq.\ (\ref{eq:electronicepanalytical})):
\begin{equation}
    \begin{split}
        \epsilon_{1...n|0}(U) &= 1 - \prod_{m = 0}^n \Big(\frac{d_m}{d_m + 1}\Big)\sum_{x'|y'}\text{Tr}[(\text{Tr}_{1...nx'}[|U\rangle \langle U|])^2]\\
        &= 1 - \prod_{m = 0}^n \Big(\frac{1}{d_m + 1}\Big)\Big[\sum_{x'|y', 0'\notin \{y'\}} d_{\{x'\neq 0'\}} + \sum_{x'|y', 0'\in \{y'\}} d_{\{x'\}} (1 + G_1^{(\{x'\})})\Big]\\
        &= 1 - \prod_{m = 0}^n \Big(\frac{1}{d_m + 1}\Big)\Big[\prod_{m = 1}^n (1 + d_m) + \prod_{m = 1}^n (1 + d_m) + \prod_{m = 1}^n (1 + d_m G_1^{(\{m\})})\Big] \\
        &= \frac{1}{3} - \frac{1}{3} \prod_{m = 1}^n \Big(\frac{1}{d_m + 1} \Big)(1 + d_{m}G_1^{(m)}).\\
    \end{split}
    \label{eq:electronicappendix}
\end{equation}
\end{widetext}

\section{Higher spin example}
\label{Appendix:Spinninehalfex}
In this section, we offer an analysis of the entanglement dynamics in the QD ensemble for higher total spin and varied nuclear spin species. First, we consider the simple case where there is a single $^{115}$In coupled to the central electron. This is followed by a study of the one-tangling power for a mixed ensemble including both $^{115}$In and $^{71}$Ga nuclei, in analogy to the analysis in the main text. 

\begin{figure}
    \centering
    \includegraphics[width=\linewidth]{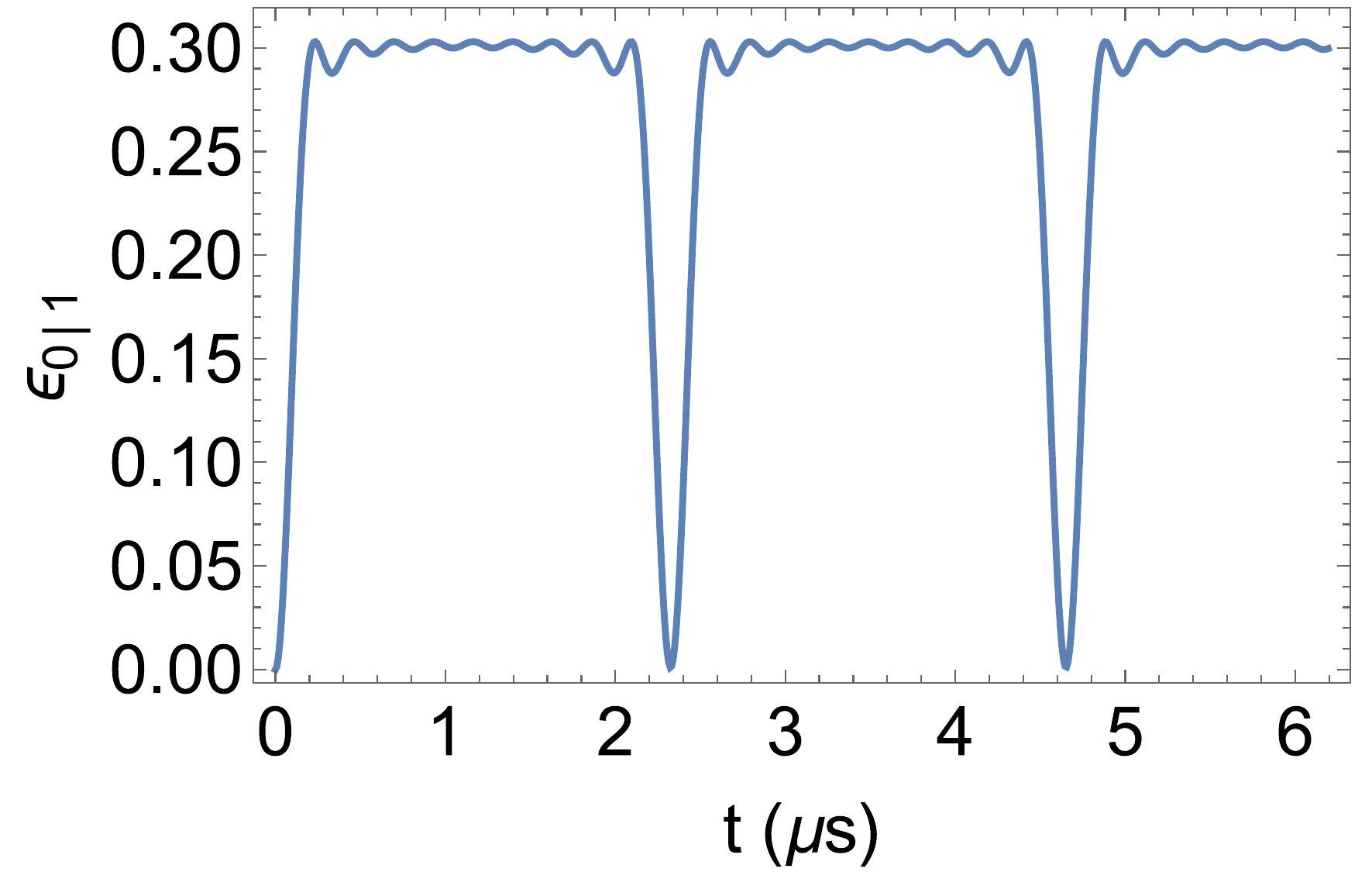}
    \caption{Free evolution one-tangling power quantifying the amount of entanglement between the QD electron and a single $^{115}$In nucleus. We assume $\omega_1/2\pi = 9.33$ MHz/T ($B = 1$ T), $a_1/2\pi = 0.43$ MHz, $a^{\mathrm{nc}}_1/2\pi = 0.034$ MHz, $\Delta_{\text{Q},1}/2\pi = 0.058$ MHz, and $\theta_1 = \frac{\pi}{3}$. This illustrates how the dips in the global maxima of the plot correspond to the spin of the nucleus entangled with the electron.}
    \label{fig:singlenucfree9/2}
\end{figure}

In general, the one-tangling power with respect to a single spin-$9/2$ nucleus and the QD electron is given as: 
\begin{equation}
\epsilon_{0|1} = \frac{10}{33}(1 - |G_1|). \\
    \label{eq:spinninehalfsinglenuc}
\end{equation}
\noindent During free evolution, where $U_{\text{free}} = |0\rangle \langle 0| \otimes e^{-i H_0 t} + |1\rangle \langle 1| \otimes e^{-i H_1 t}$, the entanglement arising between the electron and a single nuclear spin looks similar to the spin-$3/2$ case, however now there are now nine local minima within the peaks in the plot instead of three, as shown in Fig.\ \ \ref{fig:singlenucfree9/2}. This figure illustrates the dependence of the one-tangling power on the dimension of the nuclear spin in question. Note that the expression is now upper-bounded by $\frac{10}{33}\approx 0.30$ when $G_1 = 0$, in contrast to the maximum from the spin-$3/2$ case which was $\frac{4}{15}\approx 0.27$. 

\begin{figure*}
    \centering
    \includegraphics[width=\linewidth]{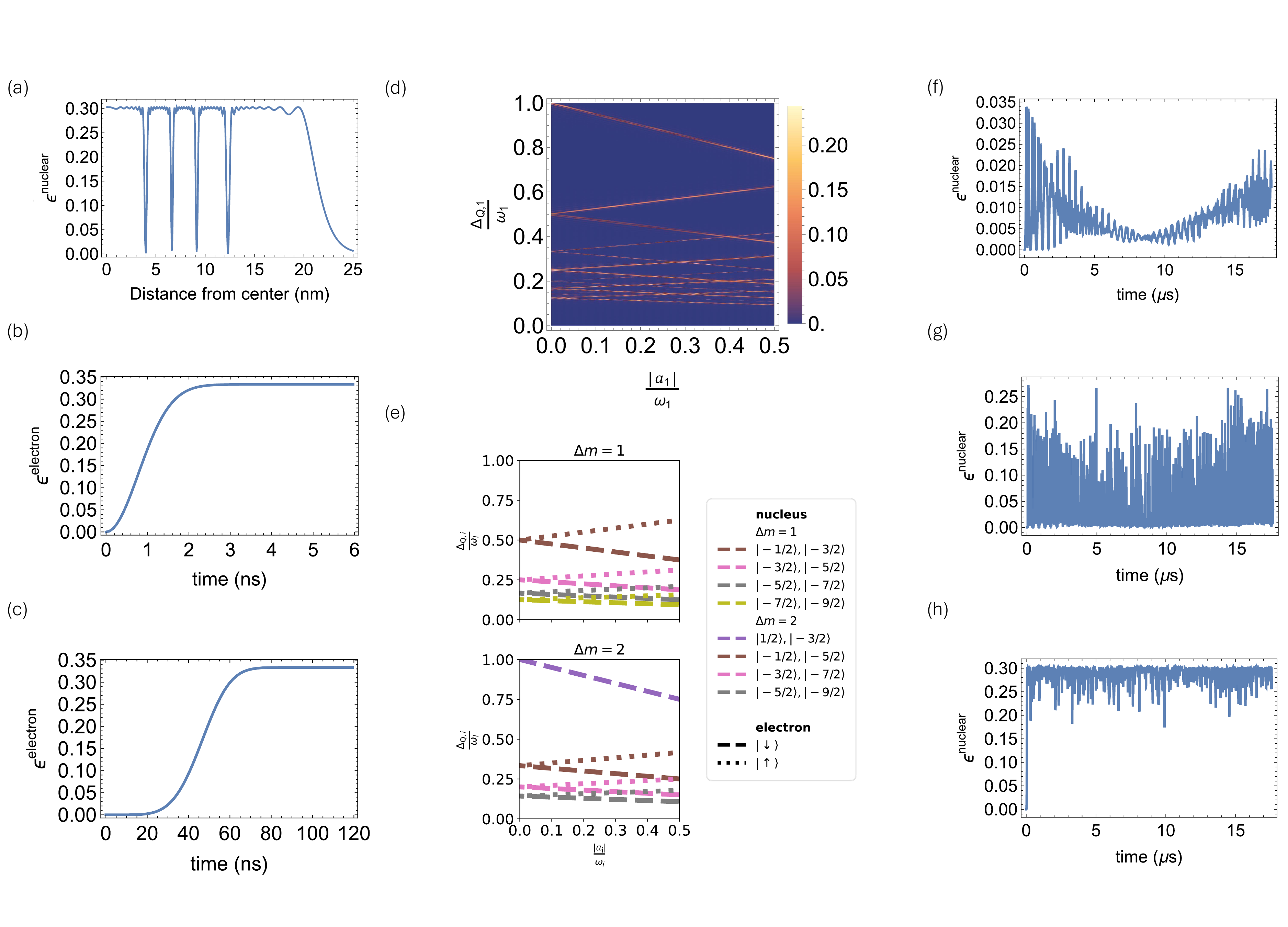}
    \caption{(a) Free evolution nuclear one-tangling power with respect to an $^{115}$In nucleus as a function of radial position from the center of the dot at time $5.32$ $\upmu$s. The Larmor frequency of this nucleus is set to $\omega_{1}/2\pi = 9.33$ MHz (assuming the external magnetic field is $B = 1$ T). We set the nuclear spin operators to be spin-$9/2$. This offers a comparison to Fig.\ \ref{fig:freeensembles}(c). (b) Free evolution electronic one-tangling power as a function of time for a nuclear bath of $80,247$ nuclei. This ensemble is made up of $40,123$ $^{71}$Ga nuclei and $40,124$ $^{115}$In nuclei. The spins in the ensemble are assigned parameters corresponding to either species. This includes $\omega_{\text{Ga}}/2\pi = 12.98$ MHz, the previously mentioned Larmor frequency for $^{115}$In, and the corresponding total dimension of these spins (either $d = 4$ or $d = 10$, respectively). This shows how the electron spin dephases at a quicker rate when it is coupled to higher spin nuclei than in the spin-$3/2$ case shown in Fig.\ \ref{fig:freeensembles}(d). (c) Electronic one-tangling power when a single unit (including two $\pi$ pulses) of the CPMG sequence has been applied over the duration $t = 120$ ns. Once again, we assume the mixed bath from (b). For Figs.\  (a)-(c), we implement the distributions of the hyperfine and quadrupolar coupling parameters from Sec.\ \ref{subsec:spatialdep} and use $a^{\mathrm{nc}}_i/2\pi = 0.0051$ MHz, as well as $\theta_i = \frac{\pi}{3}$. (d) One-tangling power for a single $^{115}$In nucleus coupled to the electron as a function of QD parameters $\Delta_{\text{Q},1}/\omega_{1}$ and $|a_1|/\omega_1$. We set $N = 1$ and implement the CPMG sequence over the duration of $t = 23.4$ $\upmu$s, and choose $a^{\mathrm{nc}}_1/2\pi = 0.0058$ MHz.(e) Density plots showing occurrences of degeneracies between eigenstates of the system for varying values of $\Delta_{\text{Q},1}/\omega_{1}$ and $|a_1|/\omega_1$ that coincide with the increased one-tangling power in (d). (f) One-tangling power for an $^{115}$In nucleus coupled to the electron as a function of time for a single iteration of the CPMG sequence applied over $t = 17.6$ $\upmu$s. (g) Single nuclear one-tangling power for an $^{115}$In nucleus coupled to the QD electron for $N = 85$ iterations of the CPMG sequence applied over $\tau = N t = 17.6$ $\upmu$s. For both (f) and (g), $a^{\mathrm{nc}}_1/2\pi = 0.056$ MHz. (h) Nuclear one-tangling power for a single nucleus when $N = 1$, $a^{\mathrm{nc}}_1/2\pi = 1.3$ MHz, and $t = 17.6$ $\upmu$s. For Figs.\  (d)-(h) we set $\theta_1 = \frac{\pi}{3}$, fix $\omega_{1}/2\pi = 9.33$ MHz, and assign the nuclear spin operators to be spin-$9/2$. }
    \label{fig:spinninehalf}
    \label{fig:nucensembleninehalf}
    \label{fig:spinninehalfCPMG}
\end{figure*}

We also consider a nuclear ensemble to study many-body system dynamics in the case where our isolated nucleus is spin-$9/2$. If we are interested in the entanglement arising between a spin-$3/2$ nucleus and the remaining spins, the nuclear one-tangling power expression will not change from Eq.\ (\ref{eq:nuclearep3/2}) when we incorporate spin-$9/2$ nuclei into the ensemble. However, if the spin in question is spin-$9/2$, then the nuclear one-tangling power takes on the same form as Eq.\ (\ref{eq:spinninehalfsinglenuc}). Plotting this using the same distribution of hyperfine and quadrupolar coupling values given in Sec.\ \ref{subsec:spatialdep} yields the result shown in Fig.\ \ \ref{fig:nucensembleninehalf}(a).  Similar to Fig.\ \ \ref{fig:singlenucfree9/2}, this plot bears resemblance to its spin-$3/2$ counterpart (Fig.\ \ \ref{fig:freeensembles}(c)) with the key distinction being the nine local minima that appear in the each of the global peaks in the plot.
Unlike the nuclear one-tangling power expression, the electronic one will always change with the addition of nuclei of differing parameters, because the electronic one-tangling power depends on all of the nuclei interacting appreciably with the electron. To study the growth of entanglement between the QD electron and the ensemble of varied nuclear spin species, we refer to the generalized electronic one-tangling power expression, Eq.\ (\ref{eq:electronicappendix}):
\begin{equation*}
    \epsilon^{\text{electronic}}_{p|q} = \frac{1}{3} -  \frac{1}{3}\prod_{i=1}^n\Big(\frac{1}{d_i + 1}\Big) \Big(1 + d_i G_1^{(i)} \Big). 
\end{equation*}
\noindent Note that $d_i$ takes on values corresponding to the dimensions of the nuclei we are considering in the ensemble. Since we are considering a mixed bath of $^{71}$Ga and $^{115}$In spins, $d_i$ is either $4$ or $10$, respectively. To realistically consider the growth of entanglement between the central electron and surrounding nuclei, we once again implement the ensemble given in Sec.\ \ref{subsec:spatialdep}, but now we include a second nuclear spin species. To do this, we assign each of the $80,247$ nuclei in the simulated ensemble a Larmor frequency corresponding to either $^{71}$Ga ($\omega_{\text{Ga}}/B = 12.98 * 2\pi$ MHz/T) or $^{115}$In ($\omega_{\text{In}}/B = 9.33 * 2\pi$ MHz/T). The dimension of the rotation operator acting on a given nucleus is also assigned based on whether it is spin-$3/2$ (for $^{71}$Ga) or spin-$9/2$ (for $^{115}$In). The rate of saturation of the electronic one-tangling power corresponds to electron spin dephasing caused by the dense ensemble of nuclei in its environment. Plugging the values associated with this nuclear ensemble along with $U_{\text{free}}$ into Eq.\ (\ref{eq:electronicappendix}) shows in Fig.\ \ \ref{fig:spinninehalf}(b) that this dephasing happens at a more rapid timescale than in the purely spin-$3/2$ case (from Fig.\ \ \ref{fig:freeensembles}(d)), at about 2.5 ns. If we consider the entanglement arising in this ensemble during the evolution describing the application of a single iteration of the CPMG sequence, we once again find in Fig.\ \ \ref{fig:spinninehalf}(c) that while noise is mitigated in contrast to the free evolution case, the addition of higher-spin nuclei increases the rate of electron spin dephasing when compared to the strictly spin-$3/2$ case shown in Fig.\ \ \ref{fig:ensemble}(a). 

We also examine the nuclear one-tangling power when the electron is subject to a single iteration of the CPMG sequence in analogy to Fig.\ \ \ref{fig:nuclearensemble}(a). Figure \ref{fig:spinninehalfCPMG}(d) reveals the narrow lines of increased entanglement between a single spin-$9/2$ nucleus and the electron in the regime of low $a^{\mathrm{nc}}_1$ as a function of Hamiltonian parameters $|a_1|/\omega_1$ and $\Delta_{\text{Q},1}/\omega_1$. It once again shows that in this regime, the nuclear spin transitions that enable entanglement generation occur when eigenstates of the system are degenerate. In particular, the allowed transitions correspond to $\Delta m = \pm 1$ and $\pm 2$ as shown in Fig.\ \ \ref{fig:spinninehalf}(e). In this case, there are naturally more degeneracies to account for in comparison to the spin-$3/2$ case, which has fewer levels. These lines once again help to inform the deliberate generation of entanglement between spins in the QD, as well as selectivity of entanglement generation between the central electron and subsets of nuclei.

We also explore means for entangling target nuclei simply by increasing the number of CPMG iterations, $N$. First, Fig.\ \ \ref{fig:spinninehalf}(f) offers a baseline for the relatively low amount of entanglement arising between an $^{115}$In nuclear spin and the QD electron when $N = 1$ and $a^{\mathrm{nc}}_1$ is small. We draw a comparison between Fig.\ \ \ref{fig:NiterationsCPMG}(b) and Fig.\ \ \ref{fig:spinninehalf}(g) by increasing $N$ to be $85$ and keeping all other system parameters the same between the plots, aside from $\omega_1$ and the spin-$9/2$ operators. Figure \ref{fig:spinninehalf}(g) shows that we can once again selectively increase the entanglement between a particular subset of nuclei and the electron in the $^{115}$In case. In addition, we observe the saturation of entanglement between a QD electron and target nucleus when $a^{\mathrm{nc}}_1/2\pi$ is increased to $1.3$ MHz in Fig.\ \ \ref{fig:spinninehalf}(h), with $a^{\mathrm{nc}}_1$, other Hamiltonian parameters, and total time duration set equal to those in Fig.\ \ \ref{fig:NiterationsCPMG}(c). 

\section{Analytical $G_1$}
\label{appendix:G1analytical}
This section includes the derivation for the analytical generalized Makhlin invariant, $G_1$, for a spin-$3/2$ nuclear spin coupled to a central QD electron. This expression holds for the equivalent cases $\theta_1 = \pi$ and $\theta_1 = 0$. If we assume the single-nuclear form of the QD Hamiltonian, Eq.\ (\ref{eq:H}), and ignore the electronic Zeeman term because it has no impact on the dynamics, we have
\begin{equation*}
\label{eq:Honespin}
\begin{split}
    H &= \omega_1 I^z_1 + \Delta_{\mathrm{Q, 1}}(I^z_1)^2 + a_1 S_z I^z_1 \\ 
    &- a^{\mathrm{nc}}_1 S_z [\cos^2 \theta_1 ((I^x_1)^2 - (I_1^y)^2) + \sin 2 \theta_1 (I^z_1 I_1^x + I_1^x I_1^z)]. \\
\end{split}
\end{equation*}
\noindent Again, we note the block-diagonal structure of $H$ and see that it can be written in the form 
\begin{equation*}
    H = |0\rangle \langle 0| \otimes H_0 + |1\rangle \langle 1 | \otimes H_1, 
\end{equation*} where 
\begin{equation*}
\begin{split}
    H_{0/1} &= \omega_1 I^z_1 + \Delta_{Q, 1}(I^z_1)^2 \pm \frac{a_1}{2} I_1^z \\
    &\mp \frac{a_1^{\mathrm{nc}}}{2}[\cos^2\theta_1 ((I^x_1)^2 - (I_1^y)^2) + \sin 2 \theta_1 (I^z_1 I^x_1 + I^x_1 I^z_1)].
\end{split}
\end{equation*}
\noindent The free evolution operator of this system is given as $U_{\mathrm{free}} = |0\rangle \langle 0| \otimes R_{n_0} + |1\rangle \langle 1 | \otimes R_{n_1}$, where $R_i = e^{- i H_i t}$. This evolution is characterized by a generalized $G_1$ of the form $G_1^{(1)} = \frac{1}{d_1^2}|\text{Tr}[R_{n_0}^{\dagger}R_{n_1}]|^2$. This simplifies to
\begin{equation}
    \begin{split}
        G_1^{(1)} &= \frac{1}{4} (c_{13}^2 + c_{24}^2 + 2 c_{13}c_{24}\cos{{a_{i}} t}). \\
        \label{eq:analyticG1}
    \end{split}
\end{equation}
\noindent Here, 
$c_{13} = c_1 - c_3$, $c_{24} = c_2 + c_4$ and \\
\begin{equation*}
\begin{split}
    c_1 &= \cos{\big(\frac{\sqrt{x_1}t}{2}\big)}\cos{\big(\frac{\sqrt{x_3}t}{2}\big)}, \\
    c_2 &= \cos{\big(\frac{\sqrt{x_2}t}{2}\big)}\cos{\big(\frac{\sqrt{x_4}t}{2}\big)}, \\
    c_3 &= \frac{(a_1^2 + 3  (a_{1}^{\mathrm{nc}})^2 - 4(\Delta_{\mathrm{Q},1} - \omega_{1})^2)\sin{(\frac{\sqrt{x_1 }t}{2})}\sin{(\frac{\sqrt{x_3}t}{2})}}{\sqrt{x_1}\sqrt{x_3}},\\ 
    c_4 &= \frac{(-a_1^2 - 3  (a_{1}^{\mathrm{nc}})^2 + 4(\Delta_{\mathrm{Q},1} + \omega_{1})^2)\sin{(\frac{\sqrt{x_2}t}{2})}\sin{(\frac{\sqrt{x_4}t}{2})}}{\sqrt{x_2}\sqrt{x_4}}, \\
\end{split}
\end{equation*}
\noindent and the $x_i$ are given by 
\begin{equation*}
    \begin{split}
        x_1 &= 3 (a^{\mathrm{nc}}_1)^2 + (a_1 - 2(\Delta_{\mathrm{Q}, 1} - \omega_1))^2,  \\
        x_2 &= 3 (a^{\mathrm{nc}}_1)^2 + (a_1 + 2(\Delta_{\mathrm{Q}, 1} + \omega_1))^2, \\
        x_3 &= 3 (a^{\mathrm{nc}}_1)^2 + (a_1 + 2(\Delta_{\mathrm{Q}, 1} - \omega_1))^2, \\
        x_4 &= 3 (a^{\mathrm{nc}}_1)^2 + (a_1 - 2(\Delta_{\mathrm{Q}, 1} + \omega_1))^2 .\\
    \end{split}
\end{equation*}
\noindent We further simplify this analytical expression for $G_1$ to solve for times at which the nucleus is maximally entangled with the QD electron. Note that because the nuclear one-tangling power only depends on the singled-out nucleus and the central electron, this expression can be used for larger ensembles as well. We find that the resonance times, $t_k$, occur when $c_3$ goes to zero, and this happens when $\Delta_{\mathrm{Q}, 1} = \omega_1 \pm \frac{1}{2}\sqrt{a_1^2 + 3(a^{\mathrm{nc}}_1)^2}$. We consider the limiting case where $\Delta_{\mathrm{Q}, 1}, a^{\mathrm{nc}}_1 \rightarrow 0$ (since $\omega_{q,1} \ll \omega_1,a_1$). We can start with the simple case of $\theta_1 = \pi$, $\Delta_{\mathrm{Q},1} = 0 = a^{\mathrm{nc}}_1$. With these values, we end up with $x_1 = x_2$ and $x_3 = x_4$. This leads to $c_1 = c_2$ and $c_3 = c_4$, so the analytical, generalized Makhlin invariant (Eq.\ (\ref{eq:analyticG1})) becomes 
\begin{equation}
    G_1^{(1)} = \frac{1}{2} \big[(c_1^2 + c_3^2) + (c_1^2 - c_3^2)\cos(a_1 t) \big], \\
\end{equation}
\noindent where now
\begin{equation*}
\begin{split}
c_1 &= \cos(\frac{\sqrt{x_1}t}{2})\cos(\frac{\sqrt{x_3}t}{2}),
\end{split}
\end{equation*}
and
\begin{equation*}
    \begin{split}
        c_3 &=\frac{a_1 - 4 \omega_{1}^2}{{\sqrt{x_1}\sqrt{x_3}}}\sin(\frac{\sqrt{x_1}t}{2})\sin(\frac{\sqrt{x_3}t}{2}).
%c_3 &= \frac{(a_1 - 4 \omega_{1}^2)\sin(\frac{\sqrt{x_1}t}{2})\sin(\frac{\sqrt{x_3}t}{2})}{\sqrt{x_1}\sqrt{x_3}}.
    \end{split}
\end{equation*}
\noindent At this point, we can simplify things further by assuming $c_3 = 0$, which happens when $\omega_1 = \frac{a_1}{2}$. This leads to 
\begin{equation}
    G_1^{(1)} = \frac{1}{2} \big[(\cos(a_1 t))^2(1 + \cos(a_1 t)) \big].
\end{equation}
\noindent We can now set $G_1 = 0$ and solve for the resonance times, which occur when
\begin{equation}
    t_k = \frac{(2k+1)\pi}{a_1} \text{ and }  \frac{(2k+1)\pi}{2 a_1}.
\end{equation}

\section{Solving for CPMG evolution}
\label{appendix:CPMGU}
In this section, we solve for the evolution operator describing the application of a single iteration of the CPMG sequence on the QD electron. As previously mentioned, the CPMG sequence is effective for treating noise induced by random nuclear spin dynamics in central spin systems of interest in this work. The application of carefully-timed $\pi$ pulses ($R_x(\pi) = e^{-i\sigma_x \pi/2} = -i \sigma_x$) work to refocus the electron spin when it experiences dephasing caused by inhomogeneities in the effective magnetic field generated by surrounding nuclei. A single sequence of the CPMG evolution is defined by three intervals of free evolution interleaved with $\pi$ pulses. This takes the form:
\begin{widetext}
\begin{equation}
    \begin{split}
        U_{CPMG} &= U_{\text{free}}(t_3)(R_{x}(\pi)\otimes \mathbb{I})U_{\text{free}}(t_2)(R_{x}(\pi)\otimes \mathbb{I})U_{\text{free}}(t_1) \\
        &= e^{-i H t_3 }(-i\sigma_x\otimes \mathbb{I})e^{-i H t_2}(-i\sigma_x\otimes \mathbb{I})e^{-i H t_1 } \\
        &= e^{-i H t_3 }(-i\sigma_x\otimes \mathbb{I})e^{-i H t_2}\Big(-i(|0\rangle \langle 1| + |1\rangle \langle 0|) \otimes \mathbb{I} \Big)\Big(|0\rangle \langle 0| \otimes R_{n_0}(t_1) + |1\rangle \langle 1| \otimes R_{n_1}(t_1)\Big)\\
        &= -e^{-i H t_3 } (\sigma_x \otimes \mathbb{I} )\Big(|0\rangle \langle 0| \otimes R_{n_0}(t_2) + |1\rangle \langle 1| \otimes R_{n_1}(t_2)\Big) \Big(|0\rangle \langle 1|\otimes R_{n_1}(t_1) + |1\rangle \langle 0| \otimes R_{n_0}(t_1) \Big)\\
        &= -e^{-i H t_3 }\Big((|0\rangle \langle 1| + |1\rangle \langle 0|) \otimes \mathbb{I} \Big)\Big(|0\rangle \langle 1| \otimes R_{n_0}(t_2)R_{n_1}(t_1) + |1\rangle \langle 0| \otimes R_{n_1}(t_2)R_{n_0}(t_1)  \Big) \\
        &= \Big(|0\rangle \langle 0| \otimes R_{n_0}(t_3) + |1\rangle \langle 1| \otimes R_{n_1}(t_3)\Big) \Big( |0\rangle \langle 0| \otimes R_{n_1}(t_2)R_{n_0}(t_1) + |1\rangle \langle 1| \otimes R_{n_0}(t_2)R_{n_1}(t_1)   \Big)\\
        &= |0\rangle \langle 0| \otimes R_{n_0}(t_3)R_{n_1}(t_2)R_{n_0}(t_1) + |1\rangle \langle 1|\otimes R_{n_1}(t_3)R_{n_0}(t_2)R_{n_1}(t_1) \\
        &= |0\rangle \langle 0| \otimes R_{n_0}^{\text{CPMG}} + |1\rangle \langle 1| \otimes R_{n_1}^{\text{CPMG}}.
    \end{split}
\end{equation}
\end{widetext}
\noindent where $U_{\text{free}}$ is given by (\ref{eq:Ufree}), $R_{n_j}^{\text{CPMG}}$ are the compilations of free evolution rotation operators making up the effective rotation on a particular nuclear spin during a single iteration of the CPMG sequence, and $t_3 = t_1 = \frac{t}{4}$ and $t_2 = \frac{t}{2}$ by definition of the sequence. Note the convenient block-diagonal form of $U_{\text{CPMG}}$ that allows us to apply the generalized one-tangling power expressions, Eqs.\ (\ref{eq:nuclearappendix}) and (\ref{eq:electronicappendix}). The generalized Makhlin invariant describing the evolution applied to the electron and a single nucleus in this case can be found numerically by applying $G_1^{(1)} =  \frac{1}{d_1^2}|\text{Tr}[(R_{n_0}^{\text{CPMG}})^{\dagger}R_{n_1}^{\text{CPMG}}]|^2$.

\renewcommand{\arraystretch}{1.5}
\begin{table*}[ht]
\centering
\begin{tabular}{||c||c|c|c|c||}
\hline
\textbf{$\Delta m$} & \textbf{Electron Spin} & \textbf{Nuclear Transition} & \textbf{Condition on $a_i, \Delta_{\mathrm{Q},i}$} & \textbf{Condition on $a_i^{\mathrm{nc}}, \theta_i$} \\
\hline

% Δm = 1
\multirow{6}{*}{$1$}
& \multirow{3}{*}{$\ket{\uparrow}$}
& $\ket{3/2} \leftrightarrow \ket{1/2}$ & $\frac{\Delta_{\mathrm{Q},i}}{\omega_i} = -\frac{1}{4}\frac{a_i}{\omega_i} - \frac{1}{2}$ & $-a_i^{\mathrm{nc}}\cos^2\theta_i = 0$ \\
& & $\ket{1/2} \leftrightarrow \ket{-1/2}$ & $\frac{a_i}{\omega_i} = -2$ & $\mp \frac{1}{2}\sqrt{3}a_i^{\mathrm{nc}}\sin2\theta_i - \frac{3}{4}a_i^{\mathrm{nc}}\cos^2\theta_i = 0$ \\
& & $\ket{-1/2} \leftrightarrow \ket{-3/2}$ & $\frac{\Delta_{\mathrm{Q},i}}{\omega_i} = \frac{1}{4}\frac{a_i}{\omega_i} + \frac{1}{2}$ & $-a_i^{\mathrm{nc}}\cos^2\theta_i = 0$ \\
& \multirow{3}{*}{$\ket{\downarrow}$}
& $\ket{3/2} \leftrightarrow \ket{1/2}$ & $\frac{\Delta_{\mathrm{Q},i}}{\omega_i} = \frac{1}{4}\frac{a_i}{\omega_i} - \frac{1}{2}$ & $a_i^{\mathrm{nc}}\cos^2\theta_i = 0$ \\
& & $\ket{1/2} \leftrightarrow \ket{-1/2}$ & $\frac{a_i}{\omega_i} = 2$ & $\pm \frac{1}{2}\sqrt{3}a_i^{\mathrm{nc}}\sin2\theta_i + \frac{3}{4}a_i^{\mathrm{nc}}\cos^2\theta_i = 0$ \\
& & $\ket{-1/2} \leftrightarrow \ket{-3/2}$ & $\frac{\Delta_{\mathrm{Q},i}}{\omega_i} = -\frac{1}{4}\frac{a_i}{\omega_i} + \frac{1}{2}$ & $a_i^{\mathrm{nc}}\cos^2\theta_i = 0$ \\
\hline

% Δm = 2
\multirow{4}{*}{$2$}
& \multirow{2}{*}{$\ket{\uparrow}$}
& $\ket{3/2} \leftrightarrow \ket{-1/2}$ & $\frac{\Delta_{\mathrm{Q},i}}{\omega_i} = -\frac{1}{2}\frac{a_i}{\omega_i} - 1$ & $\sqrt{3}a_i^{\mathrm{nc}}\sin2\theta_i - a_i^{\mathrm{nc}}\cos^2\theta_i = 0$ \\
& & $\ket{1/2} \leftrightarrow \ket{-3/2}$ & $\frac{\Delta_{\mathrm{Q},i}}{\omega_i} = \frac{1}{2}\frac{a_i}{\omega_i} + 1$ & $\sqrt{3}a_i^{\mathrm{nc}}\sin2\theta_i + a_i^{\mathrm{nc}}\cos^2\theta_i = 0$ \\
& \multirow{2}{*}{$\ket{\downarrow}$}
& $\ket{3/2} \leftrightarrow \ket{-1/2}$ & $\frac{\Delta_{\mathrm{Q},i}}{\omega_i} = \frac{1}{2}\frac{a_i}{\omega_i} - 1$ & $\sqrt{3}a_i^{\mathrm{nc}}\sin2\theta_i - a_i^{\mathrm{nc}}\cos^2\theta_i = 0$ \\
& & $\ket{1/2} \leftrightarrow \ket{-3/2}$ & $\frac{\Delta_{\mathrm{Q},i}}{\omega_i} = -\frac{1}{2}\frac{a_i}{\omega_i} + 1$ & $\sqrt{3}a_i^{\mathrm{nc}}\sin2\theta_i + a_i^{\mathrm{nc}}\cos^2\theta_i = 0$ \\
\hline

% Δm = 3
\multirow{2}{*}{$3$}
& $\ket{\uparrow}$ & $\ket{3/2} \leftrightarrow \ket{-3/2}$ & $\frac{a_i}{\omega_i} = -2$ & $\sqrt{3}a_i^{\mathrm{nc}}\sin2\theta_i = 0$ \\
& $\ket{\downarrow}$ & $\ket{3/2} \leftrightarrow \ket{-3/2}$ & $\frac{a_i}{\omega_i} = 2$ & $\sqrt{3}a_i^{\mathrm{nc}}\sin2\theta_i = 0$ \\
\hline

\end{tabular}
\caption{Conditions on parameters for degeneracies between nuclear spin states of a spin-$3/2$ nucleus in a quantum dot. Transitions are grouped by change in magnetic quantum number, $\Delta m$, and electron spin state.}
\label{tab:degeneracyconditions}
\end{table*}

\begin{figure*}
    \centering
    \includegraphics[width=\linewidth]{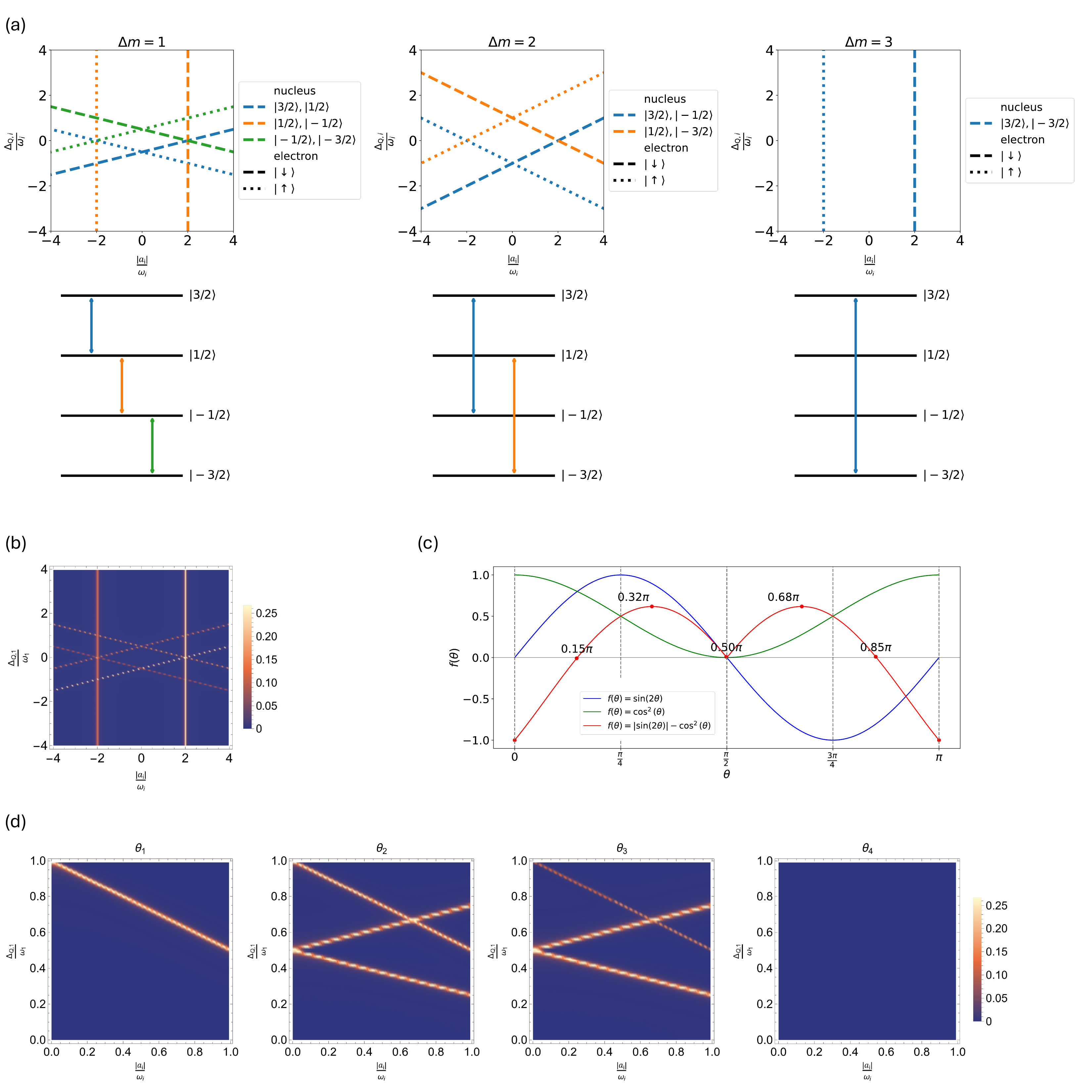}
    \caption{(a) Degeneracies corresponding to the different transitions within the spin $3/2$ nucleus. Parameter range chosen to allow degeneracy conditions to be plotted for the full set of transitions listed in Table \ref{tab:degeneracyconditions}. (b) One-tangling power between a single spin-$3/2$ $^{71}$Ga nucleus and the QD electron subject to a single iteration of the CPMG sequence when the Hamiltonian of the system is given in Sec.\ \ref{sec:diffncH}. (b) The one-tangling power plotted for the Hamiltonian with the simpler non-collinear term, Eq.\ (\ref{eq:diffnc}), for a wide range of parameter values to capture more of the system degeneracies. Parameters used to form this plot includes $\omega_1/2\pi = 12.98$ MHz and $a^{\mathrm{nc}}_1/2\pi = 0.058$ MHz. (c) Plot of the $\theta$-dependent functions that govern the dynamics of the nuclear transitions. Shown are the functions $\sin(2\theta)$, $\cos^2(\theta)$ and the difference between the two. The marks corresponding to maxima and minima of the function $|\sin(2\theta)|-\cos^2(\theta)$, as well as $|\sin(2\theta)|=\cos^2(\theta)$. (d) One-tangling power under a single unit of the CPMG sequence for a single nucleus coupled to the QD electron as a function of $\Delta_{\text{Q},1}/\omega_1$ and $|a_1|/\omega_1$ with $\omega_1/2\pi=12.98$ MHz and $a^{\mathrm{nc}}_1/2\pi=0.058$ MHz and optimized over time $t=15.6\upmu$s. $\theta$ varies in each plot as follows: $\theta_1 = 0$, $\theta_2 = 0.15\pi$, $\theta_3 = 0.32\pi$, $\theta_4=0.5\pi$.}
    \label{fig:alldegensfigs}
    \label{fig:diffnc}
\end{figure*}

\section{Degeneracies}
\label{Appendix:Degeneracies}
The conditions under which degeneracies occur between the spin states of a spin-$3/2$ nucleus in a quantum dot depend on the quadrupolar and collinear hyperfine parameters of the system Hamiltonian. To determine these conditions, we consider the Hamiltonian in Eq.\ (\ref{eq:Hi}) which is approximately diagonal in the $I_i^z$ basis when the non-collinear hyperfine coupling $a_i^{\mathrm{nc}}$ is small. These conditions are derived by equating the energy eigenvalues of the relevant states and solving for the parameters $a_i$ and $\Delta_{\mathrm{Q},i}$. For instance, the condition for degeneracy between the states $\ket{\uparrow3/2}$ and $\ket{\uparrow1/2}$ is found by solving:
\begin{equation}
    H\ket{\uparrow3/2} = H\ket{\uparrow1/2}.\\
\end{equation}
Equating the corresponding diagonal elements gives: 
\begin{align}
    \frac{3a_i}{4}+\frac{9\Delta_{\mathrm{Q},i}}{4}+\frac{3\omega_i}{2} &= \frac{a_i}{4}+\frac{\Delta_{\mathrm{Q},i}}{4}+\frac{\omega_i}{2}\nonumber,\\
    \frac{\Delta_{\mathrm{Q},i}}{\omega_i}&=-\frac{1}{4}\frac{a_i}{\omega_i}-\frac{1}{2}.
\end{align}
An additional condition on $a_i^{\mathrm{nc}}$ and $\theta_i$ is obtained by requiring the off-diagonal element to vanish:
\begin{equation}
    -a_i^{\mathrm{nc}}\cos^2{\theta_i}=0.
\end{equation}
This condition is satisfied by the assumption that $a_{nc}$ is small.

A complete list of all degeneracy conditions for pairs of nuclear spin states (for fixed electron spin) is given in Table \ref{tab:degeneracyconditions}. The transitions are categorized by the change in magnetic quantum number $\Delta m$. Figure \ref{fig:alldegensfigs} (a) visualizes these conditions in terms of the relationship between the parameters $a_i$ and $\Delta_{\mathrm{Q},i}$ and the corresponding nuclear transitions.

This analysis focuses on transitions in which the electron spin remains fixed, as these define the relevant subspaces for the entanglement analysis presented in the main text. Similar degeneracy conditions can be derived for transitions involving electron spin flips. 

\subsection{Non-collinear Interaction}
The plots in Fig.\ \ref{fig:alldegensfigs} (a) show the conditions on the quadrupolar and collinear hyperfine term under which degenerate states occur. Not all of these degeneracies give rise to entanglement as this is determined by the structure of the operators driving the non-collinear interaction. 

The non-collinear term in Eq.\ (\ref{eq:Hi}) allows both $\Delta m=1$ and $\Delta m=2$ transitions, excluding the $\ket{1/2},\ket{-1/2}$ transition. This behavior is reflected in the one-tangles plots for the system under this Hamiltonian, shown in Fig.\ \ref{fig:degeneracies}. 

Further control over which transitions and degeneracies lead to entanglement can be achieved by tuning the parameter $\theta_i$, which corresponds to the strain within the quantum dot. The two sets of transitions induced by the evolution of our Hamiltonian result from the non-collinear interaction given by: 
\begin{align}
    H_{\mathrm{nc}} &= S_z \sum_i a_i^{\mathrm{nc}} [\cos^2\theta_i((I_i^x)^2 -(I_i^y)^2)\nonumber\\
    &+ \sin2\theta_i(I_i^zI_i^x + I_i^xI_i^z)],
\end{align}
This is also the term responsible for generating entanglement in the system, which the is measured using the one-tangles formalism. Specifically:
\begin{itemize}
    \item the term $\cos^2{\theta_i}((I_i^x)^2-(I_i^y)^2)$ allows the $\Delta m=2$ transitions in the spin-$3/2$ system.
    \item the term $\sin{2\theta_i}(I_i^zI_i^x+I_i^xI_i^z)$ gives us the $\Delta m=1$ transitions, with the exception of the $\ket{1/2},\ket{-1/2}$ transition.
\end{itemize}
By varying the parameter $\theta$ (Fig.\ \ref{fig:alldegensfigs} (c)), we can control which transition is activated and the relative strength between the two transition types. These effects are summarized as follows:

\begin{itemize}
    \item $\theta_1=n\pi$: Only the $\Delta m=2$ transitions are activated. 
    \item $\theta_2 = n\pi \pm \frac{1}{2} \arccos (\frac{3}{5})$: Both transitions are activated and generate comparable one-tangles.
    \item $\theta_3 = \frac{n\pi}{2}\pm\frac{1}{2}\arctan(2)$: the $\Delta m=1$ transitions dominate and lead to stronger entanglement.
    \item $\theta_4 = \frac{\pi}{2} +n\pi$: Neither transition gives rise to significant one-tangles.
\end{itemize}
We can see the effect of changing $\theta$ on the one-tangles in Fig.\ \ref{fig:alldegensfigs} (d) where we have plotted the one-tangling power for different values of theta corresponding to the solutions listed above.

\subsection{Different non-collinear term}
\label{sec:diffncH}
If we instead study a system with a different non-collinear term, such as the one from \cite{jackson2022optimal, ruskuc2022nuclear, shofer2024tuning, appel2025many}, which has a Hamiltonian of the form in Eq.\ (\ref{eq:Hgen}) where:
\begin{equation}
\begin{split}
    H_{0/1} &= \omega_e + \sum_i \omega_i I_i^z + \Delta_{\mathrm{Q},i}(I_i^z)^2 \pm \frac{a_i}{2} I_i^z \mp \frac{a_i^{\mathrm{nc}}}{2}I_i^x. \\
\end{split}
\label{eq:diffnc}
\end{equation}
We observe from the one-tangle plots in Fig.\ \ref{fig:diffnc} that now only the $\Delta m=1$ transitions give rise to entanglement. In this case, the $\ket{1/2},\ket{-1/2}$ transition also contributes to entanglement generation. 

\section{Alternative derivation of generalized one-tangling power}
In this section, we offer shorter derivations for the generalized one-tangling power expressions from Appendices \ref{appendix:nuclearonetangles} and \ref{appendix:electroniconetangles}. This method exploits the block-diagonal structure of the unitary evolution operators of interest and involves directly computing the reduced state purity with respect to the spin we are interested in. The procedure is laid out in the following way. First, we compute the reduced, time-evolved density operator corresponding to the individual spin that we are interested in studying, $\rho_q(t)$. We then square this quantity and take the trace to compute the subsystem purity: $\tau_{p|q} = 1 - \text{Tr}[\rho_{q}(t)^2]$. Then, to calculate the one-tangling power, we average over $\tau_{p|q}$ with respect to the space of every spin in the system. For spin-$1/2$ systems, this would look like integrating over the Bloch sphere associated with every qubit in the system, but we compute this for the general case that involves an arbitrarily large bath of $n$ nuclei of any total spin $\sum_i d_i$. 

\subsection{Nuclear one-tangling power}
\label{appendix:simplerderivationnuc}
We begin with the general initial state, $\rho(0) = \rho_{e}(0)\otimes \rho_{\text{N}}(0)$, where $\rho_e(0) = |\psi_e(0)\rangle \langle \psi_e(0)|$ corresponds to the initial state of the electron and $\rho_N(0) = |\psi_{\text{N}}(0)\rangle \langle \psi_{\text{N}}(0)| = |\psi_{1}(0)\rangle \langle \psi_1(0)|\otimes...\otimes|\psi_{n}(0)\rangle \langle \psi_n(0)|$ captures the initial states of all of the $n$ nuclei in the ensemble. The time-evolved composite system is described by $\rho(t) = U^{\dagger}\rho(0)U$. We can compute the reduced, time-evolved state with respect to any spin in the system by tracing out the rest of the spins from this composite density operator. We first compute the subsystem density operator $\rho_{n}(t)$ via $\rho_n(t) = \text{Tr}_{e...n-1}[U^{\dagger}\rho(t)U]$ in order to derive the generalized nuclear one-tangling power. Assuming the block-diagonal form of $U = |0\rangle \langle 0|\otimes R_{n_0} + |1\rangle \langle 1|\otimes R_{n_1}$ leads to:
\begin{widetext}
\begin{equation}
\begin{split}
    \rho_n(t) &= \text{Tr}_{e...n-1}[U^{\dagger}\rho(t)U]\\
    &= (|0\rangle \langle 0| \otimes R_{n_0}^{\dagger} + |1\rangle \langle 1|\otimes R_{n_1}^{\dagger})\rho(t)(|0\rangle \langle 0| \otimes R_{n_0} + |1\rangle \langle 1|\otimes R_{n_1})\\
    &= \text{Tr}_{e...\text{n-1}}[(|0\rangle \langle 0| \otimes R_{n_0}^{\dagger} + |1\rangle \langle 1|\otimes R_{n_1}^{\dagger})(|\psi_e(t)\rangle \langle \psi_e(t)|\otimes |\psi_{\text{N}}(t)\rangle \langle \psi_{\text{N}}(t)|)(|0\rangle \langle 0| \otimes R_{n_0} + |1\rangle \langle 1|\otimes R_{n_1})]\\
    &= \text{Tr}_{e...\text{N-1}}[|0\rangle \langle 0| \psi_e(t)\rangle \langle \psi_e(t)|0\rangle \langle 0|\otimes R_{n_0}^{\dagger}|\psi_{\text{N}(t)}\rangle \langle \psi_{\text{N}}(t)|R_{n_0}\\
    &+ |0\rangle \langle 0| \psi_e(t)\rangle \langle \psi_e(t)|1\rangle \langle 1|\otimes R_{n_0}^{\dagger}|\psi_{\text{N}(t)}\rangle \langle \psi_{\text{N}}(t)|R_{n_1}\\
    &+ |1\rangle \langle 1| \psi_e(t)\rangle \langle \psi_e(t)|0\rangle \langle 0|\otimes R_{n_1}^{\dagger}|\psi_{\text{N}(t)}\rangle \langle \psi_{\text{N}}(t)|R_{n_0}\\
    &+ |1\rangle \langle 1| \psi_e(t)\rangle \langle \psi_e(t)|1\rangle \langle 1|\otimes R_{n_1}^{\dagger}|\psi_{\text{N}(t)}\rangle \langle \psi_{\text{N}}(t)|R_{n_1}]. \\
\end{split}
\end{equation}
\noindent Taking the partial trace with respect to the electron leaves the terms along the diagonal of the electronic subspace:
\begin{equation}
    \begin{split}
    \rho_n(t) &= |\langle 0|\psi_e(t)\rangle |^2 \text{Tr}_{1...n-1}[R_{n_0}^{\dagger}|\psi_{\text{N}(t)}\rangle \langle \psi_{\text{N}}(t)|R_{n_0}] + |\langle 1|\psi_e(t)\rangle |^2 \text{Tr}_{1...n-1}[R_{n_1}^{\dagger}|\psi_{\text{N}(t)}\rangle \langle \psi_{\text{N}}(t)|R_{n_1}].\\
    \end{split}
\end{equation}
\noindent At this point, we label the constants $\langle 0|\psi_e(t)\rangle = c_0$ and $\langle 1|\psi_e(t)\rangle = c_1$. We also write the nuclear rotation operators in terms of the operators that act on the bipartitioned nucleus tensored with all the others, e.g., $R_{n_0} = R_{n_0}^{n}\otimes R_{n_0}^{\{n-1\}}$ (where $R_{n_0}^{\{n-1\}} = R_{n_0}^{1}\otimes...\otimes R_{n_0}^{n-1}$). The expression becomes
\begin{equation}
    \begin{split}
    \rho_n(t) &= |c_0|^2 \text{Tr}_{1...n-1}[(R_{n_0}^{n}\otimes R_{n_0}^{\{n-1\}})^{\dagger}|\psi_{\text{N}(t)}\rangle \langle \psi_{\text{N}}(t)|R_{n_0}^{n}\otimes R_{n_0}^{\{n-1\}}] \\
    &+ |c_1|^2 \text{Tr}_{1...n-1}[(R_{n_1}^{n}\otimes R_{n_1}^{\{n-1\}})^{\dagger}|\psi_{\text{N}(t)}\rangle \langle \psi_{\text{N}}(t)|R_{n_1}^{n}\otimes R_{n_1}^{\{n-1\}}].\\
    \end{split}
\end{equation}
\noindent Next we can take out the bipartitioned nuclear operators and wavefunction from the partial trace, and evaluate the trace terms. We also denote $|\psi_{\text{N}}(t)\rangle = |\psi_{n}(t)\rangle \otimes |\psi_{\{n-1\}}(t)\rangle$. This gives
\begin{equation}
    \begin{split}
    \rho_n(t) &= |c_0|^2 (R_{n_0}^{n})^{\dagger}|\psi_{n}(t)\rangle \langle \psi_{n}(t)|R_{n_0}^{n}\text{Tr}_{1...n-1}[(R_{n_0}^{\{n-1\}})^{\dagger}|\psi_{\{n-1\}}(t)\rangle \langle \psi_{\{n-1\}}(t)|R_{n_0}^{\{n-1\}}] \\
    &+ |c_1|^2 (R_{n_1}^{n})^{\dagger}|\psi_{n}(t)\rangle \langle \psi_{n}(t)|R_{n_1}^{n}\text{Tr}_{1...n-1}[(R_{n_1}^{\{n-1\}})^{\dagger}|\psi_{\{n-1\}}(t)\rangle \langle \psi_{\{n-1\}}(t)|R_{n_1}^{\{n-1\}}] \\
    &= |c_0|^2 (R_{n_0}^{n})^{\dagger}|\psi_{n}(t)\rangle \langle \psi_{n}(t)|R_{n_0}^{n} + |c_1|^2 (R_{n_1}^{n})^{\dagger}|\psi_{n}(t)\rangle \langle \psi_{n}(t)|R_{n_1}^{n}. \\
    \end{split}
\end{equation}
\noindent Next, we square $\rho_n(t)$ to find 
\begin{equation}
    \begin{split}
        (\rho_n(t))^2 &= |c_0|^4 (R_{n_0}^{n})^{\dagger}|\psi_n(t)\rangle \langle \psi_n (t)|R_{n_0}^{n} + |c_1|^4 (R_{n_1}^{n})^{\dagger}|\psi_n(t)\rangle \langle \psi_n (t)|R_{n_1}^{n}\\
        &+ |c_0|^2|c_1|^2 (R_{n_0}^{n})^{\dagger}|\psi_n(t)\rangle \langle \psi_n (t)|R_{n_0}^{n}(R_{n_1}^{n})^{\dagger}|\psi_n(t)\rangle \langle \psi_n (t)|R_{n_1}^{n}\\
        &+ |c_1|^2|c_0|^2 (R_{n_1}^{n})^{\dagger}|\psi_n(t)\rangle \langle \psi_n (t)|R_{n_1}^{n}(R_{n_0}^{n})^{\dagger}|\psi_n(t)\rangle \langle \psi_n (t)|R_{n_0}^{n}.\\
    \end{split}
\end{equation}
\noindent Taking the trace, we obtain
\begin{equation}
    \begin{split}
    \text{Tr}[(\rho_n(t))^2] &= |c_0|^4 + |c_1|^4 + 2|c_0|^2|c_1|^2 |\text{Tr}[\langle \psi_n (t)|(R_{n_0}^{n})^{\dagger}R_{n_1}^{n}|\psi_n(t)\rangle]|^2. \\
    \end{split}
\end{equation}
\noindent Solving for $c_0$ and $c_1$ based on the general expression for the electron's initial state $|\psi_{e}(0)\rangle = \cos\frac{\theta_{e}}{2} |0\rangle + e^{i \phi_{e}} \sin \frac{\theta_{e}}{2}|1\rangle$ gives $c_0 = \langle 0|\psi_{e}(0)\rangle = \cos\frac{\theta_e}{2}$, and $c_1 = \langle 1|\psi_{e}(0)\rangle = e^{i \phi_e}\sin\frac{\theta_e}{2}$. This leaves
\begin{equation}
    \begin{split}
    \text{Tr}[(\rho_n(t))^2] &= \Big(\cos\frac{\theta_e}{2}\Big)^4 + \Big(\sin\frac{\theta_e}{2}\Big)^4 + 2\Big(\cos \frac{\theta_e}{2}\Big)^2 \Big(\sin \frac{\theta_e}{2}\Big)^2 |\text{Tr}[\langle \psi_n (t)|(R_{n_0}^{n})^{\dagger}R_{n_1}^{n}|\psi_n(t)\rangle]|^2. \\
    \end{split}
    \label{eq:trtermsnuc}
\end{equation}
\noindent The one-tangling power is obtained by averaging over one minus this quantity. We average via
\begin{equation}
\begin{split}
    \epsilon^{\text{nuclear}} &= \langle\langle 1 - \text{Tr}[(\rho_n(t))^2]\rangle\rangle \\
    &= \frac{3}{4\pi}\frac{1}{V_{\text{N}}}\int_e\int_{\text{N}}dV (1 - \text{Tr}[(\rho_n(t))^2]) \\
    &= 1 - \frac{3}{4\pi}\frac{1}{V_{\text{N}}}\int_e\int_{\text{N}}dV \text{Tr}[(\rho_n(t)^2]).
\end{split}
\label{eq:nuclearintegral}
\end{equation}
\noindent Integrating over the first term in Eq.\ (\ref{eq:trtermsnuc}) yields
\begin{equation}
\begin{split}
    \frac{3}{4\pi}\frac{1}{V_\text{N}}\int_0^1\int_0^{\pi}\int_{0}^{2\pi} r_e^2\sin\theta_e dr_e d\theta_e d\phi_e \Big(\cos\frac{\theta_e}{2}\Big)^4 \int_{\text{N}} dV = \frac{1}{3}. \\
\end{split}
\end{equation}
\noindent The same result applies to the second term as well. For the third term, we have:
\begin{equation}
\begin{split}
    &\frac{3}{4\pi}\frac{1}{V_\text{N}}\int_0^1\int_0^{\pi}\int_{0}^{2\pi} r_e^2 \sin\theta_e dr_e d\theta_e d\phi_e \Big(\cos\frac{\theta_e}{2}\Big)^2 \Big(\sin\frac{\theta_e}{2}\Big)^2  \int_{S^{2\text{N}-1}}dV |\text{Tr}[\langle \psi_n (t)|(R_{n_0}^{n})^{\dagger}R_{n_1}^{n}|\psi_n(t)\rangle]|^2\\
    &= \frac{1}{3}\frac{1}{V_{\text{N}}}\int_{S^{2\text{N}-1}}dV |\text{Tr}[\langle \psi_n (t)|(R_{n_0}^{n})^{\dagger}R_{n_1}^{n}|\psi_n(t)\rangle]|^2.
\end{split}
\end{equation}
\noindent This integral has already been solved by Pedersen, M\o{}ller, and M\o{}lmer in their derivation of the average fidelity of a unitary transformation \cite{pedersen2007fidelity}. With their result, we find (implicitly including the normalization):
\begin{equation}
    \int_{S^{2 d_n - 1}} dV \big|\text{Tr}[\langle \psi_{n}(0)|(R_{n_0}^{n})^{\dagger}R_{n_1}^{n}|\psi_{n}(0)\rangle]\big|^2 = \frac{1}{d_n(d_n + 1)}\big(d_n + |\text{Tr}[(R_{n_0}^{n})^{\dagger}R_{n_1}^{n}]|^2] \big),
\end{equation}
\noindent where we identify the last term as the generalized Makhlin invariant corresponding to each electron-nuclear interaction, which takes the form: $|\text{Tr}[(R_{n_0}^{n})^{\dagger}R_{n_1}^{n}]|^2 = d_n^{2}G_1^{(n)}$. Plugging this result and the results for the previous terms in the nuclear one-tangling power expression, Eq.\ (\ref{eq:nuclearintegral}) leads to 
\begin{equation}
\begin{split}
    \epsilon^{\text{nuclear},n} &= 1 - \frac{2}{3} - \frac{1}{3}\frac{1}{d_n(d_n + 1)}\big(d_n + d_n^2G_1^{(n)} \big)\\
    &= \frac{1}{3}\frac{d_n}{d_n + 1} (1 - G_1^{(n)}),
\end{split}
\end{equation}
\noindent Note that this applies to any of the $i$ nuclei in the ensemble and can be written in general as
\begin{equation}
    \epsilon^{\text{nuclear},i} = \frac{1}{3}\frac{d_i}{d_i + 1} (1 - G_1^{(i)}),
\end{equation}
\noindent which agrees with the previously derived expression, Eq.\ (\ref{eq:nuclearappendix}).
\subsection{Electronic one-tangling power}
\label{appendix:simplerderivationelec}
In this section, we are interested in computing $1 - \text{Tr}[\rho_{e}(t)^2]$ in order to derive the one-tangling power with respect to an electron coupled to a surrounding nuclear ensemble. So if we trace out all nuclei, we have $\rho_e(t) = \text{Tr}_{\mathrm{N}}\big[U^{\dagger}\rho(t)U\big]$. We again assume the block-diagonal form of $U = |0\rangle \langle 0|\otimes R_{n_0} + |1\rangle \langle 1|\otimes R_{n_1}$. This leads to 
\begin{equation}
\begin{split}
        \rho_e(t) &= \text{Tr}_{\mathrm{N}}\big[(|0\rangle \langle 0|\otimes R_{n_0}^{\dagger} + |1\rangle \langle 1|\otimes R_{n_1}^{\dagger}) \rho(t)(|0\rangle \langle 0|\otimes R_{n_0} + |1\rangle \langle 1|\otimes R_{n_1}) \big]\\
        &= \text{Tr}_{\mathrm{N}}\big[(|0\rangle \langle 0|\otimes R_{n_0}^{\dagger} + |1\rangle \langle 1|\otimes R_{n_1}^{\dagger})(|\psi_e(0)\rangle \langle \psi_e(0)|\otimes |\psi_N(0)\rangle \langle \psi_N(0)|)(|0\rangle \langle 0|\otimes R_{n_0} + |1\rangle \langle 1|\otimes R_{n_1}) \big]\\
        &= \text{Tr}_{\mathrm{N}}\big[|0\rangle \langle 0|\psi_e(0)\rangle \langle \psi_e(0)|0\rangle \langle 0|\otimes R_{n_0}|\psi_N(0)\rangle \langle \psi_N(0)|R_{n_0}^{\dagger} \\
        &+ |0\rangle \langle 0|\psi_e(0)\rangle \langle \psi_e(0)|1\rangle \langle 1|\otimes  R_{n_0}|\psi_N(0)\rangle \langle \psi_N(0)|R_{n_1}^{\dagger}\\
        &+ |1\rangle \langle 1|\psi_e(0)\rangle \langle \psi_e(0)|0\rangle \langle 0|\otimes R_{n_1} |\psi_N(0)\rangle \langle \psi_N(0)|R_{n_0}^{\dagger} \\
        &+ |1\rangle \langle 1|\psi_e(0)\rangle \langle \psi_e(0)|1\rangle \langle 1|\otimes R_{n_1}|\psi_N(0)\rangle \langle \psi_N(0)|R_{n_1}^{\dagger}\big]\\
        &= |0\rangle \langle 0|\psi_e(0)\rangle \langle \psi_e(0)|0\rangle \langle 0|\text{Tr}_{\mathrm{N}}\big[R_{n_0}|\psi_N(0)\rangle \langle \psi_N(0)|R_{n_0}^{\dagger}\big] \\
        &+ |0\rangle \langle 0|\psi_e(0)\rangle \langle \psi_e(0)|1\rangle \langle 1|\text{Tr}_{\mathrm{N}}\big[  R_{n_0}|\psi_N(0)\rangle \langle \psi_N(0)|R_{n_1}^{\dagger}\big]\\
        &+ |1\rangle \langle 1|\psi_e(0)\rangle \langle \psi_e(0)|0\rangle \langle 0|\text{Tr}_{\mathrm{N}}\big[ R_{n_1} |\psi_N(0)\rangle \langle \psi_N(0)|R_{n_0}^{\dagger}\big] \\
        &+ |1\rangle \langle 1|\psi_e(0)\rangle \langle \psi_e(0)|1\rangle \langle 1|\text{Tr}_{\mathrm{N}}\big[R_{n_1}|\psi_N(0)\rangle \langle \psi_N(0)|R_{n_1}^{\dagger}\big]. \\
\end{split}
\end{equation}
\noindent In the lines above, we have made use of the fact that the trace is only taken over the nuclear part of the initial state. Next, we can use the cyclic property of the trace and the unitarity of the nuclear rotations, $R_{n_0}^{\dagger}R_{n_0} = \mathbb{I}$ to simplify further and obtain
\begin{equation}
    \begin{split}
        \rho_e(t) &= |0\rangle \langle 0|\psi_e(0)\rangle \langle \psi_e(0)|0\rangle \langle 0|\text{Tr}_{\mathrm{N}}\big[|\psi_N(0)\rangle \langle \psi_N(0)|\big] + |0\rangle \langle 0|\psi_e(0)\rangle \langle \psi_e(0)|1\rangle \langle 1|\text{Tr}_{\mathrm{N}}\big[|\psi_N(0)\rangle \langle \psi_N(0)|R_{n_1}^{\dagger}R_{n_0} \big]\\
        &+ |1\rangle \langle 1|\psi_e(0)\rangle \langle \psi_e(0)|0\rangle \langle 0|\text{Tr}_{\mathrm{N}}\big[|\psi_N(0)\rangle \langle \psi_N(0)|R_{n_0}^{\dagger}R_{n_1}\big] + |1\rangle \langle 1|\psi_e(0)\rangle \langle \psi_e(0)|1\rangle \langle 1|\text{Tr}_{\mathrm{N}}\big[|\psi_N(0)\rangle \langle \psi_N(0)|\big] \\
        &= |0\rangle \langle 0|\psi_e(0)\rangle \langle \psi_e(0)|0\rangle \langle 0|\text{Tr}_{\text{N}}[\langle \psi_N(0)|\psi_N(0)\rangle] + |0\rangle \langle 0|\psi_e(0)\rangle \langle \psi_e(0)|1\rangle \langle 1|\text{Tr}_{\mathrm{N}}\big[|\psi_N(0)\rangle \langle \psi_N(0)|R_{n_1}^{\dagger}R_{n_0}\big]\\
        &+ |1\rangle \langle 1|\psi_e(0)\rangle \langle \psi_e(0)|0\rangle \langle 0|\text{Tr}_{\mathrm{N}}\big[|\psi_N(0)\rangle \langle \psi_N(0)|R_{n_0}^{\dagger} R_{n_1} \big] + |1\rangle \langle 1|\psi_e(0)\rangle \langle \psi_e(0)|1\rangle \langle 1|\text{Tr}_{\mathrm{N}}\big[\langle \psi_N(0)|\psi_N(0)\rangle \big]. \\
    \end{split}
\end{equation}
\noindent We know that $|\psi_{\text{N}}\rangle$ is normalized, and so $\text{Tr}_{\mathrm{N}}\big[\langle \psi_N(0)|\psi_N(0)\rangle] = 1$. We can also once again label the constants $c_0 = \langle 0|\psi_e(0)\rangle$ and $c_1 = \langle 1|\psi_e(0)\rangle$. With this, we are left with
\begin{equation}
    \begin{split}
        \rho_e(t) &= |c_0|^2|0\rangle \langle 0| + |c_1|^2|1\rangle \langle 1| \\
        &+ c_0c_1^* |0\rangle \langle 1|\cdot \text{Tr}_{\mathrm{N}}\big[ \langle \psi_N(0)|R_{n_1}^{\dagger}R_{n_0}|\psi_N(0)\rangle\big] \\
        &+ c_1c_0^*|1\rangle \langle 0| \cdot \text{Tr}_{\mathrm{N}}\big[ \langle \psi_N(0)|R_{n_0}^{\dagger} R_{n_1} |\psi_N(0)\rangle\big]. \\
    \end{split}
\end{equation}
Now we can square $\rho_e(t)$ to obtain
\begin{equation}
    \begin{split}
        (\rho_e(t))^2 &= |c_0|^4|0\rangle \langle 0| + |c_1|^4|1\rangle \langle 1| \\
        &+ (|c_0|^2 + |c_1|^2)c_0c_1|0\rangle \langle 1| \text{Tr}[\langle \psi_{\text{N}}(0)|R_{n_1}^{\dagger}R_{n_0} |\psi_{\text{N}}(0)\rangle] \\
        &+ (|c_0|^2 + |c_1|^2)c_1c_0^*|1\rangle \langle 0|\text{Tr}[\langle \psi_{\text{N}}(0)| R_{n_0}^{\dagger}R_{n_1}|\psi_{\text{N}}(0)\rangle ]\\
        &+ |c_0|^2|c_1|^2|1\rangle \langle 1| \big|\text{Tr}[\langle \psi_{\text{N}}(0)|R_{n_0}^{\dagger}R_{n_1}|\psi_{\text{N}}(0)\rangle ]\big|^2.\\
        \label{eq:rhoesq}
    \end{split}
\end{equation}
\noindent Taking the trace of $(\rho_e(t))^2$ gives
\begin{equation}
    \begin{split}
        \text{Tr}[(\rho_e(t))^2] &=  |c_0|^4 + |c_1|^4| + |c_0|^2|c_1|^2 \big|\text{Tr}[\langle \psi_{\text{N}}(0)|R_{n_0}^{\dagger}R_{n_1}|\psi_{\text{N}}(0)\rangle]\big|^2. \\
        \label{eq:trrhoesq}
    \end{split}
\end{equation}
\noindent We can again plug in the general expression for the electron's initial state $|\psi_{e}(0)\rangle = \cos\frac{\theta_{e}}{2} |0\rangle + e^{i \phi_{e}} \sin \frac{\theta_{e}}{2}|1\rangle $. This gives $c_0 = \langle 0|\psi_{e}(0)\rangle = \cos\frac{\theta_e}{2}$, and $c_1 = \langle 1|\psi_{e}(0)\rangle = e^{i \phi_e}\sin\frac{\theta_e}{2}$. Plugging the electronic coefficients into Eq.\ (\ref{eq:trrhoesq}) leads to
\begin{equation}
    \begin{split}
        \text{Tr}[(\rho_e(t))^2] &= \Big(\cos\frac{\theta_e}{2}\Big)^4 + \Big(\sin\frac{\theta_e}{2}\Big)^4 + \Big(\cos\frac{\theta_e}{2}\Big)^2\Big(\sin\frac{\theta_e}{2}\Big)^2 \big|\text{Tr}[\langle \psi_{\text{N}}(0)|R_{n_0}^{\dagger}R_{n_1}|\psi_{\text{N}}(0)\rangle]\big|^2. \\
        \label{eq:trrhosquaredintegrationterms}
    \end{split}
\end{equation}
\noindent Now we can average over one minus this quantity to solve for the electronic one-tangling power:
\begin{equation}
\begin{split}
    \epsilon_{e|\text{N}} &= \langle\langle 1 - \text{Tr}[(\rho_e(t))^2]\rangle\rangle \\
    &= \frac{3}{4\pi}\frac{1}{V_{\text{N}}}\int_e\int_{\text{N}}dV (1 - \text{Tr}[(\rho_e(t))^2]) \\
    &= 1 - \frac{3}{4\pi}\frac{1}{V_{\text{N}}}\int_e\int_{\text{N}}dV \text{Tr}[(\rho_e(t)^2]),
\end{split}
\label{eq:epsilonintegral}
\end{equation}
\noindent where $V_{\text{N}}$ is the product of the volumes of each $d_i$-dimensional sphere describing the nuclei in the ensemble: $V_{\text{N}} = \prod_{i = 1}^n V_{d_i} = \frac{\pi^{d_i/2}R^{d_i}}{\Gamma(1 + d_i/2)}$ (assuming $R =1$). If we plug Eq.\ (\ref{eq:trrhosquaredintegrationterms}) into the expression above, and examine the first term in the integral only, we find 
\begin{equation}
    \begin{split}
        \frac{3}{4\pi}\frac{1}{V_{\text{N}}}\int_{e}\int_{\text{N}} dV \Big(\cos\frac{\theta_e}{2}\Big)^4 &= \frac{3}{4\pi}\int_0^1 \int_0^{\pi}\int_0^{2\pi} r_e^2 \sin \theta_e dr_e d\theta_e d\phi_e \Big(\cos\frac{\theta_e}{2}\Big)^4 \frac{1}{V_{\text{N}}}\int_{\text{N}} dV_{\text{N}} \\
        &= \frac{3}{4\pi}\cdot \frac{2 \pi}{3} \int_0^{\pi}\Big(\cos\frac{\theta_e}{2}\Big)^4 \sin\theta_e d\theta_e \\
        &= \frac{1}{3}. \\
        \label{eq:firstepsilonterm}
    \end{split}
\end{equation}
\noindent Averaging over the second term in Eq.\ (\ref{eq:trrhosquaredintegrationterms}) yields the same result. As for the third and only nontrivial term, we have:
\begin{equation}
        \frac{3}{4\pi}\frac{1}{V_{\text{N}}}\int_{e}\int_{\text{N}} dV \Big(\cos\frac{\theta_e}{2}\Big)^2\Big(\sin\frac{\theta_e}{2}\Big)^2 \big|\text{Tr}[\langle \psi_{\text{N}}(0)|R_{n_0}^{\dagger}R_{n_1}|\psi_{\text{N}}(0)\rangle]\big|^2 = \frac{1}{3} \frac{1}{V_{\text{N}}}\int_N dV \big|\text{Tr}[\langle \psi_{\text{N}}(0)|R_{n_0}^{\dagger}R_{n_1}|\psi_{\text{N}}(0)\rangle]\big|^2. \\
    \label{eq:inteop}
\end{equation}
\noindent We once again refer to \cite{pedersen2007fidelity} for the solution to this integral. In order to employ this result, we first note that the expression above can be written in terms of products nuclear rotation operators acting on each individual nuclear wavefunction: $\big|\text{Tr}[\langle \psi_{\text{N}}(0)|R_{n_0}^{\dagger}R_{n_1}|\psi_{\text{N}}(0)\rangle]\big|^2 = \prod_{i = 1}^n \big|\text{Tr}[\langle \psi_{i}(0)|(R_{n_0}^{(i)})^{\dagger}R_{n_1}^{(i)}|\psi_{i}(0)\rangle]\big|^2.$ From here, we find (implicitly including the normalization):
\begin{equation}
\begin{split}
    \prod_{i = 1}^n \int_{S^{2 d_i - 1}} dV \big|\text{Tr}[\langle \psi_{i}(0)|(R_{n_0}^{(i)})^{\dagger}R_{n_1}^{(i)}|\psi_{i}(0)\rangle]\big|^2 = \prod_{i = 1}^n \frac{1}{d_i(d_i + 1)}\big(d_i + |\text{Tr}[R_{n_0}^{\dagger}R_{n_1}]|^2] \big),
\end{split}
\end{equation}
\noindent Again we recognize the appearance of the generalized Makhlin invariants: $|\text{Tr}[(R_{n_0}^{(i)})^{\dagger}R_{n_1}^{(i)}]|^2 = d_i^{2}G_1^{(i)}$. Plugging this along with the results from Eqs.\ (\ref{eq:firstepsilonterm}) and (\ref{eq:inteop}) into Eq.\ (\ref{eq:epsilonintegral}) allows us to recover the generalized electronic one-tangling power expression as shown in Eq.\ (\ref{eq:electronicappendix}):
\begin{equation}
\begin{split}
    \epsilon^{\text{electronic}} &=  1 - \Big[\frac{1}{3} + \frac{1}{3} + \frac{1}{3}\prod_{i = 1}^n \frac{1}{d_i(d_i + 1)} (d_i + d_i^2 G_1^{(i)}) \Big] \\
    &= \frac{1}{3} - \frac{1}{3}\prod_{i = 1}^n \frac{1}{d_i + 1}(1 + d_i G_1^{(i)}).
\end{split}
\end{equation}
\end{widetext}

\bibliography{references}% Produces the bibliography via BibTeX.

\end{document}